\documentclass[prd,twocolumn,preprintnumbers,showpacs,floatfix%superscriptaddress
]{revtex4}
\usepackage[colorlinks=true,linkcolor=blue,citecolor=blue,urlcolor=blue]{hyperref}
\usepackage{gensymb,comment}
\usepackage{amsmath,amssymb,epsfig,bm,mathrsfs,color}
\usepackage{slashed}

\newcommand{\be}{\begin{equation}}
\newcommand{\ee}{\end{equation}}
\newcommand{\bea}{\begin{eqnarray}}
\newcommand{\eea}{\end{eqnarray}}
\newcommand{\half}{\frac1 2}

\newcommand{\ep}{\epsilon}

\newcommand{\vecq}{{\bm q}}
\newcommand{\vecp}{{\bm p}}

\newcommand{\veck}{{\bm k}}

\newcommand{\vectau}{{\bm \tau}}
\newcommand{\vecrho}{{\bm \rho}}

\newcommand{\ie}{{\it i.e.}}
\newcommand{\eg}{{\it e.g.}}

\newcommand{\MeV}{{\rm MeV}}
\newcommand{\aap}{Astron.\& Astrophys.}
\newcommand{\mnras}{Mon. Not. RAS}
\newcommand{\apjl}{ApJ Lett.}

\usepackage[normalem]{ulem}  % \sout{old text} for strikeout  

\begin{document}

\title{ Bulk viscosity of baryonic matter with trapped neutrinos}

\author{Mark Alford} \email{alford@wuphys.wustl.edu}
\affiliation{Department of Physics, Washington University, St.~Louis,
  Missouri 63130, USA}

\author{ Arus Harutyunyan} \email{arus@bao.sci.am}
\affiliation{Byurakan Astrophysical Observatory,
  Byurakan 0213, Armenia\\
  Yerevan State University, Alek Manukyan 1, Yerevan 0025, Armenia}

\author{ Armen Sedrakian}
\email{sedrakian@fias.uni-frankfurt.de}
\affiliation{Frankfurt Institute for Advanced Studies, D-60438
  Frankfurt am Main, Germany\\
 Institute of Theoretical Physics, University of Wroclaw,
50-204 Wroclaw, Poland}

\date{22 November 2019} 

\begin{abstract}
  We study bulk viscosity arising from weak current Urca processes in
  dense baryonic matter at and beyond nuclear saturation density.  We
  consider the temperature regime where neutrinos are trapped and
  therefore have nonzero chemical potential.  We model the nuclear
  matter in a relativistic density functional approach, taking into
  account the trapped neutrino component. We find that the resonant
  maximum of the bulk viscosity would occur at or below the neutrino
  trapping temperature, so in the neutrino trapped regime the bulk
  viscosity decreases with temperature as $T^{-2}$, this decrease
  being interrupted by a drop to zero at a special temperature where
  the proton fraction becomes density-independent and the material
  scale-invariant.  The bulk viscosity is larger for matter with lower
  lepton fraction, i.e., larger isospin asymmetry. We find that bulk
  viscosity in the neutrino-trapped regime is smaller by several
  orders of magnitude than in the neutrino-transparent regime, which
  implies that bulk viscosity in neutrino-trapped matter is probably
  not strong enough to affect significantly the evolution of neutron
  star mergers. This also implies weak damping of gravitational waves emitted by the oscillations of the postmerger remnant in the high-temperature, neutrino-trapped phase of evolution.

\end{abstract}
%\pacs{21.65.+f, 21.30.Fe, 26.60.+c}

\maketitle

\section{Introduction}
\label{sec:intro}

The recent detection of gravitational waves by the LIGO-Virgo
collaboration, in coincidence with electromagnetic counterparts, has brought
into focus the study of the physics of binary neutron star
mergers~\cite{Abbott2017}. In these events, a postmerger object is
formed which either evolves into a stable neutron star or collapses to
a black hole, once it cannot be supported by the differential
rotation.  As seen in numerical simulations
\cite{Perego:2019adq,Hanauske:2019qgs,Hanauske:2017oxo,Kastaun:2016elu,Bernuzzi:2015opx,Foucart:2015gaa,kiuchi:2012mk,sekiguchi:2011zd,Ruiz:2016,East:2016,Most2019,Bauswein2019,Baiotti:2016qnr,Faber2012:lrr}
{there are} significant {density} oscillations in the postmerger
remnant, which can generate observable gravitational waves.

These oscillations will be damped eventually by dissipative processes
on characteristic secular timescales controlled by thermodynamics of
background matter and the kinetics of the relevant dissipative
process. Bulk viscosity is known to be one of the dissipative
processes that could efficiently damp certain classes of oscillations
of general relativistic equilibria.

Studies of damping mechanisms in the context of binary neutron star
mergers are still at an embryonic stage. Reference~\cite{Alford2017} has
suggested that modified Urca processes can produce significant bulk
viscous dissipation on timescales of order a few milliseconds, i.e.,
on timescales that are relevant to neutron star merger and
postmerger evolution. This might then affect the emitted
gravitational signal. It remains less clear whether shear viscosity
and thermal conduction could play a significant
role~\cite{Alford2017}. In parallel, the electrical conductivity was
computed in the relevant regime to assess its impact on the evolution
of the electromagnetic fields. It was shown that the Hall effect could
be important on characteristic timescales of the merger and
postmerger evolution~\cite{Harutyunyan:2016a,Harutyunyan:2018}.

To quantify the amount of bulk viscous dissipation a more detailed
analysis is required that takes into account the realistic temperature
and density conditions encountered in this context.  The aim of this
work is to obtain the bulk viscosity of dense matter created in
neutron star mergers in the temperature and density regime
characteristic for such events. Specifically, we will consider the
dominant weak-interaction processes of Urca type at temperatures that
are above the trapping temperature $T_{\rm tr}\simeq 5$ MeV
\cite{Roberts:2012um,Alford:2018lhf}. In this regime, the
  neutrinos have a mean free path that is significantly shorter than
the stellar size, and consequently a nonzero chemical potential.
This affects the composition of the background baryonic matter. Thus,
compared to the extensively studied case of cold neutron
  stars, the key new features that arise in the neutron-star merger
and postmerger context is the higher temperatures at which the weak
reactions take place and the significantly different background
composition of baryonic matter.

The bulk viscosity of baryonic matter has been studied extensively in
the low-temperature limit for purely nucleonic
matter~\cite{Sawyer1979ApJ,Sawyer1980ApJ,Sawyer1989,Haensel1992PhRvD,Dong2007,Alford2010JPhG,Kolomeitsev2015},
including possible leptonic contributions \cite{Alford:2010jf}, for
hyperonic
matter~\cite{Jones2001PhRvD,Lindblom2002,Dalen2004PhRvC,Haensel2002,Chatterjee2006,Chatterjee2008ApJ},
and for quark matter
\cite{Madsen1992,Drago2005,Alford:2006gy,Alford:2007rw,Manuel:2007pz,SaD2007a,SaD2007b,Alford:2008pb,Sinha2009,Wang2010PhRvD,Huang2010}
including the effects of an interface with the nuclear matter envelope
\cite{AlfordHan2015}. For a review see \cite{Schmitt2017}. At temperatures $T\lesssim 1$\,MeV the physics of
bulk viscosity is affected by the superfluidity of
baryons~\cite{Haensel2000,Haensel2001,Gusakov2007}, but we will assume
that the temperatures are always larger than the critical temperatures
for the pairing of various baryons, as our focus is on the regime
where neutrinos are trapped, i.e., temperatures $T\ge T_{\rm tr} $.\
{The neutrino transparent case is discussed in Ref.~\cite{Alford:2019zzz} and Appendix~\ref{app:bulk_visc0}.}

Physical conditions that are similar to those we are aiming to study
are encountered also in proto-neutron stars born in supernova
explosions. In this case, the matter is much more isospin symmetrical,
but the rest of the physics is quite analogous. We will cover
this case as well having in mind the possibility of observations of
oscillations of a proto-neutron star, should a supernova explosion
occur within the detectable range.

In the present study we will assume that thermal conduction is
efficient enough to keep matter isothermal while it is being
compressed and uncompressed. A rough estimate (Ref.~\cite{Alford2017},
Eq.~(2)) gives the timescale for relaxation of thermal gradients order
of 1 sec $\times (z_{\rm typ}/{\rm km})^2 (T/10\, {\rm MeV})^2$, where
$z_{\rm typ}$ is the typical scale of thermal gradients. Thus, for
thermal gradients on a distance scale of about 30m or less, the
relaxation time would be 1~ms or less, i.e., the characteristic
time-scale of binary neutron star merger.  So, for density
fluctuations on this distance scale the assumption of isothermal
matter is the relevant one.  In particular, if turbulent flow arises
in the merger then this could give flows and density variations on
such distance scale. On the scales over which thermal conduction is
inefficient, the matter should be treated as isoentropic. The
formalism presented below can be simply adapted to this case. We
anticipate that for an adiabatic calculation the bulk viscosity will
be of the order as found here. We will return to this problem in a
separate work.

This work is organized as follows. In Sec.~\ref{sec:urca_rates} we
discuss the rates of the two relevant
processes. Section~\ref{sec:bulk} derives the corresponding formulas
for the bulk viscosity of matter.  In Sec.~\ref{sec:num_results} we
first describe the properties of the background matter derived on the
basis of the density functional theory at a finite temperature which
accounts for a neutrino component with nonzero chemical potential
(Sec.~\ref{sec:DFT}).  This is followed by a discussion of
perturbed quantities and bulk viscosity in
Sec.~\ref{sec:bulk_visc}. Our conclusions are collected in
Sec.~\ref{sec:conclusions}.

In this work we use the natural (Gaussian) units with 
$\hbar = c = k_B = 1$,   %$e^2 = \alpha = 1/137$
and the metric $g_{\mu\nu} = \textrm{diag}(1, -1, -1, -1)$.

\begin{widetext}
\section{Urca process rates}
\label{sec:urca_rates}

We will consider the simplest composition of baryonic matter
consisting of neutrons ($n$), protons ($p$), electrons ($e$), muons
($\mu$), and neutrinos at densities in the range $0.5n_0$ to $3 n_0$
($n_0 \simeq 0.16$ fm$^{-3}$) and temperatures in the range $T_{\rm
  tr}\simeq 5$ to 50\,MeV. Other constituents and forms of matter have
been proposed, but we will focus on the standard scenario for this
regime, which can serve as a starting point for future explorations of
more complex phases of baryonic matter. Note that positrons do not
appear in matter in substantial amounts because the electron chemical
potential is of the order of 100~MeV, see Sec. \ref{sec:DFT}. The weak
processes {involving} positrons will be suppressed roughly by a factor
$\exp(-\mu_e/T) \simeq 0.1$ at $T=50$ and 0.01 at $T=30$ MeV.

In the dynamically evolving environment of a neutron star merger,
fluid elements undergo rhythmic cycles of compression and
decompression, which can lead to bulk viscous dissipation if the rate
at which the proton fraction relaxes toward its equilibrium value
(``beta equilibrium'') is comparable to the frequency of the
compression cycles.  To analyze this, we consider the simplest beta
equilibration processes, the Urca processes:
%------------------------------------------------------
\bea\label{eq:Urca_1}
n\rightleftarrows p + e^-+\bar{\nu}_e,\\
\label{eq:Urca_2}
p + e^- \rightleftarrows n+\nu_e.
\eea
%------------------------------------------------------
The first process is the $\beta$-decay of a neutron and the second
one is the electron capture on a proton.
 If the matter is in
$\beta$-equilibrium, then the chemical potentials of particles obey the relation 
%------------------------------------------------------
\bea\label{eq:eq_condition1}
\mu_n=\mu_p+\mu_e + \mu_{\bar\nu}, \\
\label{eq:eq_condition2}
\mu_p+\mu_e = \mu_n+\mu_{\nu},
\eea
%------------------------------------------------------
where the neutrino and antineutrino chemical potentials are related by
$\mu_{\bar\nu} = -\mu_{\nu}$, which leaves us with a single 
relation. As noted above, the matter
  can be driven out of $\beta$-equilibrium by a cycle of 
  compression and rarefaction, and this can be characterized via a
  nonzero value of the chemical potential that measures the deviation
  from $\beta$-equilibrium
%------------------------------------------------------
\be
\mu_\Delta\equiv\mu_n+\mu_{\nu}-\mu_p-\mu_e.
\label{eq:mu-Delta}
\ee
%------------------------------------------------------
The rate at which $\mu_\Delta$ relaxes to zero is a measure of
the speed at which the chemical constitution of the matter
adjusts to a change in pressure. We start with the computation 
of the $\beta$-equilibration rate assuming a given value of
$\mu_\Delta\neq 0$.
 
%------------------------------------------------------
 The squared matrix element of processes \eqref{eq:Urca_1} and 
\eqref{eq:Urca_2} is given by the well-known expression~\cite{Greiner2000gauge}
%-----------------------------------------------------------
\be\label{eq:matrix_el_full}
\sum \vert {\cal M}_{Urca}\vert^2 = 32 \tilde G^2 (k\cdot p') (p\cdot k'),
\ee 
%-----------------------------------------------------------
where $p'$ and $p$ refer to the four-momenta of the neutron and
proton, $k$ and $k'$ to the four-momenta of neutrino (antineutrino)
and electron, respectively, and the sum includes summation over the
spins of neutron, proton, and electron.  Note that each of baryon
four-momenta is dotted into a four-lepton momentum.  We consider only
the Standard Model neutrinos (antineutrinos) which are left-handed
(right-handed) only, therefore they have only one projection of
helicity that has to be counted. In the following, we will use the
nonrelativistic limit of the matrix element \eqref{eq:matrix_el_full}
because in the temperature and density range that we consider the
baryons are nonrelativistic to an accuracy of about 10\%.

Thus we keep only the contribution of timelike parts of the scalar
 products in the matrix element,
%-----------------------------------------------------------
\be\label{eq:matrix_el}
 \vert {\cal M}_{Urca}\vert^2 = 32\tilde G^2p_0p'_0k_0k'_0,
\ee
%-----------------------------------------------------------
where index 0 refers to the timelike component of a four-vector,
$\tilde{G}^2\equiv G_F^2\cos^2 \theta_c (1+3g_A^2)$, where
$G_F=1.166\cdot 10 ^{-5}$ GeV$^{-2}$ is the Fermi coupling constant,
$\theta_c$ is the Cabibbo angle ($\cos\theta_c=0.974$) and $g_A=1.26$
is the axial-vector coupling constant.

\subsection{The rates of the processes $n \rightleftarrows p + e^-+\bar{\nu}_e$}
\label{sec:urca1} 
 
The $\beta$-equilibration rate for the neutron decay
$n\rightarrow p + e^-+\bar{\nu}_e$ is given by 
%------------------------------------------------------
\bea \label{eq:Gamma1p_def}
\Gamma_{1p} (\mu_\Delta) &=& 
\int\!\! \frac{d^3p'}{(2\pi)^32p'_0} \int\!\!
\frac{d^3p}{(2\pi)^32p_0} \int\!\! \frac{d^3k'}{(2\pi)^32k'_0}
\int\!\! \frac{d^3k}{(2\pi)^32k_0}\sum \vert {\cal M}_{Urca}\vert^2 \nonumber\\
& \times &f(p')
 [1-f(k')][1-f(k)][1-f(p)] (2\pi)^4\delta^{(4)}(p+k+k'-p'),
\eea
%------------------------------------------------------
where $f(p) = \{\exp[(E_p-\mu)]+1\}^{-1}$ etc.  are the Fermi
distributions of particles, with energies $E_p$ and chemical potential
$\mu$.  Similarly, the rate of the inverse process, \ie,
$p + e^-+\bar{\nu}_e\rightarrow n$ is given by
%------------------------------------------------------
\bea \label{eq:Gamma1n_def}
\Gamma_{1n} (\mu_\Delta) &=& 
\int\!\! \frac{d^3p'}{(2\pi)^32p'_0} \int\!\!
\frac{d^3p}{(2\pi)^32p_0} \int\!\! \frac{d^3k'}{(2\pi)^32k'_0}
\int\!\! \frac{d^3k}{(2\pi)^32k_0} \sum \vert {\cal M}_{Urca}\vert^2 \nonumber\\
& \times &
f(k')f(k)f(p)[1-f(p')] (2\pi)^4\delta^{(4)}(p+k+k'-p').
\eea
%------------------------------------------------------
Some of the phase-space integrals in Eqs.~\eqref{eq:Gamma1p_def} and
\eqref{eq:Gamma1n_def} can be carried out analytically; the details
are relegated to Appendix~\ref{app:rates}. We find
%------------------------------------------------------
\bea \label{eq:Gamma1p_final}
\Gamma_{1p}(\mu_\Delta) &=&
 \tilde{G}^2 \frac{m^{*2}T^6}{8\pi^5}\int_{-\alpha_e+\alpha_{\nu}}^{\infty} 
 dy\ g\left(y-\mu_\Delta/T\right)\int_0^{y+\alpha_e-\alpha_{\nu}} dz
  \ln \Bigg\vert \frac{1+\exp(-y_0)} {1+\exp\left(-y_0-y+\mu_\Delta/T\right)}\Bigg\vert\nonumber\\
 & \times& \int_{x_{\rm min}}^{x_{\rm max}}  
 dx (x-\alpha_{\nu}) (y+\alpha_e-x)f(x-y)[1-f(x)],\\
%------------------------------------------------------
 \label{eq:Gamma1n_final}
\Gamma_{1n}(\mu_\Delta) &=&
 \tilde{G}^2 \frac{m^{*2}T^6}{8\pi^5}\int_{-\alpha_e+\alpha_{\nu}}^{\infty} 
 dy\ \left[1+g\left(y-\mu_\Delta/T\right)\right]
 \int_0^{y+\alpha_e-\alpha_{\nu}} dz
  \ln \Bigg\vert \frac{1+\exp(-y_0)} {1+\exp(-y_0-y+\mu_\Delta/T)}\Bigg\vert\nonumber\\
 & \times & \int_{x_{\rm min}}^{x_{\rm max}}  
 dx (x-\alpha_{\nu}) (y+\alpha_e-x)f(x)[1-f(x-y)],
\eea 
%------------------------------------------------------
where $\alpha_j\equiv \mu_j^*/T$ and index $j=n,p,e,\nu$ labels the
constituents of matter with $\mu_j^*$ being the effective chemical
potentials of particles (see Sec.~\ref{sec:DFT}), $m^\ast$ stands for the
{\it nonrelativistic}
effective mass of a nucleon~\footnote{Here we do not distinguish the
  neutron and proton effective masses, but a straightforward
  generalization would give $m^{*2} \to m_n^* m_p^*$, where the indices $n$
  and $p$ refer to neutrons and protons. }, the Fermi and Bose
functions of nondimensional arguments have the form
$f(x)=(e^x+1)^{-1}$ and $g(x)=(e^x-1)^{-1}$,
%------------------------------------------------------
\bea\label{eq:y_0_mu}
y_0 = \frac{m^*}{2Tz^2}\bigg(\alpha_n-\alpha_p+y
-z^2\frac{T}{2m^*}-\frac{\mu_\Delta}{T}\bigg)^2-\alpha_p,
\eea
%------------------------------------------------------
and the integration limits $x_{\rm min}$ and $x_{\rm max}$ are given
by $x_{\rm min/max}=(y+\alpha_e+\alpha_{\nu}\mp z)/2$.  The
integration variables $y$ and $z$ are the transferred energy and
momentum, respectively, normalized by the temperature, and the
variable $x$ is the normalized-by-temperature antineutrino energy
computed from the chemical potential, \ie,
$x=(\ep_{\bar\nu}+\mu_{\nu})/T$ (recall that antineutrino chemical
potential is $-\mu_\nu$).  Note that in our rate calculations we
numerically evaluate the full three-dimensional integral. We do not
use the Fermi surface approximation (assuming that all momenta lie
close to the Fermi momentum) because it is no longer valid at the
temperatures of interest to us.

%------------------------------------------------------ 

When the system is in beta equilibrium, $\mu_\Delta=0$ and the rates
of the direct and inverse processes are equal
$\Gamma_{1n}=\Gamma_{1p}\equiv \Gamma_{1}$ in exact
$\beta$-equilibrium. For small departures from $\beta$-equilibrium,
$\mu_\Delta \ll T$, only the terms that are linear in the departure
$\mu_\Delta$ are of interest; the coefficients of the expansion
involve the derivatives
%------------------------------------------------------
\bea \label{eq:lambda1}
\lambda_1 & \equiv & \frac{\partial\Gamma_{1p}(\mu_\Delta)}
{\partial\mu_\Delta}\bigg\vert_{\mu_\Delta=0}-
\frac{\partial\Gamma_{1n}(\mu_\Delta)}
{\partial\mu_\Delta}\bigg\vert_{\mu_\Delta=0}
=\frac{\Gamma_1}{T},
\eea
%------------------------------------------------------
where $\Gamma_1$ is the rate in $\beta$-equilibrium 
(as defined above) and is given explicitly by
%------------------------------------------------------
\bea \label{eq:Gamma_1_general}
\Gamma_1 &=&
 \frac{m^{*2} \tilde{G}^2}{8\pi^5} T^6
 \int_{-\alpha_e+\alpha_{\nu}}^\infty
 dy~g(y)\int_0^{y+\alpha_e-\alpha_{\nu}} dz
 \ln \bigg\vert \frac{1+\exp\left(-y_0\right)}{1+\exp\left(-y_0-y\right)}\bigg\vert
\nonumber\\ 
&\times&\int_{x_{\rm min}}^{x_{\rm max}}  dx (x-\alpha_{\nu})(y+\alpha_e-x)f(x-y)[1-f(x)].
\eea 
%----------------------------------------------------------
Note that if in some density-temperature range neutrinos are trapped
and are degenerate, \ie, $\mu_\nu\gg T$, then one can approximate
$\ep_\nu\simeq \mu_\nu$, $y\simeq 1$, therefore
$x-y\simeq 2\mu_\nu/T \gg 1$, and the electron Fermi function $f(x-y)$
in Eq.~\eqref{eq:Gamma_1_general} vanishes.  If one were to
extrapolate the neutrino-trapped rate to low temperature, one would
find that $\Gamma_1=\lambda_1=0$ in this limit.

In the case of neutrino-transparent matter, one should
drop the antineutrino distribution $f(x)$ and substitute $\mu_\nu=0$
 in the neutron-decay rate $\Gamma_{1p}$, while the rate of
the inverse process $\Gamma_{1n}$ vanishes. 
In this case, the $\lambda_1$ parameter is given by
%----------------------------------------------------------
\bea \label{eq:lambda_1_no_nu}
\lambda_1 &=& 
\frac{m^{*2} \tilde{G}^2 }{8\pi^5}T^5 \int_{-\alpha_e}^\infty dy~g(y) 
\int_0^{y+\alpha_e} dz~\bigg\{ 
 \ln  \bigg\vert \frac{1+\exp\left(-y_0\right)} 
{1+\exp\left(-y_0-y\right)}\bigg\vert
[1+g(y)]\nonumber\\
&&-  f(y_0+y)-\left[  f(y_0+y)  - f(y_0) \right]
\frac{m^*}{z^2T}\bigg(\alpha_n-\alpha_p+y
-z^2\frac{T}{2m^*}\bigg) \bigg\}\nonumber\\
&&\times \int_{x_{\rm min}}^{x_{\rm max}} 
 dx~x(y+\alpha_e-x)f(x-y).
\eea
%----------------------------------------------------------
In the limit of strongly degenerate matter 
($T\lesssim 1\,\MeV$ \cite{Alford:2018lhf}) we find 
the following limits for $\Gamma_{1p}$
and $\lambda_1$ for the neutrino-transparent case
%------------------------------------------------------
\bea\label{eq:Gamma1_deg}
\Gamma_{1p} = \alpha m^{*2} \tilde{G}^2 T^5 
p_{Fe}  \theta(p_{Fp}+p_{Fe} -p_{Fn}).
\eea 
%------------------------------------------------------
where $\alpha=3
\left[\pi^2 \zeta(3) + 15 \zeta(5)\right]/16\pi^5=0.0168$,
and $p_{Fi}$ are the Fermi-momenta of the particles, and 
%------------------------------------------------------
\bea \label{eq:lambda1_deg}
\lambda_1 = \frac{17}{480\pi} m^{*2} \tilde{G}^2 
T^4 p_{Fe}\theta( p_{Fp}+p_{Fe} -p_{Fn}).
\eea 
%------------------------------------------------------

\subsection{The rates of the processes $n+\nu_e \rightleftarrows p + e^-$}
\label{sec:urca2}

The computation of the rates of the processes \eqref{eq:Urca_2} is
carried out in an analogous manner. The rates of the direct and the inverse
processes are given, respectively, by
%------------------------------------------------------
\bea \label{eq:Gamma2p_def}
\Gamma_{2p}(\mu_\Delta) &=& \frac{\tilde{G}^2}{2}
\int\!\! \frac{d^3p'}{(2\pi)^32p'_0} \int\!\!
\frac{d^3p}{(2\pi)^32p_0} \int\!\! \frac{d^3k'}{(2\pi)^32k'_0}
\int\!\! \frac{d^3k}{(2\pi)^32k_0}
\sum \vert {\cal M}_{Urca}\vert^2 \nonumber\\
& \times&
f(p')f(k)[1-f(k')][1-f(p)] (2\pi)^4\delta(p-k+k'-p'), \\
%------------------------------------------------------
 \label{eq:Gamma2n_def}
\Gamma_{2n}(\mu_\Delta) &=& \frac{\tilde{G}^2}{2}
\int\!\! \frac{d^3p'}{(2\pi)^32p'_0} \int\!\!
\frac{d^3p}{(2\pi)^32p_0} \int\!\! \frac{d^3k'}{(2\pi)^32k'_0}
\int\!\! \frac{d^3k}{(2\pi)^32k_0}
\sum \vert {\cal M}_{Urca}\vert^2 \nonumber\\
& \times & f(k')f(p) [1-f(p')][1-f(k)] (2\pi)^4 \delta(p-k+k'-p').
\eea 
%------------------------------------------------------
These rates can be reduced to the following (see Appendix~\ref{app:rates}) 
%------------------------------------------------------
\bea \label{eq:Gamma2p_final}
\Gamma_{2p}(\mu_\Delta) &=& 
 \tilde{G}^2 \frac{m^{*2}T^6}{8\pi^5}\int_{-\infty}^{\infty} 
 dy~g(y-\mu_\Delta/T)
\int_{|y+\alpha_e-\alpha_{\nu}|}^{\infty} dz
  \ln \left[ \frac{1+\exp\left(-y_0\right)} 
{1+\exp\left(-y_0-y+\mu_\Delta/T)\right)}\right]\nonumber\\
&\times &
\int_{\bar{x}_{\rm min}}^{\infty}  dx (x+\alpha_{\nu})
(y+\alpha_e+x)f(x)[1-f(x+y)],\\
 \label{eq:Gamma2n_final}
\Gamma_{2n}(\mu_\Delta) &=& 
 \tilde{G}^2 \frac{m^{*2}T^6}{8\pi^5}\int_{-\infty}^{\infty} 
 dy~[1+g(y-\mu_\Delta/T)]
\int_{|y+\alpha_e-\alpha_{\nu}|}^{\infty} dz
  \ln \left[ \frac{1+\exp\left(-y_0\right)} 
{1+\exp\left(-y_0-y+\mu_\Delta/T)\right)}\right]\nonumber\\
&\times&
\int_{\bar{x}_{\rm min}}^{\infty}  dx (x+\alpha_{\nu})
(y+\alpha_e+x)f(x+y)[1-f(x)],
\eea 
%------------------------------------------------------
with $\bar{x}_{\rm min}=(z-y-\alpha_e-\alpha_{\nu})/2$. One can verify 
that $\Gamma_{2p}=\Gamma_{2n}$ when $\mu_\Delta=0$.
%------------------------------------------------------
In this case, the $\lambda$-parameter is given by 
%------------------------------------------------------
\bea \label{eq:lambda2}
\lambda_2 & \equiv & \frac{\partial\Gamma_{2p}(\mu_\Delta)}
{\partial\mu_\Delta}\bigg\vert_{\mu_\Delta=0}-
\frac{\partial\Gamma_{2n}(\mu_\Delta)}
{\partial\mu_\Delta}\bigg\vert_{\mu_\Delta=0} = \frac{\Gamma_2}{T},
\eea
%--------------------------------------------------------
where 
%--------------------------------------------------------
\bea\label{eq:Gamma_2_general}
\Gamma_2  &=&
 \frac{m^{*2} \tilde G^2}{8\pi^5} T^6 \int_{-\infty}^\infty
 dy~g(y)\int_{|y+\alpha_e-\alpha_{\nu}|}^{\infty} dz
 \ln \bigg\vert \frac{1+\exp\left(-y_0\right)}
{1+\exp\left(-y_0-y)\right)}\bigg\vert\nonumber\\
&\times&
\int_{\bar{x}_{\rm min}}^{\infty}  dx (x+\alpha_{\nu})(y+\alpha_e+x)f(x)[1-f(x+y)].
\eea 
%------------------------------------------------------
If we extrapolate the neutrino-trapped result into the 
low-temperature regime (where in reality neutrinos are no longer trapped)
of highly degenerate matter where the fermionic chemical
potentials satisfy the condition $\mu_{i}\gg T$ ($i \in n, p, e, \nu$) the
rate $\Gamma_2$ given by Eq.~\eqref{eq:Gamma_2_general} reduces to
%------------------------------------------------------
\bea \label{eq:Gamma2_trap}
\Gamma_2 = \frac{m^{*2}\tilde G^2 }{12\pi^3}T^3p_{Fe}p_{F\nu}
(p_{Fe}+p_{F\nu}-| p_{Fn}-p_{Fp}|).
\eea 
%------------------------------------------------------
Thus, the computation of the parameters $\lambda_{1,2}$, which
determine the nonequilibrium relaxation rate of Urca processes at
arbitrary degeneracy of the involved fermions, nonzero temperature
and in the presence of neutrino trapping reduces to an evaluation of
three-dimensional integrals given by Eqs.~\eqref{eq:Gamma_1_general}
and \eqref{eq:Gamma_2_general}.  These results are essentially exact,
the only approximation being the neglect of the terms $O(m^*/E)$, where
$E = \sqrt{p^2+m^{*2}}$, in the tree-level weak-interaction matrix
element. Note, however, that many-body correlations in the baryonic
matter, which arise from a resummation of particle-hole diagrams are
not included yet. In other words, our results correspond to the
evaluation of the polarization tensor of baryonic matter in the
one-loop approximation.

We consider also the neutrino-transparent case, where
we have $\Gamma_{2p}=0$, and for $\lambda_2$ we find
%----------------------------------------------------------
\bea \label{eq:lambda_2_no_nu}
\lambda_2 &=& -\frac{m^{*2} \tilde{G}^2}{8\pi^5} T^5 \int_{-\infty}^\infty 
 dy~[1+g(y)]\int_{|y+\alpha_e|}^{\infty} dz~
\Bigg\{g(y) \ln \bigg\vert \frac{1+\exp\left(-y_0\right)} 
  {1+\exp\left(-y_0-y)\right)}\bigg\vert\nonumber\\
  &-& f(y_0+y) - [f(y_0+y)-f(y_0)]
\frac{m^*}{z^2T}\bigg(\alpha_n-\alpha_p+y-z^2\frac{T}{2m^*}\bigg) 
 \Bigg\} \nonumber\\
&\times & \int_{\bar{x}_{\rm min}}^{\infty}  dx~x
(y+\alpha_e+x)f(x+y).
\eea
%----------------------------------------------------------
In the limit of strongly degenerate matter 
($T\lesssim 1\,\MeV$ \cite{Alford:2018lhf}), the rate
 $\Gamma_{2n}$ is the same as in the case of $\Gamma_{1p}$
%------------------------------------------------------
\bea\label{eq:Gamma2_deg}
\Gamma_{2n} = \alpha m^{*2} \tilde{G}^2 T^5 
p_{Fe}  \theta(p_{Fp}+p_{Fe} -p_{Fn}).
\eea 
%------------------------------------------------------
Similarly, the low-temperature limit for $\lambda_2$ reads
%------------------------------------------------------
\bea \label{eq:lambda2_deg}
\lambda_2 = \frac{17}{480\pi} m^{*2}\tilde{G^2}T^4 p_{Fe}\theta( p_{Fp}+p_{Fe} -p_{Fn}),
\eea 
%------------------------------------------------------
therefore, the total rate $\lambda \equiv \lambda_1+\lambda_2$ is given by
%------------------------------------------------------
\bea \label{eq:lambda_deg}
\lambda 
= \frac{17}{240\pi} m^{*2}\tilde{G^2}T^4 p_{Fe}\theta( p_{Fp}+p_{Fe} -p_{Fn}),
\eea 
%------------------------------------------------------
which agrees with the corresponding expression given in Ref.~\cite{Haensel2000}.

\section{Bulk viscosity}
\label{sec:bulk}

The purpose of this section is to derive a microscopic formula for the
bulk viscosity.  We will assume that the matter is composed of
neutrons, protons, electrons, muons, and neutrinos.  Although there
are large-amplitude density oscillations in a merger
\cite{Alford2017,Perego:2019adq}, we will restrict our analysis to the
``subthermal'' case where the matter is only slightly perturbed from
equilibrium by density oscillations of some characteristic frequency
$\omega$.  We will not study the ``suprathermal'' case of high
amplitude oscillations, but it can only lead to an enhancement of the
bulk viscosity when the beta equilibration rate is slower than the
density oscillation frequency, and, as we will show, in the
neutrino-trapped temperature range we are in the opposite regime of
fast equilibration.  In our analysis we take into account the
contribution of muons to the thermodynamic quantities, but neglect
their contribution to the bulk viscosity, as it is subdominant to the
processes involving electrons.

In the case where neutrinos are trapped, the equilibrium
with respect to weak interactions implies the
conditions~\eqref{eq:eq_condition1}.  The charge neutrality condition implies
$n_p=n_e+n_\mu$. These two conditions are sufficient to fix the
number densities of the constituents for any given temperature $T$,
the value of the baryon number density $n_B=n_n+n_p$ and the lepton
number density $n_L=n_e+n_\nu$. Here the neutrino net density is given
by ${n}_{\nu}\equiv\tilde{n}_\nu-\tilde{n}_{\bar{\nu}}$ where
$\tilde{n}_\nu$ and $\tilde{n}_{\bar{\nu}}$ are the neutrino and
antineutrino number densities, respectively.
 
Consider now small-amplitude density oscillations in the matter, with
characteristic timescales that are long compared to the strong
interaction timescale $\sim 10^{-23}$\,s. Since the strong
interactions establish thermal equilibrium, the particle distributions
are always thermal (Fermi-Dirac or Bose-Einstein); the only deviation
from equilibrium that is induced by the oscillations is a departure
from beta equilibrium, which can be expressed in terms of a single
chemical potential $\mu_\Delta$ \eqref{eq:mu-Delta}.

The perturbed densities are written as $n_B(t)=n_{B0}+\delta n_B(t)$,
and $n_L(t)=n_{L0}+\delta n_L(t)$, with $n_{B0}$ and $n_{L0}$ being
the unperturbed background densities of baryons and leptons.  The
time-dependence of the density perturbations is taken as
$\delta n_B(t),\delta n_L(t) \sim e^{i\omega t}$. The continuity
equation $\partial n_i /\partial t + {\rm div}\, n_i \bm v =0$ then
implies
%------------------------------------------------------
\bea\label{eq:cont_i}
\delta n_i(t) =-\frac{\theta}{i\omega}\, n_{i0},
\quad i=\{B,L\},
\eea 
%------------------------------------------------------
where $\bm v$ is the bulk (hydrodynamic) velocity of matter and
 $\theta={\rm div}\, \bm v$.
(Note that we consider only linear perturbation in densities.)

The density perturbations above imply density perturbation of particle
number which can be separated into {\it equilibrium} and {\it
  nonequilibrium} parts
%------------------------------------------------------
\bea\label{eq:dens_j}
n_j(t)=n_{j0}+\delta n_j(t), && \delta n_j(t)=\delta n^{\rm eq}_j(t)+\delta n'_j(t),
\eea 
%------------------------------------------------------ 
where $j=\{n,p,e,\nu\}$ labels the particles. The variations
$\delta n^{\rm eq}_j(t)$ denote the shift of the equilibrium state for
the instantaneous values of the baryon and lepton densities $n_{B}(t)$
and $n_{L}(t)$, whereas $\delta n'_j(t)$ denote the deviations of the
corresponding densities from their equilibrium values.  Due to the
nonequilibrium shifts $\delta n'_j(t)$ the composition balance of
matter is disturbed leading to a small shift
$\mu_\Delta(t) =\delta\mu_n(t)+\delta\mu_\nu(t)-\delta\mu_p(t)-\delta\mu_e(t)$,
which can be written as
%------------------------------------------------------
\bea\label{eq:delta_mu}
\mu_\Delta(t)=(A_{nn}-A_{pn}) \delta n_n(t)+A_{\nu\nu} \delta n_\nu(t) -
(A_{pp}-A_{np})  \delta n_p(t) -A_{ee}\delta n_e(t)\equiv \sum_i s_{i}A_i\delta n_i(t)
\eea 
%------------------------------------------------------
where $s_i = +1$ for $n,\, \nu$ and $-1$ for $p,\,e$;
$A_n=A_{nn}-A_{pn}$, $A_p=A_{pp}-A_{np}$, and $A_e=A_{ee}$, $A_\nu=A_{\nu\nu}$ with
%------------------------------------------------------
\bea\label{eq:A_j} A_{ij} = \left(\frac{\partial \mu_i}{\partial
  n_j}\right)_0, \eea and index 0 denotes the static equilibrium
state. The off-diagonal elements $A_{np}$ and $A_{pn}$ are nonzero
because of the cross-species strong interaction between neutrons and
protons. Since we treat the electrons and neutrinos as
ultrarelativistic noninteracting gas, we have kept only the terms that
are diagonal in indices $i,j$, which we will further denote simply as
$A_e$ and $A_{\nu}$.  The computation of susceptibilities $A_i$ is
performed in Appendix~\ref{app:A_coeff}.

If the weak processes are turned off, then a perturbation
conserves all particle numbers, therefore
%------------------------------------------------------
\bea\label{eq:cont_j}
\frac{\partial}{\partial t} \delta {n}_j(t)+ \theta n_{j0} =0,\qquad 
\delta {n}_j(t) = -\frac{\theta}{i\omega}\, n_{j0}.
\eea 
%------------------------------------------------------  
Once the weak reactions are turned on, there is an imbalance between
the rates of weak processes given by Eqs.~\eqref{eq:Gamma_1_general}
and \eqref{eq:Gamma_2_general}. To linear order 
in $\mu_\Delta$ the imbalance is given by~\cite{Sawyer1989,Haensel1992PhRvD,Huang2010}
%------------------------------------------------------
\bea\label{eq:lambda_def}
\Gamma_p-\Gamma_n =\lambda\mu_\Delta,\quad\lambda > 0,
\eea
%------------------------------------------------------
where $\Gamma_{j}=\Gamma_{1j}+\Gamma_{2j}$ and 
$\lambda=\lambda_1+\lambda_2$ see Eqs.~\eqref{eq:lambda1} and
\eqref{eq:lambda2} (in the case of neutrino-transparent matter
one should use Eqs.~\eqref{eq:lambda1_deg} and \eqref{eq:lambda2_deg}
for $\lambda_j$).
Then instead of Eq.~\eqref{eq:cont_j} we will have the
following rate equations which take into account the loss and gain of
particles via the weak interactions
%------------------------------------------------------
\bea\label{eq:cont_n_weak}
\frac{\partial}{\partial t}\delta n_n(t) &=& 
-\theta  n_{n0} -\lambda\mu_\Delta(t),\\
\label{eq:cont_p_weak}
\frac{\partial}{\partial t}\delta n_p(t) &=& 
-\theta n_{p0} +\lambda\mu_\Delta(t).
\eea 
%------------------------------------------------------ 
We next substitute $\delta n_j(t)\sim e^{i\omega t}$, eliminate 
$\mu_\Delta$ using Eq.~\eqref{eq:delta_mu}, and employ
the relations $\delta n_p=\delta n_B-\delta n_n$,
$\delta n_e=\delta n_p$ 
and $\delta n_L =\delta n_e+\delta n_\nu$,
and Eq.~\eqref{eq:cont_i} to find 
%------------------------------------------------------
\bea\label{eq:delta_n_0ol_tr}
\delta n_n(t) &=& -\frac{i\omega n_{n0} +\lambda (A_p+A_e+A_\nu)n_{B0}-\lambda A_\nu n_{L0}}
{i\omega+\lambda A}\frac{\theta}{i\omega},\\
\label{eq:delta_p_sol_tr}
\delta n_p(t) &=& \delta n_e(t)= -\frac{i\omega n_{p0} +\lambda A_nn_{B0}+\lambda A_\nu n_{L0}}
{i\omega+\lambda A}\frac{\theta}{i\omega},\\
\label{eq:delta_nu_sol_tr}
\delta n_\nu(t) &=& -\frac{i\omega n_{\nu 0}+\lambda (A_n+A_p+A_e)n_{L0}
 -\lambda A_n n_{B0}}{i\omega+\lambda A}\frac{\theta}{i\omega},
\eea 
%------------------------------------------------------
where 
%------------------------------------------------------
\bea\label{eq:A_def}
A=\sum_j A_j %\sum_j\left(\frac{\partial\mu_j}{\partial n_j}\right)_0
=\left(\frac{\partial\mu_n}{\partial n_n}\right)_0
+\left(\frac{\partial\mu_p}{\partial n_p}\right)_0
-\left(\frac{\partial\mu_n}{\partial n_p}\right)_0
-\left(\frac{\partial\mu_p}{\partial n_n}\right)_0
+\left(\frac{\partial\mu_e}{\partial n_e}\right)_0
+\left(\frac{\partial\mu_\nu}{\partial n_\nu}\right)_0 
= -\dfrac{1}{n_B}\biggl(\dfrac{\partial \mu_\Delta}{\partial x_p} \biggr)_{\!n_B} \ ,
\eea 
%------------------------------------------------------
so $A$ is the ``beta-disequilibrium--proton-fraction'' susceptibility:
it measures how the out-of-beta-equilibrium
chemical potential is related to a change in the proton fraction.
In order to separate the nonequilibrium parts of $\delta n_j$
we need to find also the equilibrium shifts $\delta n_j^{\rm eq}$. 
According to the definition of the $\beta$-equilibrium state we have $\mu^{\rm eq}_n(t)+\mu^{\rm eq}_\nu(t)-\mu^{\rm eq}_p(t)-\mu^{\rm eq}_e(t)=0$, therefore
%------------------------------------------------------
\bea
A_n\delta n^{\rm eq}_n(t)+A_\nu\delta n^{\rm eq}_\nu(t)-
A_p\delta n^{\rm eq}_p(t)-A_e\delta n^{\rm eq}_e(t)=0.
\eea
%------------------------------------------------------
Using the relations $\delta n^{\rm eq}_n+\delta n^{\rm eq}_p=\delta n_B$,
$\delta n^{\rm eq}_e=\delta n^{\rm eq}_p$, $\delta n^{\rm eq}_e+\delta n^{\rm eq}_\nu=\delta n_L$, and
substituting also $\delta n_B$ and $\delta n_L$ from Eq.~\eqref{eq:cont_i} we find
%------------------------------------------------------
\bea\label{eq:delta_n_eq_tr}
\delta n^{\rm eq}_n(t) &=& \frac{-(A_p+A_e+A_\nu)n_{B0}
+A_\nu n_{L0}}{A} \frac{\theta}{i\omega},\\
\label{eq:delta_p_eq_tr}
\delta n^{\rm eq}_p(t) &=& \delta n^{\rm eq}_e(t)= -\frac{A_n n_{B0}
+ A_\nu n_{L0}}{A}\frac{\theta}{i\omega},\\
\label{eq:delta_nu_eq_tr}
\delta n^{\rm eq}_\nu(t) &=& \frac{-(A_n+A_p+A_e)n_{L0}
+A_n n_{B0}}{A} \frac{\theta}{i\omega}.
\eea 
%-------------------------------------------------------
Then, according to Eq.~\eqref{eq:dens_j}, we find for $\delta n'_j$
%------------------------------------------------------
\bea\label{eq:delta_n'_tr}
\delta n'_n(t) &=& \delta n'_\nu(t)
 =-\frac{ C}{A(i\omega+\lambda A)}\theta,\\
\label{eq:delta_p'_tr}
\delta n'_p(t)  &=& \delta n'_e(t)
 =\frac{ C}{A(i\omega+\lambda A)}\theta,
\eea 
%------------------------------------------------------ 
with
%------------------------------------------------------
\bea\label{eq:C_def}
C &=& n_{n0} A_n+n_{\nu 0} A_\nu - n_{p0}A_p - n_{e0}A_e
= n_B\biggl(\dfrac{\partial \mu_\Delta}{\partial n_B} \biggr)_{\!x_p} \ ,
\eea 
%------------------------------------------------------
so $C$ is the ``beta-disequilibrium--baryon-density'' susceptibility:
it measures how the out-of-beta-equilibrium
chemical potential is related to a change in the baryon density at fixed
proton fraction.
%------------------------------------------------------
Now we are in a position to compute the 
full nonequilibrium pressure which is given by
%------------------------------------------------------
\bea
p(t)=p(n_j(t))=p\left[n_{j0}+\delta n_j^{\rm eq}(t)\right]
+\delta p'(t)=p^{\rm eq}(t)+\delta p'(t),
\eea 
%------------------------------------------------------
where the nonequilibrium part of the pressure, referred to as bulk
viscous pressure, is given by
%------------------------------------------------------
\bea\label{eq:Pi}
\Pi(t)\equiv \delta p'(t) =\sum_j
\left(\frac{\partial p}{\partial n_j}\right)_0\delta n'_j(t).
\eea 
%------------------------------------------------------
Using the Gibbs-Duhem relation $dp=sdT+\sum_j n_j d \mu_j$, which is
valid also out of equilibrium, we can write \footnote{Note that the
temperature is assumed to be constant because, as argued in 
Sec.~\ref{sec:intro}, we assume that the thermal equilibration rate
is much larger than the chemical equilibration rate.}
%------------------------------------------------------
\bea
\left(\frac{\partial p}{\partial n_j}\right)_0
%=\sum_l n_{l0} \left(\frac{\partial \mu_l}{\partial n_j}\right)_0
=\sum_l n_{l0}A_{lj},
\eea
%------------------------------------------------------
from which  we can identify 
%------------------------------------------------------
\bea\label{eq:C_redef}
\left(\frac{\partial p}{\partial n_n}\right)_0 
+\left(\frac{\partial p}{\partial n_\nu}\right)_0
-\left(\frac{\partial p}{\partial n_p}\right)_0
-\left(\frac{\partial p}{\partial n_e}\right)_0=C,
\eea 
%------------------------------------------------------
where we used the symmetry relation $A_{np}=A_{pn}$. Collecting the results 
\eqref{eq:delta_n'_tr}, \eqref{eq:delta_p'_tr}, \eqref{eq:C_redef} 
we find that the bulk viscous pressure \eqref{eq:Pi} is given
by 
%------------------------------------------------------
\bea\label{eq:Pi1}
\Pi =
\frac{C^2}{A}\frac{i\omega-\lambda A}{\omega^2+\lambda^2 A^2}\theta.
\eea 
%------------------------------------------------------
The bulk viscosity is the real part of $-\Pi/\theta$,
%------------------------------------------------------
\bea\label{eq:zeta}
\zeta = \frac{C^2}{A}\frac{\lambda A}{\omega^2+\lambda^2 A^2}\ ,
\eea
%------------------------------------------------------ 
which has the classic resonant form depending on two quantities:
the prefactor $C^2/A$ which is a ratio of susceptibilities
\eqref{eq:A_def},\eqref{eq:C_def}, depending
only on the EoS, and the relaxation rate $\lambda A$
which depends on the weak interaction rate
$\lambda=\lambda_1+\lambda_2$ \eqref{eq:lambda1},\eqref{eq:lambda2}
and the susceptibility that relates $\mu_\Delta$ to the proton fraction.
Note that if we extrapolate the neutrino-trapped calculation
to the low-temperature, degenerate limit, we can compute
the susceptibility $A$ analytically, see Appendix~\ref{app:A_coeff}.

\section{Numerical results}
\label{sec:num_results}

To quantify the amount of dissipation through bulk viscosity in the
present context we need first to specify the properties of
$\beta$-equilibrated nuclear matter.  We choose to do so using the
density functional theory (DFT) approach to the nuclear matter, which
is based on phenomenological baryon-meson Lagrangians of the type
proposed by Walecka, Boguta-Bodmer and
others~\cite{glendenning2000compact,weber_book,Sedrakian2007}. We will use the
parametrization of such a Lagrangian with density-dependent
meson-nucleon coupling~\cite{Lalazissis2005} and will apply the DFT to
nuclear matter with trapped neutrinos, see also~\cite{Colucci2013}.

\subsection{Beta-equilibrated nuclear matter}
\label{sec:DFT}

The Lagrangian density of matter can be written as ${\cal L} = {\cal
  L}_N+{\cal L}_l$, where the baryonic contribution is given by 
%------------------------------------------------------------
\bea\label{eq:lagrangian_B} {\cal L}_N & = &
\sum_N\bar\psi_N\bigg[\gamma^\mu \left(i\partial_\mu-g_{\omega
N}\omega_\mu - \half g_{\rho N}\vectau\cdot\vecrho_\mu\right)
- (m_N - g_{\sigma N}\sigma)\bigg]\psi_N \\
\nonumber & + & \half \partial^\mu\sigma\partial_\mu\sigma-\half
m_\sigma^2\sigma^2 - \frac{1}{4}\omega^{\mu\nu}\omega_{\mu\nu} + \half
m_\omega^2\omega^\mu\omega_\mu -
\frac{1}{4}\vecrho^{\mu\nu}\vecrho_{\mu\nu} + \half
m_\rho^2\vecrho^\mu\cdot\vecrho_\mu, \eea 
%------------------------------------------------------------
where $N$ sums over nucleons, $\psi_N$ are the nucleonic Dirac fields
with masses $m_N$.  The meson fields $\sigma,\omega_\mu$, and
$\vecrho_\mu$ mediate the interaction among baryon fields,
$\omega_{\mu\nu}$ and $\vecrho_{\mu\nu}$ represent the field strength
tensors of vector mesons and $m_{\sigma}$, $m_{\omega}$, and
$m_{\rho}$ are their masses. The baryon-meson coupling constants are
denoted by $g_{iN}$ with $i=\sigma,\omega,\rho$. The leptonic
contribution is given by
%------------------------------------------------------------
\bea\label{eq:lagrangian_L} 
{\cal L}_l & = &
 \sum_{\lambda}\bar\psi_\lambda(i\gamma^\mu\partial_\mu -
m_\lambda)\psi_\lambda,
\eea 
%------------------------------------------------------------
where $\lambda$ sums over the leptons $e^-,\mu^-$, $\nu_e$ and
$\nu_{\mu}$, which are treated as free Dirac fields with masses
$m_\lambda$; the mass of electron neutrino is negligible and is set to
zero.  We do not consider electromagnetic fields, therefore their
contribution is dropped. The coupling constants in the nucleonic
Lagrangian are density-dependent and are parametrized according to the
relation $g_{iN}(n_B) = g_{iN}(n_{0})h_i(x)$, for $i=\sigma,\omega$,
and $g_{\rho N}(n_B) = g_{\rho N}(n_{0})\exp[-a_\rho(x-1)]$ for the
$\vecrho_\mu$-meson, where $n_B$ is the baryon density, $n_0$ is the
saturation density and $x = n_B/n_0$. The density dependence of the
couplings is encoded in the functions
%------------------------------------------------------------
\be\label{ansatz} h_i(x) =
a_i\frac{1+b_i(x+d_i)^2}{1+c_i(x+d_i)^2}. 
\ee 
%------------------------------------------------------------
This parametrization has in total eight parameters, which are
adjusted to reproduce the properties of symmetric and asymmetric
nuclear matter, binding energies, charge radii, and neutron radii of
spherical nuclei, see Table \ref{table:DD-ME2}.  We recall that the
Lagrangian of this model has only linear in meson field interaction
terms and the coupling nucleon-meson constants are
density-dependent. We will also employ below the NL3
model~\cite{Lalazissis1997} as an alternative, which has
density-independent meson-nucleon couplings but contains nonlinear in
meson fields terms.

%------------------------------------
\begin{table}[ht]
  \centering
  \caption{Meson masses $m_i$ and their couplings 
    $g_{iN}$ to the baryons in DD-ME2 parametrization. The remaining
    parameters specify the density dependence of the couplings. }
\label{table:DD-ME2}
  \begin{tabular}{cccc}
\hline
           &     $ \sigma$    & $\omega$ & $\rho$  \\
\hline
    $m_i$ [MeV] &  550.1238  &  783.0000    &  763.0000  \\
    $g_{iN}(n_{0})$ & 10.5396    & 13.0189     &  3.6836\\
    $a_i$  & 1.3881    & 1.3892 &   0.5647\\
    $b_i$  & 1.0943   & 0.9240 &   ---\\
    $c_i$  & 1.7057   &  1.4620 &   ---\\
    $d_i$  & 0.4421  & 0.4775 &    ---\\
\hline
  \end{tabular}
\end{table}
%------------------------------------

From the Lagrangian densities \eqref{eq:lagrangian_B} and
\eqref{eq:lagrangian_L} we obtain the pressure of the nucleonic
component
%--------------------------------------------------
\bea
\label{eq:P_N}
P_N & = & - \frac{m_\sigma^2}{2} \sigma_0^2 +
\frac{m_\omega^2}{2} \omega_0^2 + \frac{m_\rho^2 }{2} \rho_{03}^2
\nonumber\\
&+& \displaystyle \frac{1}{3}\sum_N \frac{2J_N+1}{2\pi^2} 
\int_0^{\infty}\!\!\! \frac{k^4 \ dk}{(k^2+m^{* 2}_N)^{1/2}}
\left[f(E^N_k-\mu_N^*)+f(E^N_k+\mu_N^*)\right],
\eea
%--------------------------------------------------
where $m_N^{*} = m_N -\sigma_0 g_{\sigma N}$ is the {\it relativistic
  (Dirac)} effective nucleon
mass, $\mu^*_N = \mu_N-g_{\omega N}\omega_0 - g_{\rho N}I_3\rho^0_3$
is the nucleon chemical potential including the time component of the
fermion self-energy, $I_3$ is the third component of nucleon isospin
and $\sigma_0$, $\omega_0$ and $\rho^0_3$ are the mean values of the
meson fields, $E^N_k = \sqrt{k^2+m^{* 2}_N}$ are the single particle
energies of nucleons.  The first three terms in this expression are
associated with mean values of the mesonic fields, whereas the last
term is the fermionic contribution which is temperature-dependent.

The leptonic contribution to the
pressure is given by 
%---------------------------------------------------------------
\bea
\label{eq:P_L}
P_L & =&\frac{g_{\lambda}}{3\pi^2} \sum_{\lambda} \int_0^{\infty}\!\!\!  \frac{k^4 \
  dk}{(k^2+m^2_\lambda)^{1/2}}
\left[f(E^{\lambda}_k-\mu_\lambda)+f(E^{\lambda}_k+\mu_\lambda)\right],
\eea 
%--------------------------------------------------
where $g_{\lambda}$ is the leptonic degeneracy factor and
$E_k^{\lambda}=\sqrt{k^2+m_\lambda^2}$ are the single particle
energies of leptons.  At nonzero temperature, the net entropy of the
matter is the sum of the nucleon contribution
%--------------------------------------------------
\bea 
S_N &=& - \sum_N\frac{2J_N+1}{2\pi^2}\int_0^{\infty} dk k^2 
\Bigg\{\Bigl[f(E_k^N-\mu_N^*)\ln f(E_k^N-\mu_N^*) \nonumber\\
&+& \bar f(E_k^N-\mu_N^*)\ln \bar f(E_k^N-\mu_N^*) \Bigr] 
+ (\mu_N^* \to -\mu_N^*)
\Biggr\}
\eea
%--------------------------------------------------
and the lepton contribution
%--------------------------------------------------
\bea
S_L &=&- \sum_{\lambda} \int_0^{\infty} \!\frac{dk}{\pi^2}
\Bigl[f(E_k^\lambda-\mu_{\lambda})\ln f(E_k^\lambda-\mu_{\lambda}) \nonumber\\
&+& \bar f(E_k^\lambda-\mu_{\lambda})\ln \bar f(E_k^\lambda-\mu_{\lambda})\Bigr],
\eea
%--------------------------------------------------
where $\bar f(y) = 1-f(y)$. The energy density of the system and other
thermodynamical parameters, for example, the free-energy can be
computed in an analogous manner. 

\end{widetext}

In the case of the neutrino-transparent medium, the composition of
hadronic matter includes neutrons, protons, electrons and muons.
Neutrinos are assumed to escape. We will use the approximation that
the beta equilibrium conditions become $\mu_n=\mu_p+\mu_e$ and
$\mu_\mu=\mu_e$. At the temperatures of interest to us there are
corrections to these expressions \cite{Alford:2018lhf} but we will
neglect them.

If the neutrinos are trapped in the matter they contribute to the
energy density and entropy of matter. We take into account the two
lightest flavors of neutrinos, electron and muon neutrinos, and their
antineutrinos~\footnote{Because we restrict our consideration to the
  Standard Model, where there are no neutrino oscillations, $\tau$
  neutrinos can be ignored because $\tau$ lepton is too heavy to
  appear in matter.}.  In this case, the chemical equilibrium
conditions read $\mu_n+\mu_{\nu_l}=\mu_p+\mu_l$ with $l=\{e,\mu\}$,
and the lepton number $n_L$ conservation implies
$n_l+n_{\nu_l}=n_{L_l}=Y_{L_l}n_B$, where the lepton fractions
$Y_{L_l}$ should be fixed for each flavor separately.  In our
numerical calculations we will consider two cases: (i) $Y_L = 0.1 $
for both flavors, which is typical for matter in binary neutron star
mergers; (ii) $Y_L\equiv Y_{L_e} = 0.4$ and $Y_{L_\mu} = 0$ which are
typical for matter in supernovae and 
proto-neutron stars~\cite{Prakash1997,Malfatti2019,Weber2019}.

The neutrino transparent case $Y_\nu = 0$ which applies for
temperatures below the transparency temperature
$T_{\rm tr}\simeq 5$~MeV is discussed in Ref.~\cite{Alford:2019zzz}
and Appendix~\ref{app:bulk_visc0}.

%-------------------------------------------------
\begin{figure}[t] 
\begin{center}
\includegraphics[width=\columnwidth, keepaspectratio]{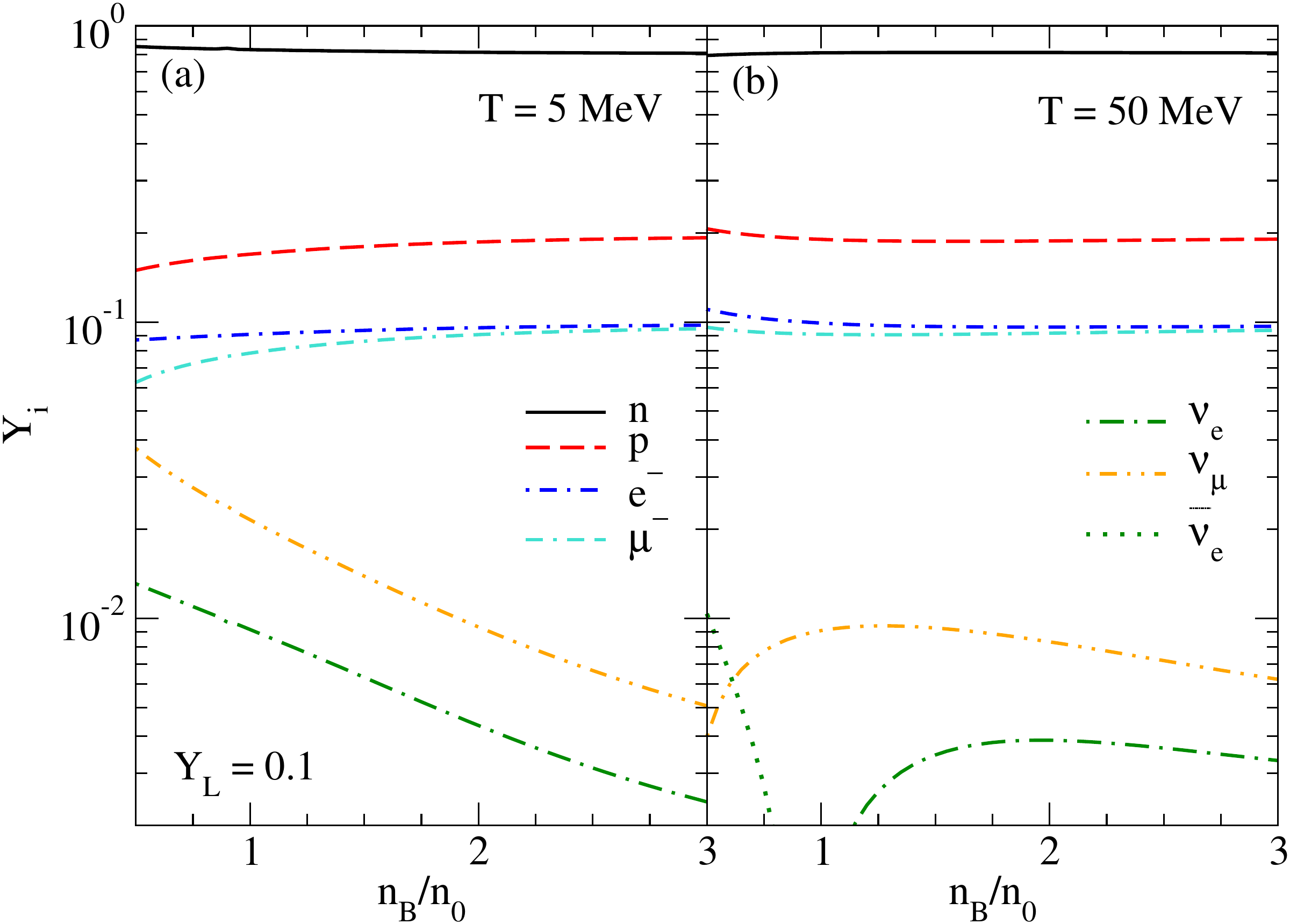}
\caption{ Particle fractions in neutron-star-merger matter,
 (lepton fraction $Y_L=0.1$). We plot $Y_i=n_i/n_B$ 
  as functions of the baryon  density $n_B$ 
  for temperatures (a) $T=5$\,MeV, and (b) $T=50$\,MeV. 
  At high temperature [panel (b)] the net electron neutrino 
  density becomes negative at sufficiently low baryon density: the
  corresponding dotted line shows the net antineutrino fraction.
}
\label{fig:fractions_dens1} 
\end{center}
\end{figure}
%-------------------------------------------------
\begin{figure}[ht] 
\begin{center}
\includegraphics[width=\columnwidth, keepaspectratio]{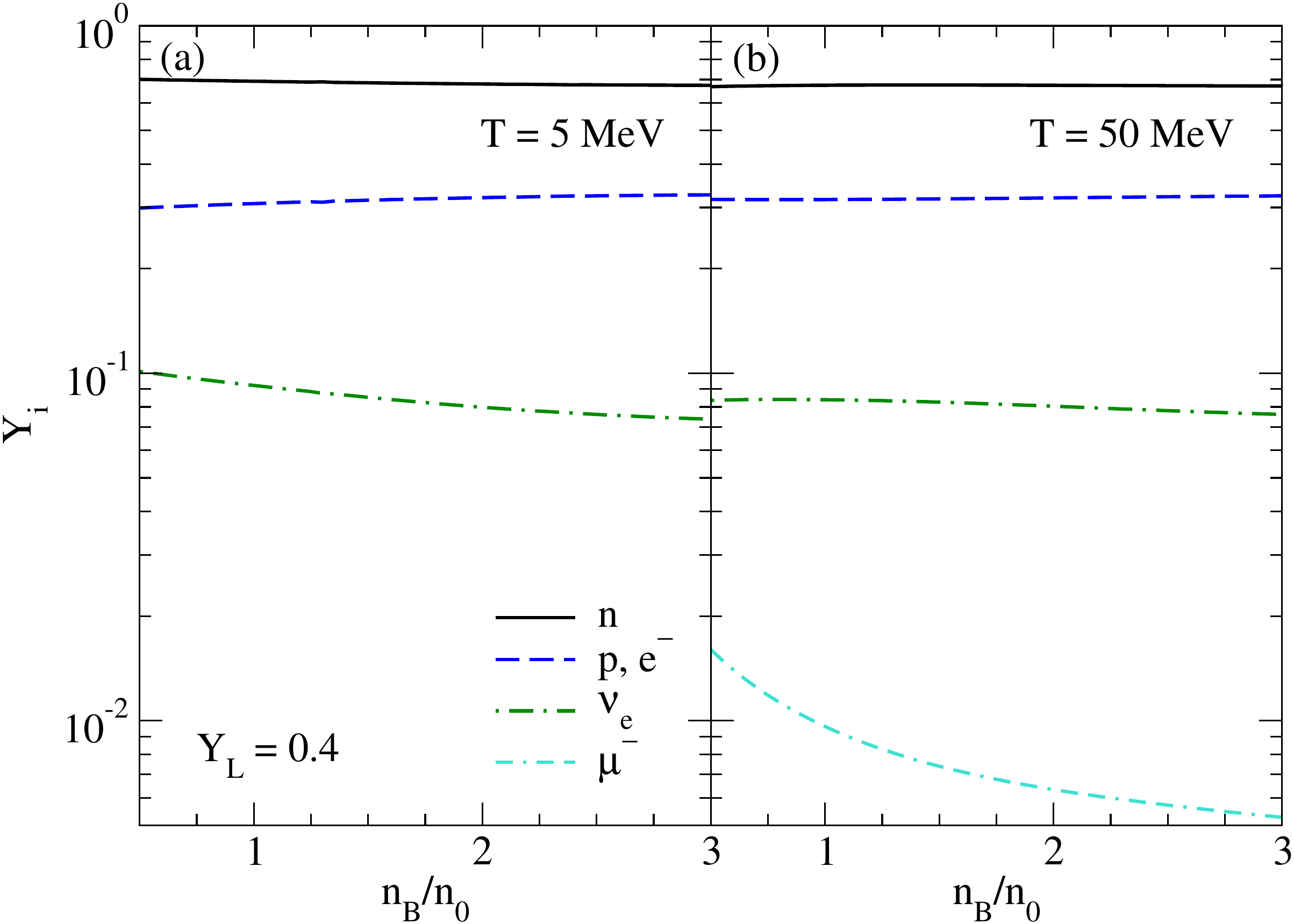}
\caption{ Particle fractions in supernova matter,
 (lepton fraction is $Y_L=0.4$ for electrons, and is zero for muons). 
 We plot $Y_i=n_i/n_B$ as functions of the baryon density $n_B$
  for temperatures (a)~$T=5$\,MeV, and (b)~$T=50$\,MeV. 
  The particle fractions show little dependence on temperature.
}
\label{fig:fractions_dens2} 
\end{center}
\end{figure}
%-------------------------------------------------

Before presenting our results on the bulk viscosity in
Sec.~\ref{sec:bulk_visc} we first discuss the thermodynamics of the
underlying relativistic density functional model of nuclear matter,
which will be used as the background equilibrium for our subsequent
perturbation analysis. We will employ the DD-ME2 parametrization of
the density functional given in Ref.~\cite{Lalazissis2005}.

%-------------------------------------------------
\begin{figure}[t] 
\begin{center}
\includegraphics[width=\columnwidth, keepaspectratio]{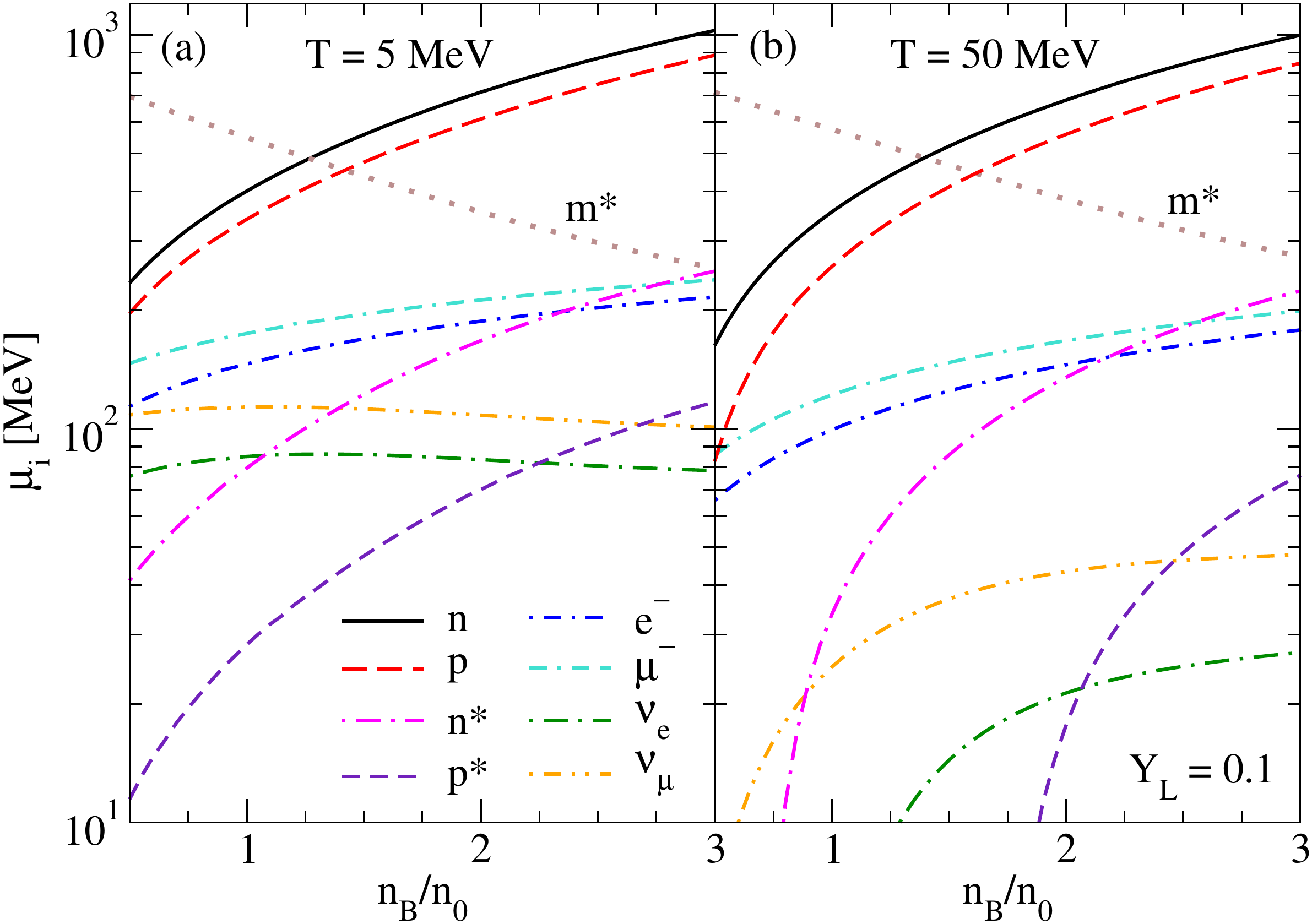}
\caption{ Chemical potentials in the neutron-star-merger matter,
  (lepton fraction $Y_L=0.1$). We plot $\mu_i$ as functions of the
  baryon density $n_B/n_0$ for temperatures (a)~$T=5$\,MeV, and
  (b)~$T=50$\,MeV.  The labels $n^*$ and $p^*$ correspond to the
  effective chemical potentials of the neutron and proton,
  respectively, defined after Eq.~\eqref{eq:P_N}.  The effective
  baryon mass $m^*\equiv m_B^*$ is shown by the dotted lines.}
\label{fig:chem_pot_dens1} 
\end{center}
\end{figure}
%-------------------------------------------------
\begin{figure}[ht] 
\begin{center}
\includegraphics[width=\columnwidth, keepaspectratio]{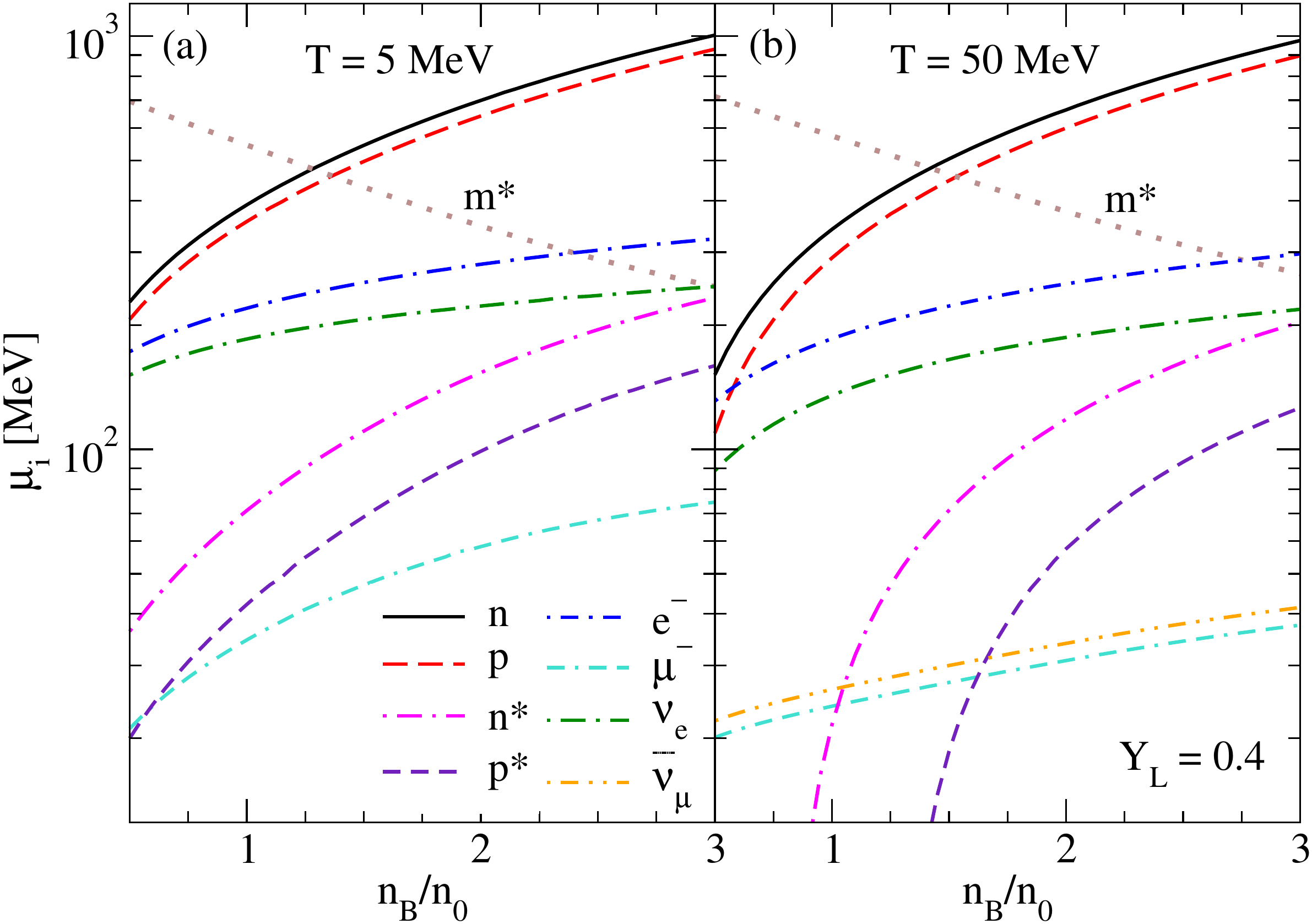}
\caption{ Chemical potentials in supernova matter, ($Y_L=0.4$ for
  electrons).  We plot $\mu_i$ as functions of the baryon density
  $n_B/n_0$ for temperatures (a)~$T=5$\,MeV, and (b)~$T=50$\,MeV.  }
\label{fig:chem_pot_dens2} 
\end{center}
\end{figure}
%-------------------------------------------------

Figures~\ref{fig:fractions_dens1} and \ref{fig:fractions_dens2} show
the particle fractions $Y_j = n_j/n_B$ as functions of the baryon
density normalized to the nuclear saturation density, which is
$n_0=0.152$ fm$^{-3}$ in the DD-ME2 model. Figures
\ref{fig:fractions_dens1} and \ref{fig:fractions_dens2} refer to the
cases of neutron star mergers and supernovae, respectively. 
The results are shown for two temperatures $T=5$ MeV
(which is of the order of $T_{\rm tr}$) [panels (a)] and $T=50$ MeV
[panels (b)], which is close to the upper limit of the temperature
range achieved in these events. Comparing the panels (a) and (b) in
Figs.~\ref{fig:fractions_dens1} and \ref{fig:fractions_dens2} we see
that the particle fractions are generally not sensitive to the
temperature for the given value of $Y_L$. The only exception is for
low lepton-fraction matter, where at low density and high temperature
(Fig.~\ref{fig:fractions_dens1}(b)), the net electron neutrino density becomes
negative, indicating that there are more (electron) antineutrinos than
neutrinos. 
As one would expect, merger (low lepton fraction) matter has much smaller
electron neutrino fraction; the electron and muon fractions are $\sim 10 \%$, so by charge neutrality $Y_p=Y_e+Y_\mu$
the proton fraction is $\sim 20\%$.

Since the bulk viscosity is related to departure from beta
equilibrium, it is instructive to examine the chemical potentials of
particles as functions of density and temperature.  These are shown as
functions of the baryon density in Figs.~\ref{fig:chem_pot_dens1} and
\ref{fig:chem_pot_dens2}.

There are two different chemical potentials for baryons: the
thermodynamic chemical potentials $\mu_n$ and $\mu_p$, which enter
into the thermodynamic relations and the $\beta$-equilibrium
condition~\eqref{eq:eq_condition1}, and the effective chemical
potentials $\mu_n^*$ and $\mu_p^*$, which enter into the baryon
distribution functions and are defined after Eq.~\eqref{eq:P_N}.  We
show also the effective nucleon mass $m^*$ with dotted
lines.

In low lepton-fraction matter at low density and high temperature
(Fig.~\ref{fig:chem_pot_dens1}(b)), we see that the neutrino chemical
potentials become negative, as expected from the particle
fraction results (Fig.~\ref{fig:fractions_dens1}(b))
which showed that there are more antineutrinos than neutrinos
in this regime.

%-------------------------------------------------
\begin{figure}[ht] 
\begin{center}
\includegraphics[width=0.9\columnwidth,keepaspectratio]{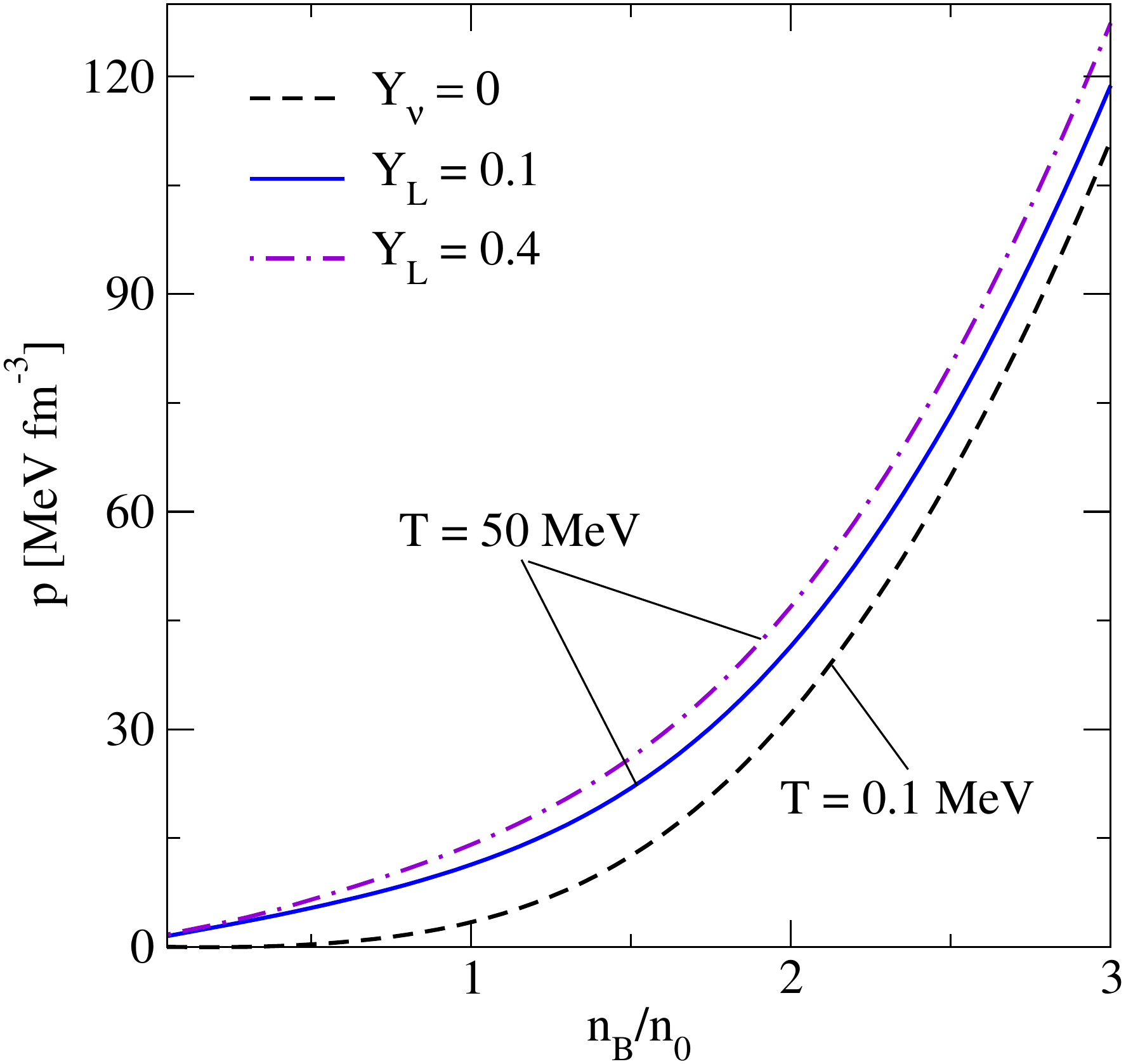}
\caption{ 
Equation of state of nuclear matter for several values of
  the lepton fraction and the temperature.}
\label{fig:eos} 
\end{center}
\end{figure}
%-------------------------------------------------

For completeness and reference, we show the equation of state 
(EoS) of the DD-ME2 model at two temperatures in 
Fig.~\ref{fig:eos}. For $T=50$~MeV the EoS is shown for 
two values of lepton fractions corresponding to   
a supernova and binary-neutron-star merger settings. The increase in the 
pressure at larger temperatures is due to the additional 
thermal contribution from baryons and the contribution from 
trapped neutrinos which is absent in the case when $T=0.1$~MeV.

%-------------------------------------------------
\begin{figure}[t] 
\begin{center}
\includegraphics[width=0.9\columnwidth,keepaspectratio]{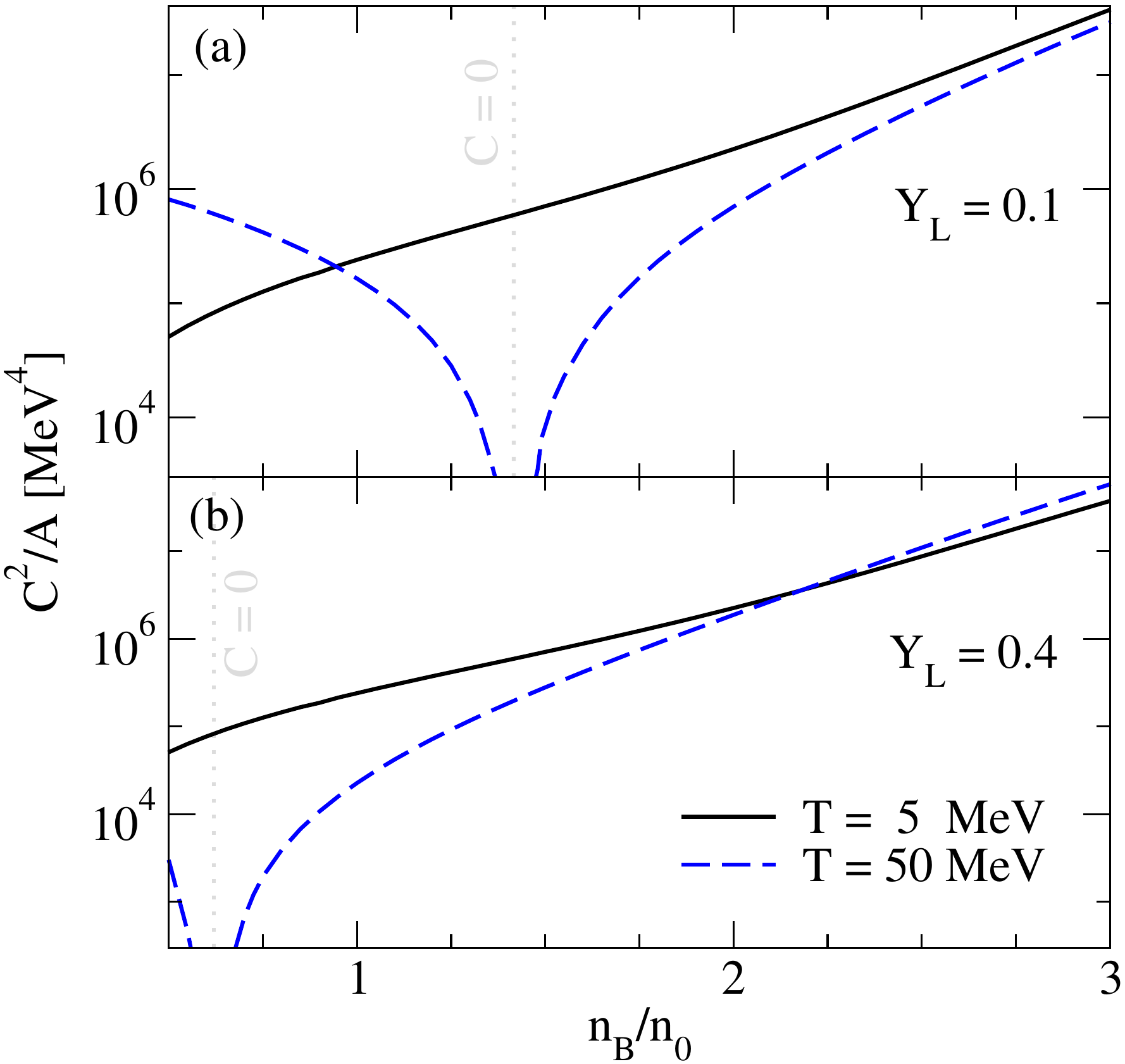}
\caption{ The ratio of beta-disequilibrium--baryon-density
  susceptibility \eqref{eq:C_def} squared $C^2$ over $A$ as a function
  of the baryon density for two values of the temperature for (a)
  $Y_L=0.1$ and (b) $Y_L=0.4$. }
\label{fig:C2A_dens} 
\end{center}
\end{figure}
%-------------------------------------------------

\subsection{Perturbations and bulk viscosity}
\label{sec:bulk_visc}

Now we turn to the discussion of perturbations on the background
equilibrium of matter presented above and concentrate on the
$\beta$-relaxation rates and the bulk viscosity.

Our numerical calculations show that the main contribution to the
beta-disequilibrium--proton-fraction susceptibility $A$
\eqref{eq:A_def} comes from neutrinos whereas electron contribution is
minor. The baryon contributions are much smaller than those of leptons
because of their finite mass, see Eq.~\eqref{eq:A_deg}. The
contribution from $\rho$ meson is negligible in the whole regime of
interest.  The susceptibility $A$ does not depend strongly on the
density and the temperature and has roughly the same order of
magnitude $A\sim 10^{-3}$ MeV$^{-2}$ in the relevant portion of the
phase diagram.

The beta-disequilibrium--proton-fraction susceptibility $C$ is an
increasing function of density and has the same order of magnitude in
the cases of neutron star mergers and supernovas.  At sufficiently
high temperatures $T\gtrsim 30$ MeV $C$ crosses zero at a
temperature-dependent critical value of the density, close to
saturation density.  The vanishing of $C$ arises when the proton
fraction in beta-equilibrated matter is independent of the density
(passing through a minimum in this case).  At the critical density the
system is scale-invariant: it can be compressed and remain in beta
equilibrium.  Thus the bulk viscosity vanishes at the critical
density. 

Figure~\ref{fig:C2A_dens} shows the ratio $C^2/A$ as a function of
density. As seen from the figure, this ratio is temperature-sensitive
only close to the point were $C$ crosses zero.

In Appendix \ref{app:beta_rates} we discuss the rates of neutron decay
and electron capture that combine to establish beta
equilibrium. Electron capture dominates because the neutron decay
process involves antineutrinos the population of which is damped by a
factor of $\exp(-\mu{_\nu}/T)$.

%-------------------------------------------------
\begin{figure}[t] 
\begin{center}
\includegraphics[width=0.9\columnwidth,keepaspectratio]{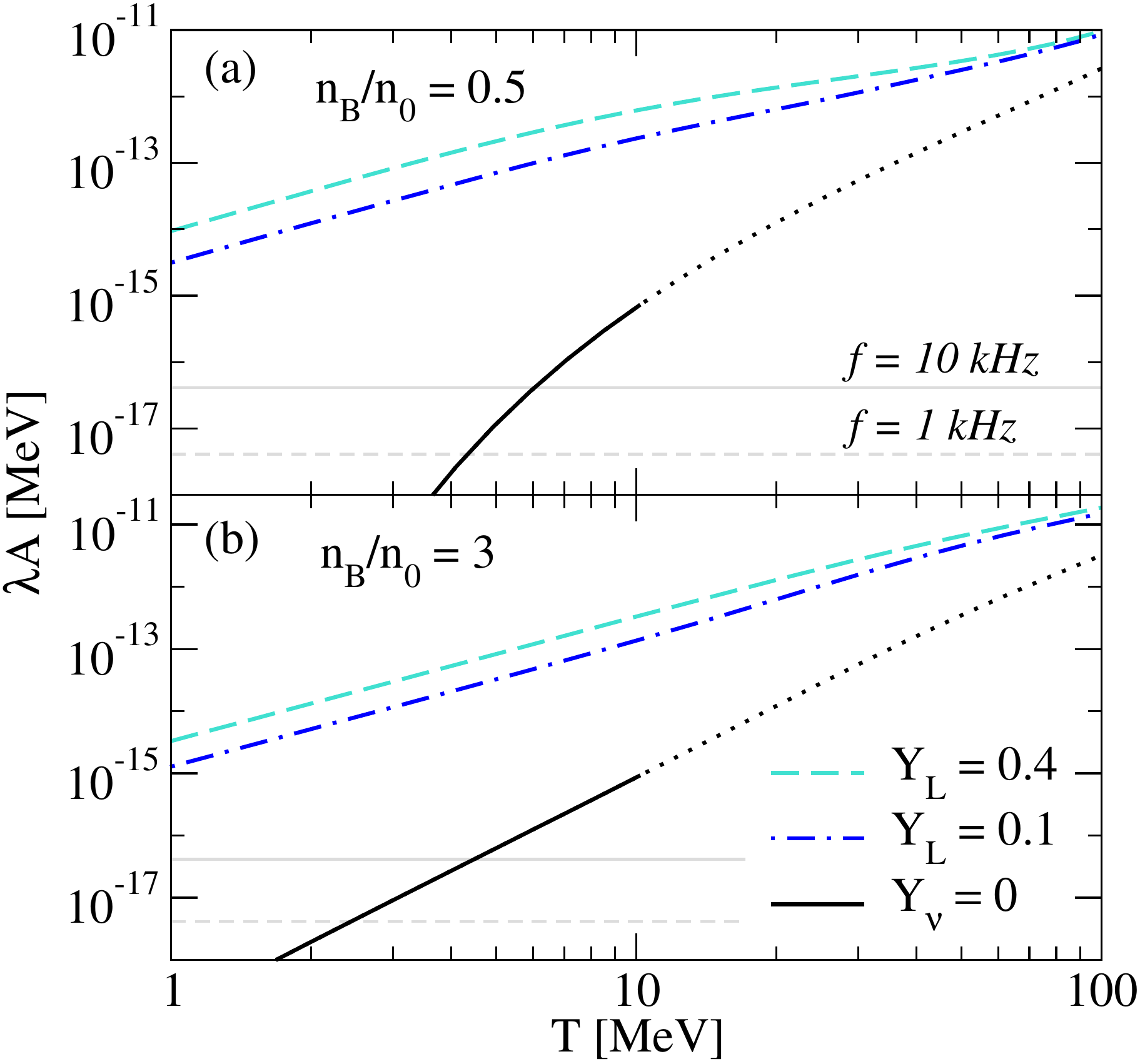}
\caption{ The beta-equilibration relaxation rate $\lambda A$ as a
  function of the temperature for fixed values of the lepton fraction
  for (a) $n_B/n_0=0.5$; (b) $n_B/n_0=3$. The horizontal lines
  correspond to fixed values of $\lambda A=\omega=2\pi f$ for the
  frequencies $f=1$ kHz (dashed line) and $f=10$ kHz (solid
  line). The black dotted lines show the extrapolation of our results 
  for $Y_\nu=0$ case to the high-temperature regime $T\ge 10$~MeV 
  where these are inapplicable. }
\label{fig:lambda_A_temp} 
\end{center}
\end{figure}
%-------------------------------------------------
\begin{figure}[t] 
\begin{center}
\includegraphics[width=0.9\columnwidth,keepaspectratio]{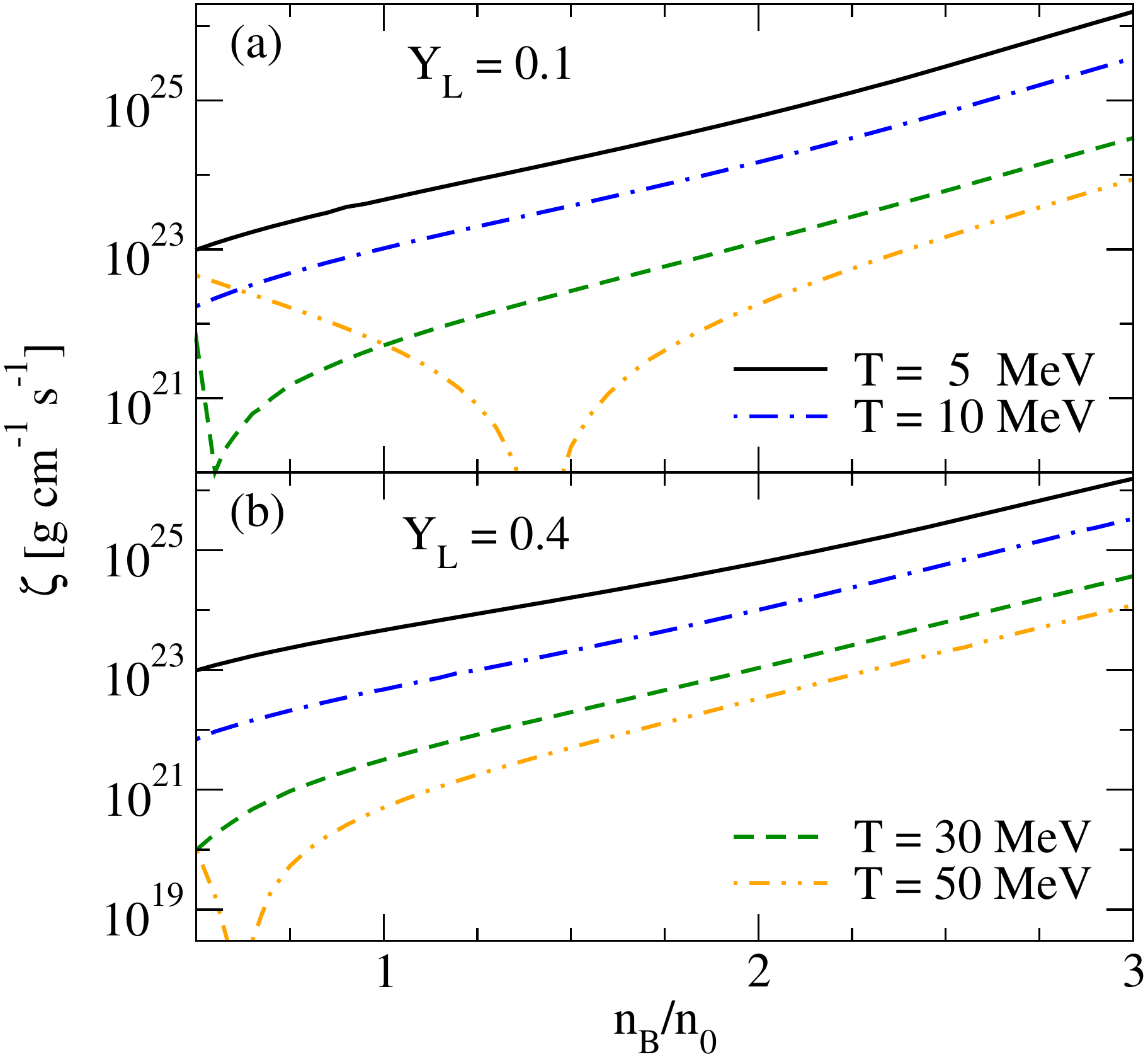}
\caption{ The density dependence of the bulk viscosity for various
  values of the temperature.  The lepton fraction is $Y_L=0.1$ for
  panel (a) and $Y_L=0.4$ for panel (b). }
\label{fig:zeta_dens} 
\end{center}
\end{figure}
%-------------------------------------------------

In Fig.~\ref{fig:lambda_A_temp} we show the beta equilibrium
relaxation rate $\lambda A$, which determines where the bulk viscosity
reaches its resonant maximum [Eq.~\eqref{eq:zeta}].  For comparison
here we show also the case of neutrino-transparent matter (solid
line).  The relaxation rate is slowest in the neutrino-transparent
case and increases with the lepton fraction in the neutrino-trapped
case.

The relaxation rate $\lambda A$ of the neutrino-trapped matter is
several orders of magnitude larger than the oscillation frequencies
$f=\omega/2\pi\lesssim 10$ kHz typical to neutron star mergers and
supernovas.  In Fig.~\ref{fig:lambda_A_temp} the horizontal lines for
different oscillations frequencies intersect the $\lambda A$ curves at
low temperatures $T\lesssim 0.1$\,MeV, indicating that the resonant
maximum occurs at low temperatures where the assumption of neutrino
trapping is no longer valid.  The neutrino-trapped regime lies at
higher temperatures, where the bulk viscosity is independent of the
oscillation frequency and takes the form
$\zeta \approx C^2/(\lambda A^2)$.

In contrast, the neutrino-transparent matter features bulk viscosity
which strongly depends on the oscillation frequency, see
Ref.~\cite{Alford:2019zzz} and Appendix~\ref{app:bulk_visc0} for
details.

We also see from Fig.~\ref{fig:lambda_A_temp}, that $\lambda A$ is
almost independent of the baryon density in the range
$0.5\leq n_B/n_0\leq 3$ for neutrino-trapped matter.  At moderately
temperatures $\lambda$ scales as $\lambda\propto T^2$ for temperatures
$T\leq 10$ MeV: this scaling is clearly seen from
Eqs.~\eqref{eq:lambda2} and \eqref{eq:Gamma2_trap}, which are
applicable as long as the fermions are semidegenerate, i.e.,
$T\le 10$~MeV in the relevant density range.

%------------------ bulk viscosity ---------------

Figure~\ref{fig:zeta_dens} shows the density dependence of the bulk
viscosity for various values of the temperature.  Because the beta
relaxation rate $\lambda A$ is almost independent of the baryon
density, the density dependence of the bulk viscosity follows that of
the susceptibility prefactor $C^2/A$, and, therefore, as noted above,
may drop to zero at a critical density where the system becomes
scale-invariant.  The critical density is present only at sufficiently
high temperatures $T\geq 30$ MeV.

\begin{figure}[t] 
\begin{center}
\includegraphics[width=\columnwidth, keepaspectratio]{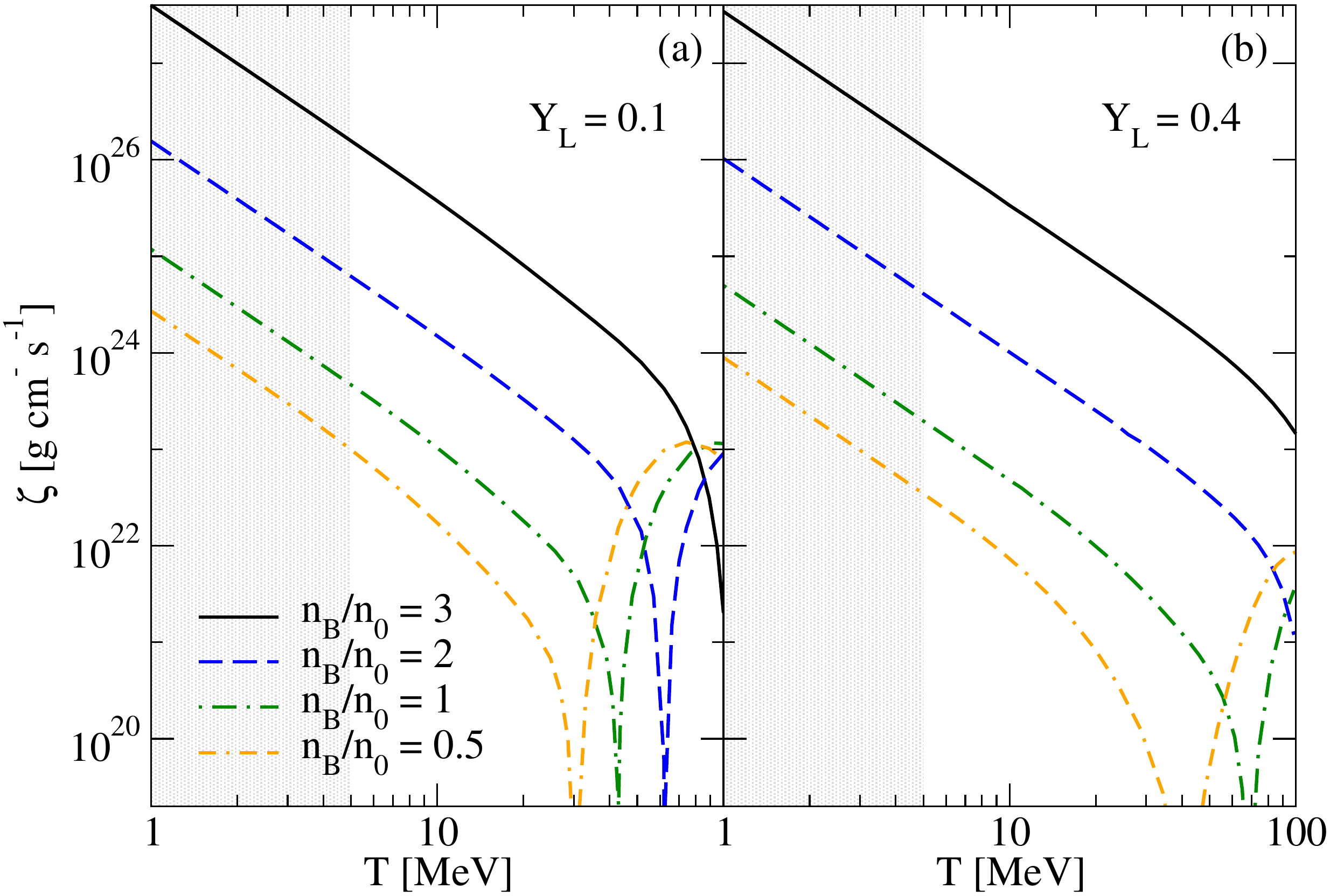}
\caption{ The temperature dependence of the bulk viscosity for DD-ME2 model or several
  values of the baryon density. The lepton fraction is fixed at
  $Y_L=0.1$ for panel (a) and at $Y_L=0.4$ for panel (b). The shaded region shows the extrapolation of our results to
  the neutrinoless regime $T\le 5$~MeV where these are supposed to be
  inapplicable. }
\label{fig:zeta_temp} 
\end{center}
\end{figure}

The temperature dependence of the bulk viscosity is shown in
Fig.~\ref{fig:zeta_temp}.  The temperature dependence of $\zeta$
arises mainly from the temperature dependence of the beta relaxation
rate $\lambda A\propto T^2$ (see Eqs.~\eqref{eq:lambda2} and
\eqref{eq:Gamma2_trap}), so the bulk viscosity decreases as
$\zeta\propto T^{-2}$ in the neutrino-trapped regime, as can be seen
also from Fig.~\ref{fig:zeta_temp}.  This scaling breaks down at
special temperatures where the bulk viscosity has zeros when the
matter becomes scale-invariant (see our discussion of
Fig.~\ref{fig:C2A_dens}).

\begin{figure}[t] 
\begin{center}
\includegraphics[width=\columnwidth, keepaspectratio]{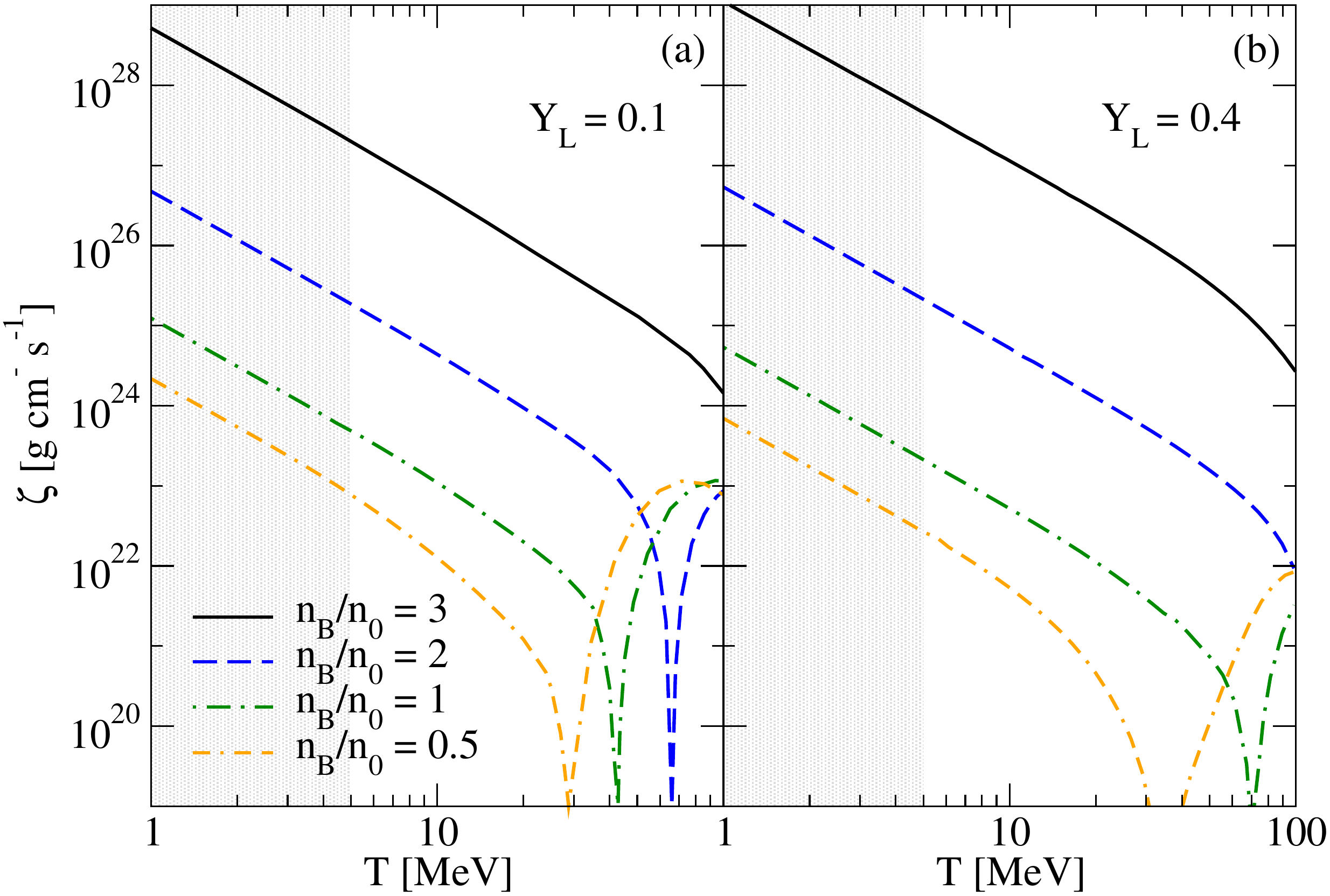}
\caption{ Same as Fig.~\ref{fig:zeta_temp} for model NL3.}
\label{fig:zeta_temp_NL3} 
\end{center}
\end{figure}

In order to check whether the high-temperature behavior found in
Fig.~\ref{fig:zeta_temp} is a universal behavior or is specific to the
EoS we used, we compared our results with those obtained in the
framework of NL3 model, see Fig.~\ref{fig:zeta_temp_NL3}. The figure
shows that the minimums arise independently of the equation of state,
and, therefore, are typical to the high-temperature regime of dense
nuclear matter.

%-------------------------------------------------
%-------------------------------------------------
% -------------------------------------------------

Comparing the results shown in panels (a) and (b) of
Figs.~\ref{fig:zeta_dens} and \ref{fig:zeta_temp} we see that the bulk
viscosity is generally by a factor of few smaller for larger lepton
fractions, which is a consequence of larger values of $\lambda A$ for
higher $Y_L$, as was seen from Fig.~\ref{fig:lambda_A_temp}. However,
the order of magnitude of the bulk viscosity is the same in both
cases.

\begin{figure}[t] 
\begin{center}
\includegraphics[width=\columnwidth, keepaspectratio]{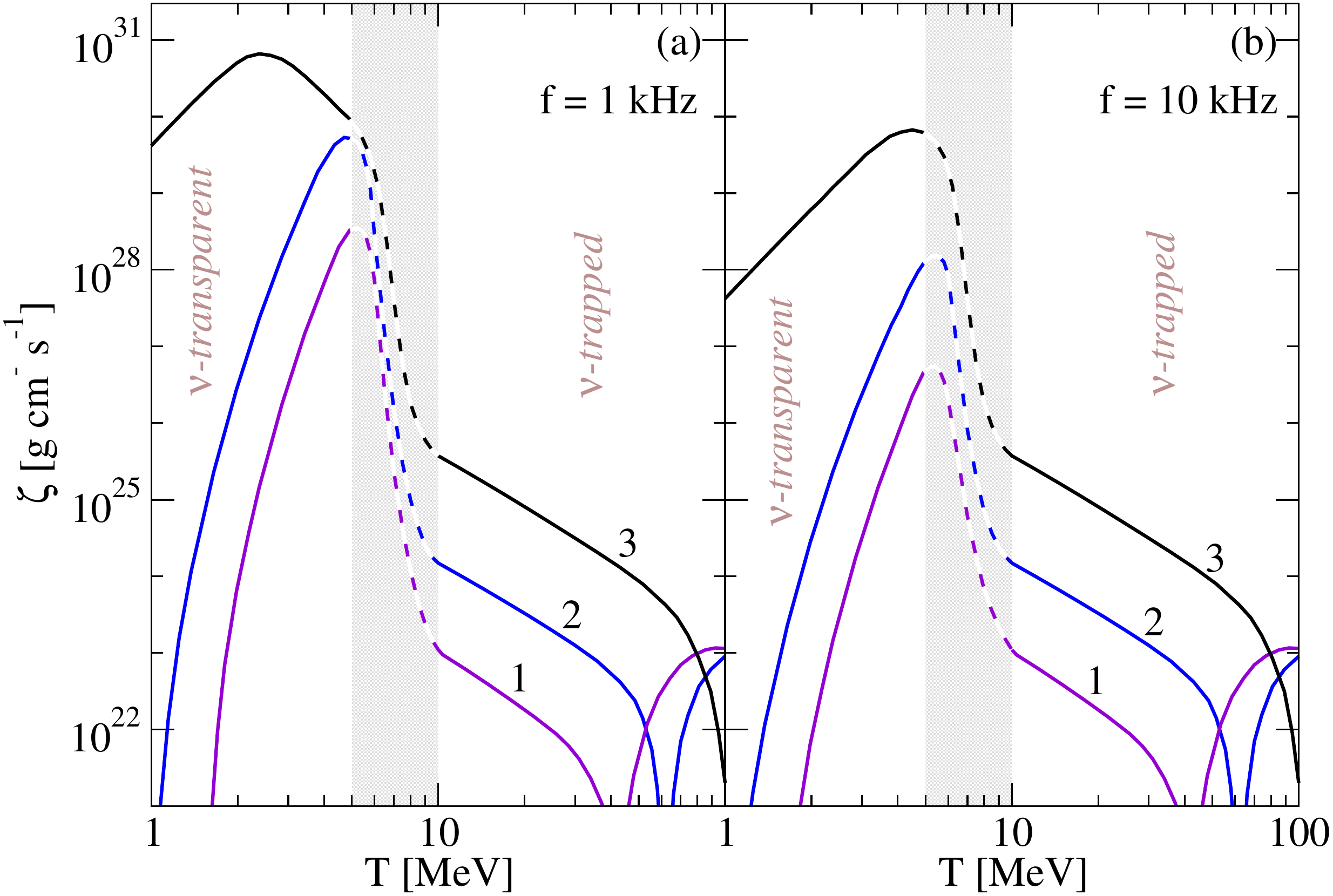}
\caption{ The bulk viscosity for DD-ME2 model for three values of the baryon density $n_B/n_0=1,2,3$ and 
for the lepton fraction $Y_L=0.1$.
In the temperature range $5\leq T\leq 10 $ MeV the 
bulk viscosity was interpolated between the results of
$\nu$-transparent ($T\leq 5$ MeV) and $\nu$-trapped 
($T\geq 10$ MeV) regimes.
The oscillation frequency is fixed at (a) $f=1$ kHz and (b) $f=10$ kHz. }
\label{fig:zeta_temp_combined} 
\end{center}
\end{figure}

\begin{figure}[!] 
\begin{center}
\includegraphics[width=\columnwidth, keepaspectratio]{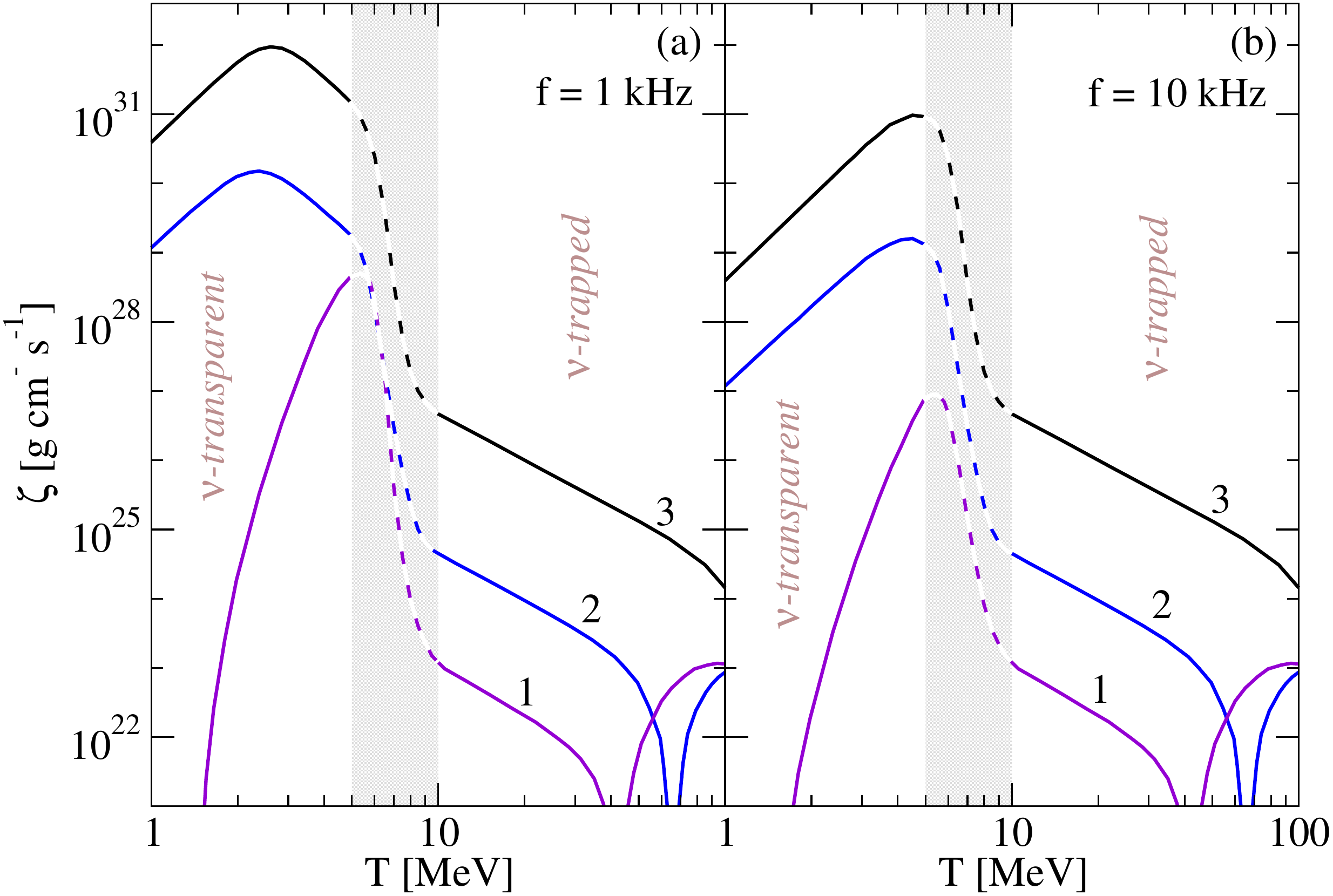}
\caption{ Same as Fig.~\ref{fig:zeta_temp_combined} 
but for model NL3.}
\label{fig:zeta_temp_combined_NL3} 
\end{center}
\end{figure}

In Figs.~\ref{fig:zeta_temp_combined} and
\ref{fig:zeta_temp_combined_NL3} we combine and compare our results
for the neutrino-trapped matter with the results for
neutrino-transparent matter (see also \cite{Alford:2019zzz}).  In the
interval $5\leqslant T\leqslant 10$ MeV we interpolate the numerical
results for the bulk viscosity between the two regimes.  We see that
the bulk viscosity in the neutrino transparent regime is larger, for
two reasons.  First, the beta relaxation rate is slower, so the
resonant peak of the bulk viscosity occurs within its regime of
validity, whereas for neutrino trapped matter the regime of validity
starts at temperatures well above the resonant maximum.  Second, the
prefactor $C^2/A$ is larger in the neutrino-transparent matter, so the
bulk viscosity reaches a higher value at its resonant maximum.

It, therefore, seems likely that bulk viscosity will have its greatest
impact on neutron star mergers in regions of the merger that are
neutrino transparent rather than neutrino trapped
\cite{Alford:2019zzz}.

%-------------------------------------------------
%-------------------------------------------------
%-------------------------------------------------

\section{Conclusions}
\label{sec:conclusions}
 
In this work, we have studied the bulk viscosity of hadronic component
of neutron stars composed of neutrons, protons, electrons, muons, and
neutrinos. The main new ingredient of our study is the trapped
neutrinos, which modify significantly the composition of the
background equilibrium matter. We have derived semianalytical
expressions for the weak interaction rates in the case of trapped
neutrinos and corresponding expressions for the bulk viscosity. Our
numerical study of the relevant quantities displays the following
features:

\smallskip\noindent (a) \underline{Electron capture dominates.}  In
neutrino-trapped matter, beta equilibration, and hence bulk viscosity,
is dominated by the electron capture process and its inverse
\eqref{eq:Urca_2}.  Neutron decay and its inverse \eqref{eq:Urca_1}
involve antineutrinos and are therefore suppressed by factors of
$\exp(-\mu_\nu/T)$.

\smallskip\noindent (b) \underline{Role of susceptibilities}.  The
beta-disequilibrium--baryon-density susceptibility $C$
\eqref{eq:C_def} plays an essential role in the bulk viscosity since
it measures the degree to which matter is driven out of beta
equilibrium when compressed at constant proton fraction (i.e., without
any weak interactions). We find that $C$ vanishes, i.e., the proton
fraction becomes independent of density, at a critical density that is
in the vicinity of saturation density at high temperatures $T\gtrsim 30$~MeV.
The subthermal bulk viscosity vanishes at this critical density
because the equilibrium value of the proton fraction is
density-independent, so compression does not drive the system out of
beta equilibrium.

\smallskip\noindent (c) \underline{Temperature dependence.}  The bulk
viscosity as a function of temperature at fixed oscillation frequency
$\omega$ shows the standard resonant form \eqref{eq:zeta}, with a
maximum when the beta relaxation rate $\lambda A$ matches $\omega$.
At the temperatures where our assumption of neutrino trapping is valid
($T\gtrsim 5\,\MeV$), we are always in the regime where beta
relaxation is fast ($\lambda A$ is greater than typical frequencies
for neutron star density oscillations, which are in the kHz range) so
$\zeta\approx C^2/(\lambda A^2)$.  Since the relaxation rate rises
with temperature (due to increasing phase space), the bulk viscosity
drops with increasing temperature as $\zeta\propto T^{-2}$.  This
scaling can be understood, by noting that the factor $\lambda A$
scales approximately as $T^{-2}$.  Given that the remaining factor
$C^2/A$ mildly depends on the temperature (except the points where it
goes to zero, see Fig.~\ref{fig:C2A_dens}), we recover the
$\zeta \propto T^{-2}$ scaling. At some temperatures and densities the
material becomes scale-invariant, so $C$ goes through zero, driving
the bulk viscosity to zero at those points.

\smallskip\noindent (d) \underline{Dependence on lepton fraction}. The
dependence of the bulk viscosity on lepton fraction can be inferred
from Figs.~\ref{fig:zeta_dens}, \ref{fig:zeta_temp}, and
\ref{fig:zeta_temp_NL3}.  It is seen that the bulk viscosity is
smaller for larger lepton fraction, specifically comparing the cases
of supernova matter with $Y_L = 0.4$ and neutron-star/binary-merger
matter with $Y_L = 0.1$ one finds that the bulk viscosity of supernova
matter is a few times smaller compared to that of neutron-star/binary-merger
matter.

\smallskip\noindent (e) \underline{Effect of neutrino trapping}.  In
astrophysical scenarios of supernova and binary-mergers, the object
formed in the aftermath of these events will cool eventually below the
neutrino trapping temperature $T_{\rm tr}$. Therefore, one may ask  how the bulk viscosity of matter changes as it passes
through $T_{\rm tr}$.  Figure~\ref{fig:zeta_temp_combined} shows the
temperature dependence of the bulk viscosity, where we combine the
results obtained in the neutrino-trapped ($T> T_{\rm tr}$) and
neutrino-transparent matter ($T <T_{\rm tr}$).  It is seen that the
bulk viscosity of matter with neutrinos is several orders of
magnitude smaller than that for the matter which is transparent to
neutrinos.

\smallskip\noindent (f) \underline{Dependence on the
  density-functional model}.  All computations have been carried out
for two alternative relativistic density functional models one being
the density-dependent DD-ME2 parametrization the other being the NL3
nonlinear parametrization. The key results show insignificant
dependence on the chosen model, therefore we conclude that our results
are largely independent of this input (see also Ref~\cite{Yakovlev2018}).

\smallskip\noindent (g) \underline{Relevance for mergers.}  It seems
likely that the presence of neutrino-trapped matter will not be an
important source of bulk viscosity in mergers.
Ref.~\cite{Alford:2019zzz} found that the bulk viscosity in neutrino
{\em transparent} matter was just enough to yield dissipation times in the 20\,ms range. We find that bulk viscosity in neutrino-trapped
matter is thousands of times smaller, so the corresponding dissipation
times are likely to be too long to affect a merger, provided other input in the analysis of Ref.~\cite{Alford:2019zzz} does not change by orders of magnitude.

In the future it would be interesting to extend our discussion to more
complicated compositions of dense matter which would include heavy
baryons such as hyperons and
delta-isobars~\cite{Li2018_PLB,Li2019ApJ}.  We also hope that this
work will help to clarify which dissipative processes are important
enough to be worth including in hydrodynamic simulations of neutron
star mergers.

\section*{Acknowledgments}

We thank Steven Harris, Kai Schwenzer, Alex Haber for discussions and
Tobias Fischer for output tables of supernova simulations and
discussions.  M. A. is supported by the U.S. Department of Energy,
Office of Science, Office of Nuclear Physics under Award No.
DE-FG02-05ER41375.  A. H. and A. S. acknowledge the partial
support of the European COST Action ``PHAROS'' (CA16214) and the State
of Hesse LOEWE-Program in Helmholtz International 
Center for FAIR.  A.S.\ acknowledges the
support by the DFG (Grant No.  SE 1836/4-1).

\appendix

\begin{widetext}

\section{Phase space integrals}
\label{app:rates}

For further computations it is convenient to write the energy
conservation in the form
$\delta(p_0\pm k_0+k'_0-p'_0)=\delta(\ep_p\pm
\ep_{\bar{\nu}/\nu}+\ep_e-\ep_n-\mu_\Delta)$,
where we added and subtracted $\mu_\Delta$ in the argument of the
$\delta$-function, and denoted by $\ep_i$ the energies of the
particles computed from their chemical potentials, \eg,
$\ep_p=p_0-\mu_p$.  To simplify the calculation of the
$\beta$-equilibration rates it is useful to introduce a so-called
``dummy'' integration, so that, by substituting
Eq.~\eqref{eq:matrix_el} into the rates~\eqref{eq:Gamma1p_def} and
\eqref{eq:Gamma2p_def} we obtain
%------------------------------------------------------
\bea \label{eq:Gamma1p}
\Gamma_{1p}(\mu_\Delta) &=& 2\tilde{G}^2 \int d^4q
I_1(q)\, I_2(q),\\ 
 \label{eq:Gamma2p}
\Gamma_{2p}(\mu_\Delta) &=& 2\tilde{G}^2 \int d^4q
I_1(q)\, I_3(q), 
\eea 
%------------------------------------------------------
where 
%------------------------------------------------------
\bea
\label{eq:I1}
I_1(q) &=& \int\!\! \frac{d^3p'}{(2\pi)^3} \int\!\!
  \frac{d^3p}{(2\pi)^3}
  [1-f(p)]f(p') (2\pi)^4\delta^{(4)}(p-p'+q) ,\\
\label{eq:I2}
I_2(q) &=& \int\!\! \frac{d^3k'}{(2\pi)^3}\int\!\!
  \frac{d^3k}{(2\pi)^3} [1-f(k')][1-f(k)] 
  \delta^{(4)}(k+k'-q),\\
\label{eq:I3}
I_3(q) &=& \int\!\! \frac{d^3k'}{(2\pi)^3}\int\!\!
  \frac{d^3k}{(2\pi)^3}f(k) [1-f(k')] 
  \delta^{(4)}(k'-k-q),
\eea
%------------------------------------------------------
with
$\delta^{(4)}(p-p'+q)=\delta(\vecp-\vecp'+\vecq)
\delta(\ep_p-\ep_{p'}+\omega-\mu_\Delta)$,
$\delta^{(4)}(k'\pm k-q)=\delta(\veck'\pm \veck-\vecq)
\delta(\ep_{k'}\pm \ep_{k}-\omega)$.
The rates of the inverse processes~\eqref{eq:Gamma1n_def} and
\eqref{eq:Gamma2n_def} can be obtained from Eqs.~\eqref{eq:Gamma1p}
and \eqref{eq:Gamma2p} by replacing $f(p_i)\to 1-f(p_i)$ for all
particles.  Thus, the problem reduces to the computation of three
$q$-dependent integrals $I_1(q)$, $I_2(q)$ and $I_3(q)$ in
Eq.~\eqref{eq:I1}-\eqref{eq:I3}.

To compute the integral $I_1(q)$ we use the following identity between 
the Fermi $f(p)$ and Bose $g(p)$ functions
%-----------------------------------------------------------
\be\label{eq:fermi_bose}
 f(p')[1-f(p)] = g(q) [f(p)-f(p')].
\ee 
%-----------------------------------------------------------
Then, integrating over neutron momentum and separating the angular
part of the remaining integral we obtain
($\tilde{\omega}= \omega-\mu_\Delta$)
%-----------------------------------------------------------
\bea I_1(q)
=(2\pi)^{-1} g(q)\int_0^\infty\! dp \, p^2[f(\epsilon_p)-f(\epsilon_{p}+\tilde{\omega})]
\int_{-1}^{1} dx\,\delta(\tilde{\omega} +\epsilon_p-\epsilon_{p+q}),
\eea 
%-----------------------------------------------------------
where $x$ is the cosine of the angle between $\bm p$ and $\bm q$.
Using the nonrelativistic spectrum for the nucleons
$\epsilon_i = p^2/2m^* -\mu_i^*$, $i=\{p,n\}$, we obtain for $I_1(q)$
%-----------------------------------------------------------
\bea 
 I_1(q) = (2\pi)^{-1}  g(q) \int_0^\infty\! dp \,
  p^2[f(\epsilon_p)-f(\epsilon_{p}+\tilde{\omega})]
\frac{m^*}{pq}\theta(1-\vert x_0\vert),
\eea 
%-----------------------------------------------------------
where $x_0$ is the zero of the argument of the $\delta$-function, \ie,
%-----------------------------------------------------------
\be
x_0 =
\frac{m^*}{pq}\left( \mu_n^*-\mu_p^* +\tilde{\omega}-\frac{q^2}{2m^*} \right).
\ee
%-----------------------------------------------------------
The step-function sets the following limit on
the momentum of a particle
%-----------------------------------------------------------
\bea 
p\geq p_{\rm  min}=
  \frac{m^*}{q}\left\vert\mu_n^*-\mu_p^* +\tilde{\omega}-\frac{q^2}{2m^*}
  \right\vert.
\eea 
%-----------------------------------------------------------
Taking now the momentum integral we finally obtain (note that $g(q)$
depends only on $\omega$)
%-----------------------------------------------------------
\bea\label{eq:I1_final}
I_1(q) = g(\tilde{\omega})\frac{m^{*2}}{2\pi q}\int_{\epsilon_{\rm min}}^{\infty} d\epsilon_p
\left[f(\epsilon_p)-f(\epsilon_{p}+\tilde{\omega})\right] 
= g(\omega-\mu_\Delta)\frac{m^{*2}T}{2\pi  q}\ln \Bigg\vert \frac{1+\exp\left(-\frac{\epsilon_{\rm
min}}{T}\right)} {1+\exp\left(-\frac{\epsilon_{\rm
min}+\omega-\mu_\Delta}{T}\right)}\Bigg\vert, 
\eea 
%-----------------------------------------------------------
where the lower limit $\epsilon_{\rm min}$ follows from $p_{\rm min}$,
i.e.,
%------------------------------------------------------
\bea\label{eq:e_min}
\ep_{\rm min} =\frac{p_{\rm  min}^2}{2m^*}-\mu_p^*
=  \frac{m^*}{2q^2}\bigg(\mu_n^*-\mu_p^* +\omega
-\mu_\Delta -\frac{q^2}{2m^*}\bigg)^2-\mu_p^*.
\eea 
%------------------------------------------------------

We now transform the second integral~\eqref{eq:I2}
% -----------------------------------------------------------
\bea\label{eq:int3}
 I_2(q) 
&=& \int\!\!
\frac{d^3k'}{(2\pi)^3}\int\!\! \frac{d^3k}{(2\pi)^3} 
[1-f(k')][1-f(k)]\delta(\ep_{k'} +\ep_k-\omega)
\delta(\veck+\veck'-\vecq)\nonumber\\
&=& \int\!\! \frac{k^2dk}{(2\pi)^5}
[1-f(\omega-\ep_k)][1-f(\ep_k)] \int_{-1}^{1}
dy\,\delta\left(k+\sqrt{k^2+q^2-2kqy}-\omega'\right),
\eea 
%------------------------------------------------------
where we neglected the electron mass and introduced the short-hand
notation $\omega'=\omega+\mu_e-\mu_{\nu}$.  The argument of the
$\delta$-function is zero at
%------------------------------------------------------
\bea
y_0 =\frac{q^2-\omega'^2 +2\omega' k}{2kq},
\eea
%------------------------------------------------------
therefore for the $\delta$-function, we obtain (note that the
$\delta$-function also implies $\omega'-k \geq0$)
%------------------------------------------------------
\bea\label{eq:delta2}
\delta\left(k+\sqrt{k^2+q^2-2kqy}-\omega'\right)
=\frac{\omega'-k}{kq}\theta(\omega'-k)\delta(y-y_0), 
\eea
%------------------------------------------------------
and
%------------------------------------------------------
\bea 
 \int_{-1}^{1}
dy\delta\left(k+\sqrt{k^2+q^2-2kqy}-\omega'\right) = 
\frac{\omega'-k}{kq}\theta(\omega'-k)\theta(1-\vert y_0\vert).
\eea 
%------------------------------------------------------
The condition $|y_0|\leq 1$ sets the following limits on the momentum
$k$
%------------------------------------------------------
\bea 
-2kq\leq q^2-\omega'^2 +2\omega' k\leq 2kq,
\eea 
%------------------------------------------------------
therefore the following two inequalities must be satisfied
simultaneously
%------------------------------------------------------
\bea \label{eq:cond}
(\omega'+q)(q-\omega'+2k)\geq 0,\qquad
(\omega'-q)(q+\omega'-2k)\geq 0.
\eea 
%------------------------------------------------------
Note that because of the condition $\omega'\geq k\geq 0$ we have
$\omega'+q\geq 0$. Then the first condition in Eq.~\eqref{eq:cond}
gives $k\geq (\omega'-q)/2$. Next, if $\omega'\geq q$, the second
condition implies $k\leq (\omega'+q)/2\leq \omega'$.  If
$\omega'\leq q$ instead, then the first condition is satisfied
automatically, and the second one gives
$k\geq(\omega'+q)/2\geq \omega'$, which is not allowed. Substituting
these results into Eq.~\eqref{eq:int3} we obtain
%------------------------------------------------------
\bea\label{eq:I2_final}
 I_2(q) = \theta(\omega'-q)\frac{1}{q(2\pi)^5}\int_{(\omega'-q)/2}^{(\omega'+q)/2} 
 kdk\,(\omega'-k)[1-f(\omega-\ep_k)][1-f({\ep}_k)].
\eea 
%------------------------------------------------------

The integral $I_3(q)$ given by Eq.~\eqref{eq:I3} is transformed as
follows
%-----------------------------------------------------------
\bea\label{eq:int3a}
 I_3(q) 
&=& \int\!\!
\frac{d^3k'}{(2\pi)^3}\int\!\! \frac{d^3k}{(2\pi)^3} 
f(k)[1-f(k')] \delta(\ep_{k'} -\ep_k-\omega)
\delta(\veck'-\veck-\vecq)\nonumber\\
&=& \int\!\! \frac{k^2dk}{(2\pi)^5}
f(\ep_k)[1-f(\ep_k+\omega)] \int_{-1}^{1}
dy\,\delta\left(-k+\sqrt{k^2+q^2+2kqy}-\omega'\right).
\eea 
%------------------------------------------------------
The argument of the $\delta$-function is zero at
%------------------------------------------------------
\bea
y_0 =-\frac{q^2-\omega'^2 -2\omega' k}{2kq},
\eea
%------------------------------------------------------
and for the $\delta$-function we obtain (note that the
$\delta$-function also implies $\omega'+k \geq0$)
%------------------------------------------------------
\bea
\delta\left(-k+\sqrt{k^2+q^2+2kqy}-\omega'\right)
=\frac{\omega'+k}{kq}\theta(\omega'+k)\delta(y-y_0),
\eea
%------------------------------------------------------
therefore
%------------------------------------------------------
\bea 
 \int_{-1}^{1}
dy\delta\left(-k+\sqrt{k^2+q^2+2kqy}-\omega'\right) = 
\frac{\omega'+k}{kq}\theta(\omega'+k)\theta(1-\vert y_0\vert).
\eea 
%------------------------------------------------------
The condition $|y_0|\leq 1$ sets the following limits on the momentum
$k$
%------------------------------------------------------
\bea 
-2kq\leq q^2-\omega'^2 -2\omega' k\leq 2kq,
\eea 
%------------------------------------------------------
therefore the following two inequalities must be satisfied simultaneously
% ------------------------------------------------------
\bea \label{eq:cond1}
(\omega'+q)(q-\omega'-2k)\leq 0,\qquad
(\omega'-q)(q+\omega'+2k)\leq 0.
\eea 
%------------------------------------------------------
Because of the condition $\omega'+ k\geq 0$ we have
$\omega'+q+2k\geq 0$. Then the second condition in
Eq.~\eqref{eq:cond1} gives $q\geq\omega'$. Next, if $\omega'\geq -q$,
the first condition implies $k\geq (q-\omega')/2\geq -\omega'$.  If
$\omega'\leq -q$ instead, then the first condition implies
$k\leq(q-\omega')/2\leq -\omega'$, which is not allowed. Substituting
these results into Eq.~\eqref{eq:int3a} we obtain
%------------------------------------------------------
\bea\label{eq:I3_final}
 I_3(q) = \theta(q-|\omega'|)\frac{1}{q(2\pi)^5}\int_{(q-\omega')/2}^{\infty} 
 kdk\, (\omega'+k) f(\ep_k)[1-f(\ep_k+\omega)].
\eea 
%------------------------------------------------------

Combining now Eqs.~\eqref{eq:Gamma1p}, \eqref{eq:Gamma2p},
\eqref{eq:I1_final}, \eqref{eq:I2_final} and \eqref{eq:I3_final}
we obtain the final formulas for $\Gamma_{1p}$ and $\Gamma_{2p}$
given in the main text by Eqs.~\eqref{eq:Gamma1p_final} and 
\eqref{eq:Gamma2p_final}. The rates of the inverse processes 
$\Gamma_{1n}$ and $\Gamma_{2n}$ given by Eqs.~\eqref{eq:Gamma1n_final} and 
\eqref{eq:Gamma2n_final} can be obtained in an analogous manner.

For the derivatives of the rates $\Gamma_{1p}$ and $\Gamma_{1n}$ we obtain 
%----------------------------------------------------------
\bea \label{eq:Gamma1p_vs_mu}
\frac{\partial\Gamma_{1p}(\mu_\Delta)}
{\partial\mu_\Delta}\bigg\vert_{\mu_\Delta=0}
&=& 
 \frac{m^{*2}\tilde{G}^2 }{8\pi^5}T^5 \int_{-\alpha_e+\alpha_\nu}^\infty dy\, g(y) 
\int_0^{y+\alpha_e-\alpha_\nu} dz\,
\bigg\{[1+g(y)] 
 \ln  \Bigg\vert \frac{1+\exp\left(-y_0\right)} 
{1+\exp\left(-y_0-y\right)}\Bigg\vert \nonumber\\
&&-  f(y_0+y)-\left[  f(y_0+y)  - f(y_0) \right]
\frac{m^*}{z^2T}\bigg(\alpha_n-\alpha_p+y
-z^2\frac{T}{2m^*}\bigg) \bigg\}\nonumber\\
&&\times \int_{x_{\rm min}}^{x_{\rm max}} 
 dx\, (x-\alpha_{\nu})(y+\alpha_e-x)[1-f(x)]f(x-y),\\
%----------------------------------------------------------
\label{eq:Gamma1n_vs_mu}
\frac{\partial\Gamma_{1n}(\mu_\Delta)}
{\partial\mu_\Delta}\bigg\vert_{\mu_\Delta=0}\nonumber &=& 
 \frac{m^{*2} \tilde{G}^2}{8\pi^5} T^5\int_{-\alpha_e+\alpha_\nu}^\infty dy\,
[1+g(y)] \int_0^{y+\alpha_e-\alpha_\nu} dz\,
\bigg\{g(y) \ln  \Bigg\vert \frac{1+\exp\left(-y_0\right)} 
{1+\exp\left(-y_0-y\right)}\Bigg\vert \nonumber\\
&& - f(y_0+y) -\left[f(y_0+y) - f(y_0) \right]
\frac{m^*}{z^2T}\bigg(\alpha_n-\alpha_p+y
-z^2\frac{T}{2m^*}\bigg) \bigg\}\nonumber\\
&& \int_{x_{\rm min}}^{x_{\rm max}} 
 dx\, (x-\alpha_{\nu})(y+\alpha_e-x) f(x)[1-f(x-y)].
\eea
%----------------------------------------------------------
In the same way we obtain for $\Gamma_{2p}$ and $\Gamma_{2n}$
%------------------------------------------------------
\bea\label{eq:Gamma2p_vs_mu}
 \frac{\partial\Gamma_{2p}(\mu_\Delta)}
{\partial\mu_\Delta}\bigg\vert_{\mu_\Delta=0}&=&
 \frac{m^{*2} \tilde{G}^2}{8\pi^5} T^5 \int_{-\infty}^\infty 
 dy\, g(y)\int_{|y+\alpha_e-\alpha_{\nu}|}^{\infty} dz\,
\Bigg\{[1+g(y)] \ln \bigg\vert \frac{1+\exp\left(-y_0\right)} 
  {1+\exp\left(-y_0-y)\right)}\bigg\vert \nonumber\\
 && -f(y_0+y)-[f(y_0+y)-f(y_0)]
\frac{m^*}{z^2T}\bigg(\alpha_n-\alpha_p+y-z^2\frac{T}{2m^*}\bigg) 
\Bigg\} \nonumber\\
   &&\times
\int_{\bar{x}_{\rm min}}^{\infty}  dx\, (x+\alpha_{\nu})
(y+\alpha_e+x)f(x)[1-f(x+y)],\\
%------------------------------------------------------
\label{eq:Gamma2n_vs_mu}
 \frac{\partial\Gamma_{2n}(\mu_\Delta)}
{\partial\mu_\Delta}\bigg\vert_{\mu_\Delta=0}&=&
 \frac{m^{*2} \tilde{G}^2}{8\pi^5} T^5 \int_{-\infty}^\infty 
 dy\,[1+g(y)]\int_{|y+\alpha_e-\alpha_{\nu}|}^{\infty} dz\,
\Bigg\{g(y) \ln \bigg\vert \frac{1+\exp\left(-y_0\right)} 
 {1+\exp\left(-y_0-y)\right)}\bigg\vert\nonumber\\
 && -f(y_0+y)- [f(y_0+y)-f(y_0)]
\frac{m^*}{z^2T}\bigg(\alpha_n-\alpha_p+y-z^2\frac{T}{2m^*}\bigg) 
\Bigg\} \nonumber\\
 &&\times \int_{\bar{x}_{\rm min}}^{\infty}  
dx\, (x+\alpha_{\nu})(y+\alpha_e+x)f(x+y)[1-f(x)].
\eea 
%------------------------------------------------------
From these expressions, it is straightforward to obtain 
Eqs.~\eqref{eq:lambda1}, \eqref{eq:lambda_1_no_nu}, \eqref{eq:lambda2}
and \eqref{eq:lambda_2_no_nu} of the main text.

\subsection{Low-temperature Urca rates}

In the case of highly degenerate matter we have $\mu_i/T\to \infty$, 
therefore $\epsilon_{\rm min}/T\to \pm \infty$. Thus we find from
Eq.~\eqref{eq:I1_final} (for $\mu_\Delta =0$)
%-----------------------------------------------------------
\bea\label{eq:I1_deg}
I_1(q)= \frac{m^{*2}\omega}{2\pi q}g(\omega)\theta
(-\epsilon_{\rm min}).
\eea 
%-----------------------------------------------------------
In terms of momenta the condition $\ep_{\rm min}\leq 0$ can be written as
$\vert p_{Fn}^2-p_{Fp}^2 -q^2 \vert \leq 2q p_{Fp} $,
where we neglected $\omega\sim T$ terms, therefore
%------------------------------------------------------
\be \label{eq:theta2}
\theta (-\epsilon_{\rm min}) =
\theta(p_{Fn}+p_{Fp}-q)\theta(q-\vert p_{Fn}-p_{Fp}\vert). 
\ee 
%------------------------------------------------------
In the case of neutrino-transparent matter, we can also set $q=p_{Fe}$,
therefore for Eq.~\eqref{eq:I1_deg} we obtain
%------------------------------------------------------
\bea\label{eq:I1_deg1}
I_1(q)= \frac{m^{*2}\omega }{2\pi q}
g(\omega)\theta( p_{Fp}+p_{Fe} -p_{Fn}).
\eea 
%------------------------------------------------------

To obtain the low-temperature limit of integral $I_2$
in the case of neutrino-transparent matter we drop the 
neutrino momentum and neutrino distribution
and approximate $\vert \bm k'\vert=p_{Fe}$
in the first equation of~\eqref{eq:int3}, which gives 
%-----------------------------------------------------------------
\bea
 I_2(q) = \int\!\!
\frac{k^2dk}{(2\pi)^3}\int\!\! \frac{dk'}{(2\pi)^3} [1-f(k')]
\int d\Omega_{k}\, \delta(\ep_{k'} +\ep_k-\omega) 
\delta(p_{Fe} -\vert \vecq\vert ).
\eea
%-----------------------------------------------------------------
Performing the integrations over $k'$ and $\Omega_k$ we obtain
%-----------------------------------------------------------------
\bea\label{eq:I2_deg}
 I_2(q) = \frac{4\pi}{(2\pi)^6}
\delta(p_{Fe} -\vert \vecq\vert)
\int_0^\infty\!\! dk\,k^{2} [1-f(\omega-\ep_k)].%[1-f(\ep_k)].
\eea
%-----------------------------------------------------------------
Substituting Eqs.~\eqref{eq:I1_deg1} and \eqref{eq:I2_deg}
into the neutron decay rate~\eqref{eq:Gamma1p} we find
%-----------------------------------------------------------------
\bea \label{eq:Gamma1_lowT}
\Gamma_{1p}
= \frac{m^{*2}\tilde{G}^2}{4\pi^5}T^5 p_{Fe} 
\theta (p_{Fp}+p_{Fe} -p_{Fn}) 
\int_0^{\infty}\!\! dx\, x^{2} 
\int_{-\infty}^{\infty} dy\, y g(y)f(x-y).
\eea
%-----------------------------------------------------------------
In the same manner we can obtain the low-$T$ result for $\Gamma_{2n}$
in the neutrino-transparent matter
%-----------------------------------------------------------------
\bea \label{eq:Gamma2_lowT}
\Gamma_{2n} 
= \frac{m^{*2} \tilde{G}^2}{4\pi^5}T^5 p_{Fe} 
\theta (p_{Fp}+p_{Fe} -p_{Fn}) 
\int_0^{\infty}\!\! dx\, x^{2}
\int_{-\infty}^{\infty} dy\, y [1+g(y)]f(x+y).
\eea
%-----------------------------------------------------------------
The integrals appearing in Eqs.~\eqref{eq:Gamma1_lowT} and
\eqref{eq:Gamma2_lowT} can be computed successively 
%------------------------------------------------------
\bea 
\int\limits_{-\infty}^{\infty} dy\ y g(y) f(x-y)
=\int_{-\infty}^{\infty} dy\, y [1+g(y)]f(x+y)
=\frac{1}{2} \frac{x^2+\pi^2}{1+e^{x}},
\eea
%------------------------------------------------------
 and
%------------------------------------------------------
\bea
 \frac{1}{2}\int_0^\infty dx\ x^2
\frac{x^2+\pi^2}{1+e^x} = \frac{3}{4}
\left[\pi^2 \zeta(3) + 15 \zeta(5)\right]
 = 20.5633,
\eea
%------------------------------------------------------
from which we obtain the low-$T$ results~\eqref{eq:Gamma1_deg}
and \eqref{eq:Gamma2_deg} of the main text.

Next, we find the low-temperature limit of
Eqs.~\eqref{eq:lambda_1_no_nu} and \eqref{eq:lambda_2_no_nu}. By
replacing $\omega\to\omega-\mu_\Delta$ in Eq.~\eqref{eq:I1_deg1} we
find
%------------------------------------------------------
\bea\label{eq:I1q_partial_deg}
\frac{\partial I_1}{\partial \mu_\Delta}\bigg\vert_{\mu_\Delta=0} =
  \Big(\frac{\omega}{T}[1+g(\omega)]-  1\Big) \frac{m^{*2}}{2\pi q}
  g(\omega)\theta( p_{Fp}+p_{Fe} -p_{Fn}).
\eea 
%------------------------------------------------------
Therefore, the low-temperature limit for $\lambda_1$ can be found
by replacing $y\to \{y[1+g(y)]-1\}/T$
in Eq.~\eqref{eq:Gamma1_lowT}. Thus
%------------------------------------------------------
\bea\label{eq:lambda1_lowT}
\lambda_1 = \frac{m^{*2}\tilde{G}^2 }{4\pi^5}T^4 p_{Fe}
\theta(p_{Fp}+p_{Fe}-p_{Fn})
\int_0^\infty dx\, x^2 \int_{-\infty}^\infty dy\,
g(y)\big\{y[1+g(y)]-1\big\}f(x-y).
\eea
%------------------------------------------------------
Similarly, $\lambda_2$ can be obtained from Eq.~\eqref{eq:Gamma2_lowT}
by replacing $y\to -[yg(y)-1]/T$
%------------------------------------------------------
\bea\label{eq:lambda2_lowT}
\lambda_2
= \frac{m^{*2}\tilde{G}^2 }{4\pi^5}T^4p_{Fe}
\theta( p_{Fp}+p_{Fe} -p_{Fn})
\int_0^\infty dx\, x^2 \int_{-\infty}^\infty dy\,
[1+g(y)][1-yg(y)]f(x+y).
\eea 
%------------------------------------------------------
The double integrals in Eqs.~\eqref{eq:lambda1_lowT} and
\eqref{eq:lambda2_lowT} are identical and are equal to $17\pi^4/120$. 
Substituting this value we obtain the final results given by 
Eqs.~\eqref{eq:lambda1_deg} and \eqref{eq:lambda2_deg}.

In the case where neutrinos remain trapped in the degenerate regime,
one finds $I_2(q)=0$, because in this limit the distribution function
of antineutrinos is exponentially suppressed. To obtain the
low-temperature limit of $I_3(q)$ we substitute $k=p_{F\nu}$ and
$\omega'=\mu_e-\mu_\nu=p_{Fe}-p_{F\nu}$ in Eq.~\eqref{eq:I3_final}
($\ep_k\equiv k-p_{F\nu}$) to find
%------------------------------------------------------
\bea\label{eq:I3_deg}
 I_3(q) = \frac{p_{Fe} p_{F\nu}}{q(2\pi)^5} \theta(1-\vert y_0\vert)
\int_{-p_{F\nu}}^\infty d\ep_k\,f(\ep_k) [1-f(\ep_k+\omega)].
\eea 
%------------------------------------------------------
The inequalities~\eqref{eq:cond1} in this approximation 
are independent of $\ep_k$ and imply
%------------------------------------------------------
\bea 
\vert p_{Fe}-p_{F\nu}\vert\leq q\leq p_{Fe}+p_{F\nu}.
\eea 
%------------------------------------------------------
Combining now Eqs.~\eqref{eq:Gamma2p}, \eqref{eq:I1_deg} and
\eqref{eq:I3_deg} and approximating $p_{F\nu}/T\to \infty$ for $\Gamma_2$ we obtain
%------------------------------------------------------
\bea
\Gamma_2 = \frac{m^{*2}\tilde{G}^2 }{8\pi^5}T^3p_{Fe}p_{F\nu} 
\int_{|p_{Fe}-p_{F\nu}|}^{p_{Fe}+p_{F\nu}} 
dq\, \theta(p_{Fn}+p_{Fp}-q) \theta(q-|p_{Fn}-p_{Fp}|)
\int_{-\infty}^\infty dx\, f(x)
\int_{-\infty}^\infty dy\, y g(y) f(-x-y).\quad
\eea 
%------------------------------------------------------
The last two integrals give $2\pi^2/3$, and for the
$q$-integral we have
%------------------------------------------------------
\bea
&&\int_{|p_{Fe}-p_{F\nu}|}^{p_{Fe}+p_{F\nu}} 
dq~\theta(p_{Fn}+p_{Fp}-q) \theta(q-|p_{Fn}-p_{Fp}|)\nonumber\\
&&= (p_{Fe}+p_{F\nu}-| p_{Fn}-p_{Fp}|)
\theta(p_{Fn}+p_{Fp}-p_{Fe}-p_{F\nu})
\theta(p_{Fe}+p_{F\nu}-|p_{Fn}-p_{Fp}|)
\theta(|p_{Fn}-p_{Fp}|-|p_{Fe}-p_{F\nu}|)\nonumber\\
&&+(p_{Fn}+p_{Fp}-| p_{Fe}-p_{F\nu}|)
\theta(p_{Fe}+p_{F\nu}-p_{Fn}-p_{Fp})
\theta(p_{Fn}+p_{Fp}-|p_{Fe}-p_{F\nu}|)
\theta(|p_{Fe}-p_{F\nu}|-|p_{Fn}-p_{Fp}|)\nonumber\\
&&+(p_{Fe}+p_{F\nu}-|p_{Fe}-p_{F\nu}|)
\theta(p_{Fn}+p_{Fp}-p_{Fe}-p_{F\nu})
\theta(|p_{Fe}-p_{F\nu}|-|p_{Fn}-p_{Fp}|)\nonumber\\
&&+(p_{Fn}+p_{Fp}-| p_{Fn}-p_{Fp}|)
\theta(p_{Fe}+p_{F\nu}-p_{Fn}-p_{Fp})
\theta(|p_{Fn}-p_{Fp}|-|p_{Fe}-p_{F\nu}|).
\eea 
%------------------------------------------------------
In neutron star matter we have typically $p_{Fn}+p_{Fp}\geq p_{Fe}+p_{F\nu}
\geq |p_{Fn}-p_{Fp}|\geq |p_{Fe}-p_{F\nu}|$, and we obtain the final result
given by Eq.~\eqref{eq:Gamma2_trap}.

\section{Computation of susceptibilities $A_j$}
\label{app:A_coeff}

To compute the susceptibilities $A_{ij}$ given by Eq.~\eqref{eq:A_j}
we use the following formula for the particle densities
%------------------------------------------------------
\bea\label{eq:dens}
n_i =\frac{g_i}{2\pi^2}\int_0^\infty p^2dp\, [f_i(p)-\bar{f}_i(p)],
\eea 
%------------------------------------------------------
where $g_i$ is the degeneracy factor, and  $f(p)$ and $\bar{f}(p)$
are the distribution functions for particles and antiparticles, respectively.
For neutrons, protons, and electrons we have $g_i=2$, 
and for neutrinos $g_\nu=1$.

Differentiating the left and right sides of Eq.~\eqref{eq:dens} 
with respect to $n_j$ and exploiting the expressions 
%------------------------------------------------------
\bea\label{eq:fermi_i}
\frac{\partial f_i}{\partial n_j} =-f_i(1-f_i)\frac{1}{T}
\left(\frac{m^*}{\sqrt{m^{*2}+p^2}}\frac{\partial m^*}{\partial n_j} 
-\frac{\partial \mu^*_i}{\partial n_j}\right),\qquad
\frac{\partial \bar{f}_i}{\partial n_j} =-\bar{f}_i(1-\bar{f}_i)\frac{1}{T}
\left(\frac{m^*}{\sqrt{m^{*2}+p^2}}\frac{\partial m^*}{\partial n_j} 
+\frac{\partial \mu^*_i}{\partial n_j}\right),
\eea
%------------------------------------------------------
in the case of baryons we obtain
%------------------------------------------------------
\bea\label{eq:matrix_eq}
\delta_{ij}=-\left(\frac{\partial m^*}{\partial n_j}\right)
{I}_{1i}^- +\left(\frac{\partial
\mu^*_i}{\partial n_j}\right) {I}_{0i}^+,
\eea
%------------------------------------------------------
where
%------------------------------------------------------
\bea\label{eq:I_rel}
{I}^{\pm}_{q i}= \frac{1}{\pi^2 T}\int_0^\infty p^2 dp
\left(\frac{m^*}{\sqrt{m^{*2}+p^2}}\right)^q
[f_i(1-f_i)\pm \bar{f}_i(1-\bar{f}_i)],\quad i=\{n,p\}.
\eea 
%------------------------------------------------------
The average values of the meson fields are given by~\cite{Chatterjee2007}
%------------------------------------------------------
\bea\label{eq:mean_fields}
g_\omega\omega_0 = \left(\frac{g_\omega}{m_\omega}\right)^2 
(n_n+n_p),\qquad
g_\rho \rho_{03} = \frac{1}{2}\left(\frac{g_\rho}{m_\rho}\right)^2
(n_p -n_n),
\eea
%------------------------------------------------------
which gives (recall that $\mu^*_i = \mu_i-g_{\omega}\omega_0 - g_{\rho}\rho_{03}I_{3i}$)
%------------------------------------------------------
\bea\label{eq:b_ij_def}
B_{ij}\equiv \frac{\partial \mu^*_i}{\partial n_j} = 
A_{ij} -\left(\frac{g_\omega}{m_\omega}\right)^2 -
 I_{3i}I_{3j}\left(\frac{g_\rho}{m_\rho}\right)^2.
\eea
%------------------------------------------------------
The scalar field is given by 
%------------------------------------------------------
\bea\label{eq:sigma}
g_\sigma \sigma =m-m^*=-\frac{g_\sigma}{m_\sigma^2}
\frac{\partial U(\sigma)}{\partial \sigma}+\frac{1}{\pi^2}
\left(\frac{g_\sigma}{m_\sigma}\right)^2\sum_{i=n,p}\int_0^\infty 
p^2dp\frac{m^*}{\sqrt{p^2+m^{*2}}} [f_i(p)+\bar{f}_i(p)],
\eea 
%------------------------------------------------------
with $U(\sigma)$ being the self-interaction potential 
of the scalar field, therefore
%------------------------------------------------------
\bea
\frac{\partial m^*}{\partial n_j}=
\frac{g_\sigma}{m_\sigma^2}
\frac{\partial^2 U(\sigma)}{\partial \sigma^2}
\frac{\partial\sigma}{\partial n_j}+
\left(\frac{g_\sigma}{m_\sigma}\right)^2
\left(\frac{\partial m^*}{\partial n_j}\right)
\left({I}_{2n}^+ +{I}_{2p}^+\right) -
\left(\frac{g_\sigma}{m_\sigma}\right)^2
\left(B_{nj}{I}_{1n}^- +B_{pj}{I}_{1p}^-\right)\nonumber\\
-\left(\frac{g_\sigma}{m_\sigma}\right)^2
\left(\frac{\partial m^*}{\partial n_j}\right)
\sum_{i=n,p}\frac{1}{\pi^2}\int_0^\infty p^2dp\frac{p^2}
{(p^2+m^{*2})^{3/2}} [f_i(p)+\bar{f}_i(p)].
\eea 
%------------------------------------------------------
The last term is suppressed in the nonrelativistic limit
and can be neglected, after which we obtain 
%------------------------------------------------------
\bea\label{eq:mass_diff}
\frac{\partial m^*}{\partial n_j}=-
\frac{\left(\frac{g_\sigma}{m_\sigma}\right)^2
\left(B_{nj}{I}_{1n}^-+ B_{pj}{I}_{1p}^- \right)}
{1-\left(\frac{g_\sigma}{m_\sigma}\right)^2
\left({I}_{2n}^+ +{I}_{2p}^+\right) +
\frac{1}{m_\sigma^2}\frac{\partial^2 U}{\partial \sigma^2}}.
\eea 
%------------------------------------------------------
Substituting this into Eq.~\eqref{eq:matrix_eq} we obtain 
the following equations for coefficients $B_{ij}$
%------------------------------------------------------
\bea\label{eq:matrix_eq1}
B_{ij} {I}_{0i}^+ -\gamma\left(B_{nj}{I}_{1n}^-+ 
B_{pj}{I}_{1p}^-\right){I}_{1i}^- =\delta_{ij},
\eea
%------------------------------------------------------
where 
%------------------------------------------------------
\bea\label{eq:gamma_def}
\gamma = \frac{1}{{I}_{2n}^+ +{I}_{2p}^+ -\beta},
\qquad \beta =\left(\frac{m_\sigma}{g_\sigma}\right)^2
\left(1+\frac{1}{m_\sigma^2}\frac{\partial^2 U}{\partial \sigma^2}\right).
\eea
%------------------------------------------------------
In the case of $i\neq j$ we find from Eq.~\eqref{eq:matrix_eq1} 
%------------------------------------------------------
\bea
B_{np}=\gamma B_{pp}\frac{I_{1p}^-I_{1n}^-}
{I_{0n}^+ -\gamma I_{1n}^{-2}},\qquad
B_{pn}=\gamma B_{nn}\frac{I_{1n}^-I_{1p}^-}
{I_{0p}^+ -\gamma I_{1p}^{-2}}.
\eea
%------------------------------------------------------
Substituting these expressions into 
Eq.~\eqref{eq:matrix_eq1} for $i=j$ we obtain
%------------------------------------------------------
\bea\label{eq:b_diag}
B_{nn}=
\frac{I_{0p}^+ -\gamma I_{1p}^{-2}}{I_{0n}^{+}I_{0p}^{+}
-\gamma I_{0p}^+ I_{1n}^{-2}-\gamma I_{0n}^+ I_{1p}^{-2}},\qquad
B_{pp}=
\frac{I_{0n}^+-\gamma I_{1n}^{-2}}{I_{0n}^{+}I_{0p}^{+}
-\gamma I_{0p}^+ I_{1n}^{-2}-\gamma I_{0n}^+ I_{1p}^{-2}},
\eea
%------------------------------------------------------
and
%------------------------------------------------------
\bea
\label{eq:b_mix}
B_{np}=B_{pn}=
\frac{\gamma I_{1p}^-I_{1n}^-}{I_{0n}^{+}I_{0p}^{+}
-\gamma I_{0p}^+ I_{1n}^{-2}-\gamma I_{0n}^+ I_{1p}^{-2}}.
\eea
%------------------------------------------------------
In the nonrelativistic limit we will use the expansion
$m/\sqrt{m^2+p^2}\simeq 1-p^2/2m^2$ in the integrals~\eqref{eq:I_rel}. 
We will further drop the contribution of antiparticles because
it is not important for the regime of interest. Then 
${I}_{0i}^{+}={I}_{0i}^{-}\simeq \tilde{I}_{2i}$,
${I}_{1i}^{+}={I}_{1i}^{-}\simeq \tilde{I}_{2i}-\tilde{I}_{4i}/2m^{*2}$, and 
${I}_{2i}^{+}={I}_{2i}^{-}\simeq \tilde{I}_{2i}-\tilde{I}_{4i}/m^{*2}$, where
%------------------------------------------------------
\bea\label{eq:I_nonrel}
\tilde{I}_{q i}= \frac{1}{\pi^2 T}\int_0^\infty p^q dp \,f_i(1-f_i).
\eea 
%------------------------------------------------------
Then in the nonrelativistic limit we find for Eq.~\eqref{eq:gamma_def}
%------------------------------------------------------
\bea
\gamma =\frac{1}{\left(\tilde{I}_{2n}+\tilde{I}_{2p} \right)}
+\frac{1}{\left(\tilde{I}_{2n}+\tilde{I}_{2p}\right)^2}
\left(\frac{\tilde{I}_{4n}+\tilde{I}_{4p}}{m^{*2}}+\beta\right),
\eea
%------------------------------------------------------
and
%------------------------------------------------------
\bea
I_{0n}^{+}I_{0p}^{+}-\gamma I_{0p}^+ I_{1n}^{-2}-\gamma 
I_{0n}^+ I_{1p}^{-2}=-\beta\frac{\tilde{I}_{2p} 
\tilde{I}_{2n}}{\tilde{I}_{2n}+\tilde{I}_{2p}}.
\eea
%------------------------------------------------------
Then 
%------------------------------------------------------
\bea
B_{nn}=-\frac{1}{\beta} +\frac{1}{\tilde{I}_{2n}+\tilde{I}_{2p}}
\frac{\tilde{I}_{2p}}{\tilde{I}_{2n}}
  +\frac{1}{m^{*2}\beta}\left(\frac{\tilde{I}_{2p}}{\tilde{I}_{2n}}
\frac{\tilde{I}_{4n}+\tilde{I}_{4p}}{\tilde{I}_{2n}+\tilde{I}_{2p}}-
\frac{\tilde{I}_{4p}}{\tilde{I}_{2n}}\right),\\
B_{pp}=-\frac{1}{\beta} +\frac{1}{\tilde{I}_{2n}+\tilde{I}_{2p}}
\frac{\tilde{I}_{2n}}{\tilde{I}_{2p}}
  +\frac{1}{m^{*2}\beta}\left(\frac{\tilde{I}_{2n}}{\tilde{I}_{2p}}
\frac{\tilde{I}_{4n}+\tilde{I}_{4p}}{\tilde{I}_{2n}+\tilde{I}_{2p}}-
\frac{\tilde{I}_{4n}}{\tilde{I}_{2p}}\right),\nonumber\\
B_{np}=B_{pn} = -\frac{1}{\beta} -\frac{1}{\tilde{I}_{2n}+\tilde{I}_{2p}}
 +\frac{1}{2m^{*2}\beta}\left(\frac{\tilde{I}_{4n}}{\tilde{I}_{2n}}+
 \frac{\tilde{I}_{4p}}{\tilde{I}_{2p}} -2\frac{\tilde{I}_{4n}+
 \tilde{I}_{4p}} {\tilde{I}_{2n}+\tilde{I}_{2p}}\right).
\eea
%------------------------------------------------------
We next obtain the two combinations relevant to the bulk viscosity
%------------------------------------------------------
\bea
B_{nn}-B_{pn} = \frac{1}{\tilde{I}_{2n}} -
\frac{1}{2m^{*2}\beta}\left(\frac{\tilde{I}_{4p}}{\tilde{I}_{2p}}-
\frac{\tilde{I}_{4n}}{\tilde{I}_{2n}}\right),\qquad
B_{pp}-B_{np} = \frac{1}{\tilde{I}_{2p}} -
\frac{1}{2m^{*2}\beta}\left(\frac{\tilde{I}_{4n}}{\tilde{I}_{2n}}-
\frac{\tilde{I}_{4p}}{\tilde{I}_{2p}}\right).
\eea
%------------------------------------------------------
Substituting the expression for $\beta$ from Eq.~\eqref{eq:gamma_def} 
and recalling Eq.~\eqref{eq:b_ij_def} we obtain 
%------------------------------------------------------
\bea\label{eq:A_n_fianl}
A_{n}=
\frac{1}{\tilde{I}_{2n}}+\frac{1}{2}\left(\frac{g_\rho}{m_\rho}\right)^2
+\frac{1}{2m^{*2}}\left(\frac{g_\sigma}{m_\sigma}\right)^2
\left(1+\frac{1}{m_\sigma^2}\frac{\partial^2 U}{\partial\sigma^2}\right)^{-1}
\left(\frac{\tilde{I}_{4n}}{\tilde{I}_{2n}} - \frac{\tilde{I}_{4p}}{\tilde{I}_{2p}}\right),\\
\label{eq:A_p_final}
A_{p}=
\frac{1}{\tilde{I}_{2p}}+\frac{1}{2}\left(\frac{g_\rho}{m_\rho}\right)^2
+\frac{1}{2m^{*2}}\left(\frac{g_\sigma}{m_\sigma}\right)^2
\left(1+\frac{1}{m_\sigma^2}\frac{\partial^2 U}{\partial\sigma^2}\right)^{-1}
\left(\frac{\tilde{I}_{4p}}{\tilde{I}_{2p}} - \frac{\tilde{I}_{4n}}{\tilde{I}_{2n}}\right).
\eea
%------------------------------------------------------
For leptons we have simply
%------------------------------------------------------
\bea\label{A_j_int_lep}
A_{e} =
\frac{1}{\tilde{I}_{2e}},\quad A_\nu =\frac{2}{\tilde{I}_{2\nu}}.
\eea 
%------------------------------------------------------
where the lepton energies in the integral 
are taken as $\ep_p=p-\mu_{L}$. Then 
%------------------------------------------------------
\bea\label{eq:A_fianl}
A=\sum_{i}A_i=
\frac{1}{\tilde{I}_{2n}}+\frac{1}{\tilde{I}_{2p}}+
\frac{1}{\tilde{I}_{2e}}+\frac{2}{\tilde{I}_{2\nu}}
+\left(\frac{g_\rho}{m_\rho}\right)^2,
\eea
%------------------------------------------------------
and
%------------------------------------------------------
\bea
n_nA_n-n_pA_p=\frac{n_n}{\tilde{I}_{2n}}-\frac{n_p}{\tilde{I}_{2p}}
+\frac{n_n-n_p}{2}\left(\frac{g_\rho}{m_\rho}\right)^2
+\frac{n_n+n_p}{2m^{*2}}\left(\frac{g_\sigma}{m_\sigma}\right)^2
\left(1+\frac{1}{m_\sigma^2}\frac{\partial^2 U}{\partial\sigma^2}\right)^{-1}
\left(\frac{\tilde{I}_{4n}}{\tilde{I}_{2n}} - \frac{\tilde{I}_{4p}}{\tilde{I}_{2p}}\right).
\eea
%------------------------------------------------------
Taking into account also Eq.~\eqref{eq:mean_fields} and 
the nonrelativistic limit of Eq.~\eqref{eq:sigma} we obtain
%------------------------------------------------------
\bea\label{eq:C_fianl}
C= \frac{n_n}{\tilde{I}_{2n}}-\frac{n_p}{\tilde{I}_{2p}}-\frac{n_e}{\tilde{I}_{2e}}
+2\frac{n_\nu}{\tilde{I}_{2\nu}}-g_\rho \rho_{03} +\frac{g_\sigma \sigma}{2m^{*2}}
\left(1+\frac{1}{\sigma m_\sigma^2}\frac{\partial U}{\partial\sigma}\right)
\left(1+\frac{1}{m_\sigma^2}\frac{\partial^2 U}{\partial\sigma^2}\right)^{-1}
\left(\frac{\tilde{I}_{4n}}{\tilde{I}_{2n}} - \frac{\tilde{I}_{4p}}{\tilde{I}_{2p}}\right).
\eea
%------------------------------------------------------

The terms containing $U(\sigma)$ vanish in the case of DD-ME2 model
and are numerically very small in the case of NL3 model.
We find also, that the terms $\propto g_\rho$ in Eqs.~\eqref{eq:A_fianl} 
and \eqref{eq:C_fianl} are negligible in comparison to the first four terms.
The last term in Eq.~\eqref{eq:C_fianl} is comparable to the rest of the terms.

In the case of degenerate matter the susceptibilities can be computed analytically
%------------------------------------------------------
\bea\label{eq:A_deg}
A &=& \frac{\pi^2}{m^*}\left(\frac{1}{p_{Fn}}+\frac{1}{p_{Fn}}\right)+\frac{\pi^2}{p_{Fe}^2}
+\frac{2\pi^2}{p_{F\nu}^2}+\left(\frac{g_\rho}{m_\rho}\right)^2,\\
\label{eq:C_deg}
C &=& \frac{p_{Fn}^2-p_{Fp}^2}{3m^*}+\frac{p_{F\nu}-p_{Fe}}{3}
+\frac{p_{Fn}^2 - p_{Fp}^2}{2m^{*2}}g_\sigma \sigma -g_\rho \rho_{03}.
\eea
%------------------------------------------------------
which agree with the results of Refs.~\cite{Chatterjee2007,Chatterjee2008ApJ}.

\end{widetext}

\section{Beta equilibration rates}
\label{app:beta_rates}

%-------------------------------------------------
\begin{figure}[t] 
\begin{center}
\includegraphics[width=0.8\columnwidth,keepaspectratio]{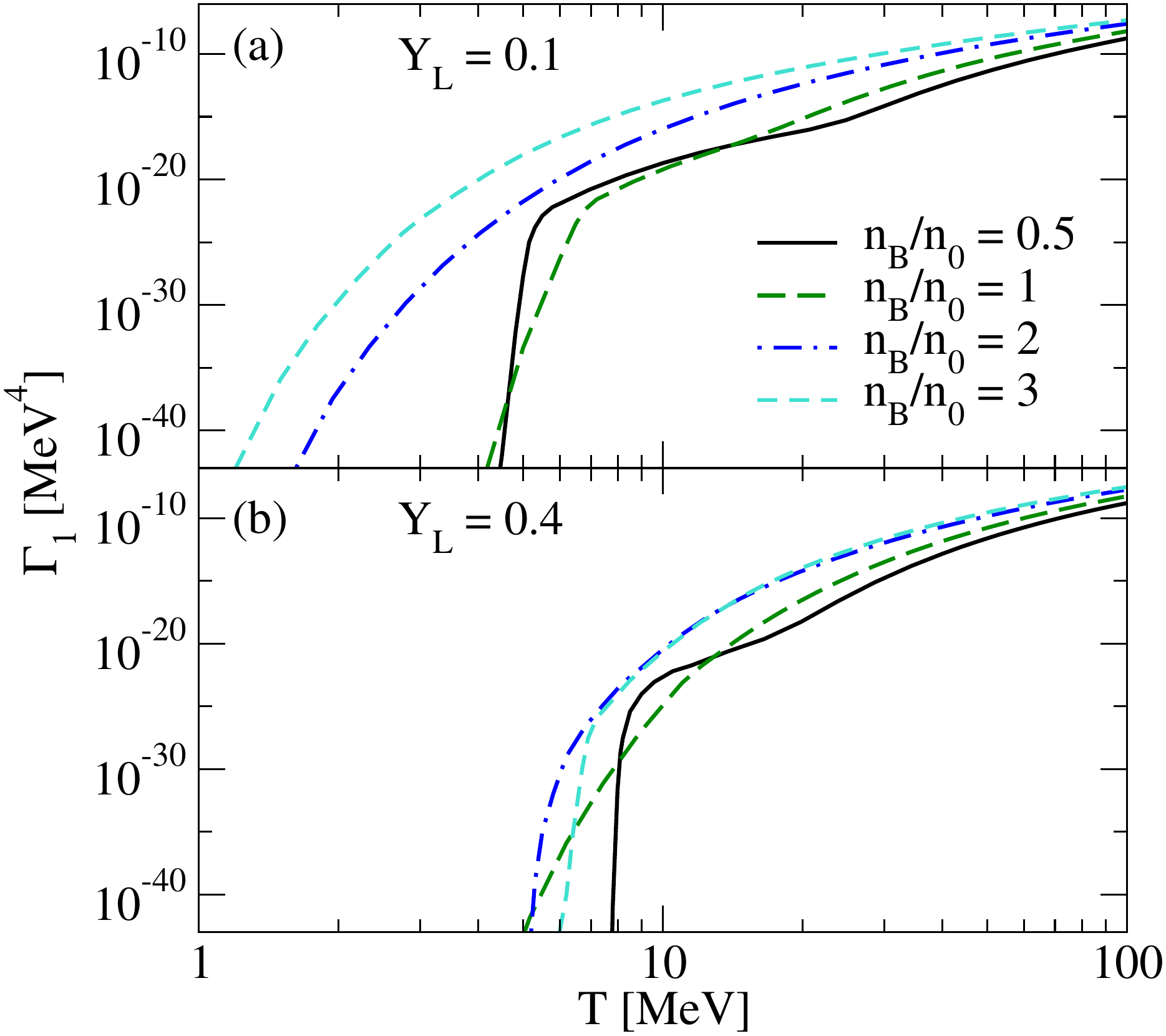}
\caption{ The neutron decay rate $\Gamma_1$ as a function of the
  temperature for fixed values of the baryon density. The lepton
  fraction is fixed at $Y_L=0.1$ for (a) and at $Y_L=0.4$ for (b). }
\label{fig:gamma1_temp} 
\end{center}
\end{figure}
%-------------------------------------------------

\begin{figure}[t] 
\begin{center}
\includegraphics[width=0.8\columnwidth,keepaspectratio]{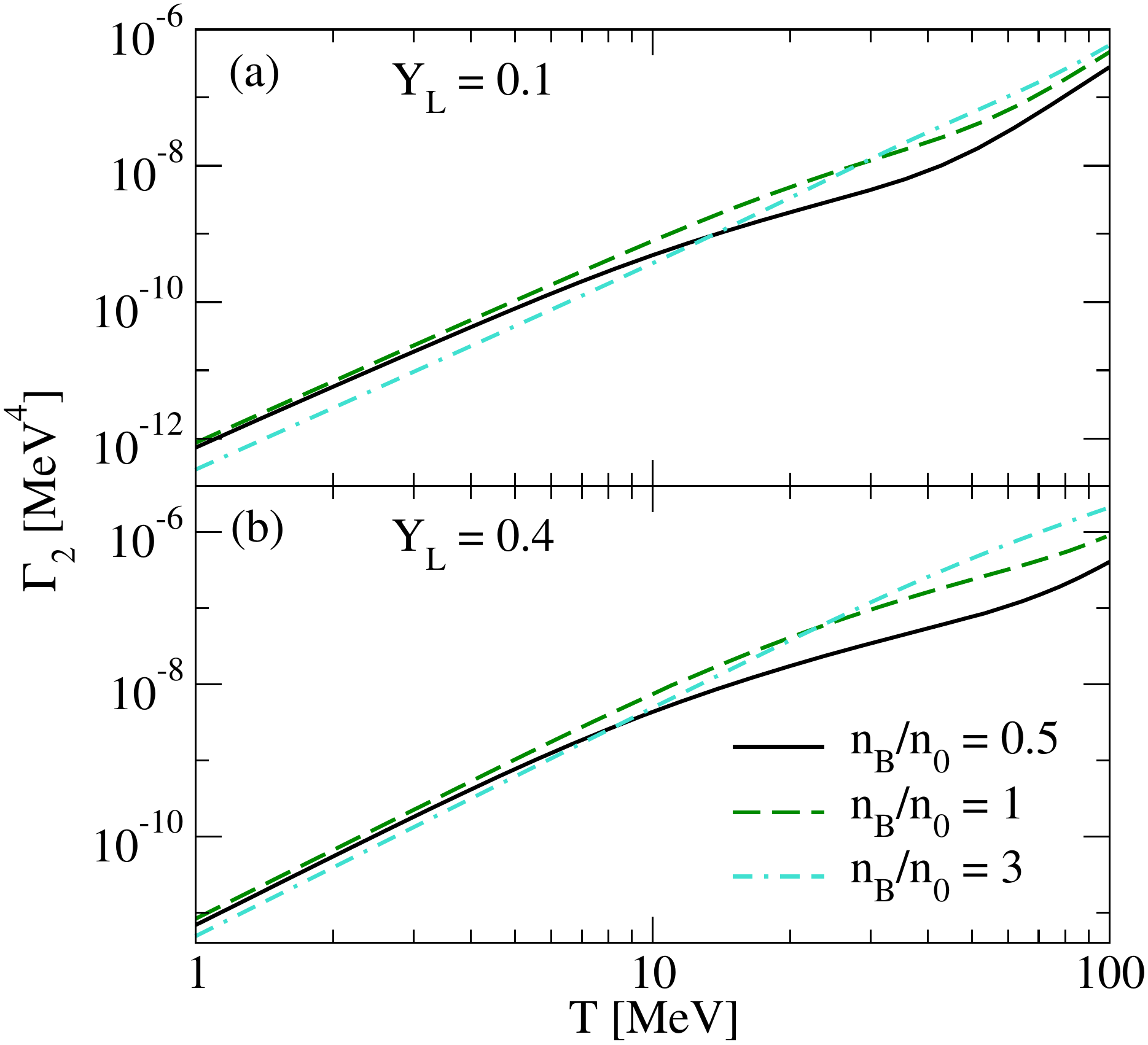}
\caption{ The electron capture rate $\Gamma_2$ as a function of the
  temperature for fixed values of the baryon density. The lepton
  fraction is fixed at $Y_L=0.1$ for (a) and at $Y_L=0.4$ for (b). }
\label{fig:gamma2_temp} 
\end{center}
\end{figure}

In this appendix we discuss the numerical results for the $\beta$-equilibration
rates $\Gamma_1$ and $\Gamma_2$ given by
Eqs.~\eqref{eq:Gamma_1_general} and
\eqref{eq:Gamma_2_general}. Figures~\ref{fig:gamma1_temp} and
\ref{fig:gamma2_temp} show the temperature dependence of the neutron
decay and the electron capture rates, respectively, for various values
of the density and lepton fraction. 

Figures~\ref{fig:gamma1_temp} and \ref{fig:gamma2_temp} 
demonstrate that the electron capture process and its
inverse \eqref{eq:Urca_2} dominate the beta equilibration and hence the
bulk viscosity. The electron capture rate $\Gamma_2$ is always many 
orders of magnitude larger than the neutron decay rate $\Gamma_1$. 
This is because we are studying neutrino-trapped matter, 
where the trapped species is typically neutrinos rather than antineutrinos
(Figs.~\ref{fig:fractions_dens1},\,\ref{fig:fractions_dens2})
so any process involving antineutrinos will be suppressed by factors
of $\exp(-\mu{_\nu}/T)$. 

As seen from the upper panels of the figures, $\Gamma_1$ is a rapidly
increasing function of the temperature and exponentially vanishes at
low temperatures because of vanishing antineutrino density in the
degenerate matter. The threshold of temperature below which $\Gamma_1$
practically vanishes is located at higher temperatures for higher
lepton fractions, because the suppression of the antineutrino density
is stronger for larger $Y_L$.

%-------------------------------------------------
%-------------------------------------------------

The temperature dependence of the electron capture rate differs
significantly from that of $\Gamma_1$. Indeed, $\Gamma_2$ increases
with the temperature according to a power-law $\Gamma_2\propto T^3$ up
to the temperatures $T\simeq 10$ MeV, as seen from the low-temperature
limit given by Eq.~\eqref{eq:Gamma2_trap}.  This scaling breaks down
at higher temperatures $T\geq 10$ MeV and sufficiently low densities
$n_B\leq n_0$, where the finite temperature effects become important
in the evaluation of the integral~\eqref{eq:Gamma_2_general}.  We have
checked numerically that the exact result for $\Gamma_2$~\eqref{eq:Gamma_2_general} 
tends to its low-temperature limit~\eqref{eq:Gamma2_trap} as $T\leq 1$~MeV.  

Comparing the upper and lower panels in Figs.~\ref{fig:gamma1_temp}
and \ref{fig:gamma2_temp}, we see that $\Gamma_1$ is smaller for
larger lepton fraction, whereas $\Gamma_2$ shows the opposite
behavior.  The reason for this behavior is clear: the neutron decay
rate $\Gamma_1$ is proportional to the antineutrino (number) density, 
whereas the electron capture rate $\Gamma_2$ is proportional to the 
neutrino density. Because the electron neutrino fraction increases with the
increase of $Y_L$, as was seen from Figs.~\ref{fig:fractions_dens1}
and \ref{fig:fractions_dens2}, the antineutrino population becomes
more suppressed at higher $Y_L$, thus leading to smaller $\Gamma_1$ and
larger $\Gamma_2$.

\section{Bulk viscosity of neutrino-transparent matter}
\label{app:bulk_visc0}

\begin{figure}[t] 
\begin{center}
\includegraphics[width=\columnwidth, keepaspectratio]{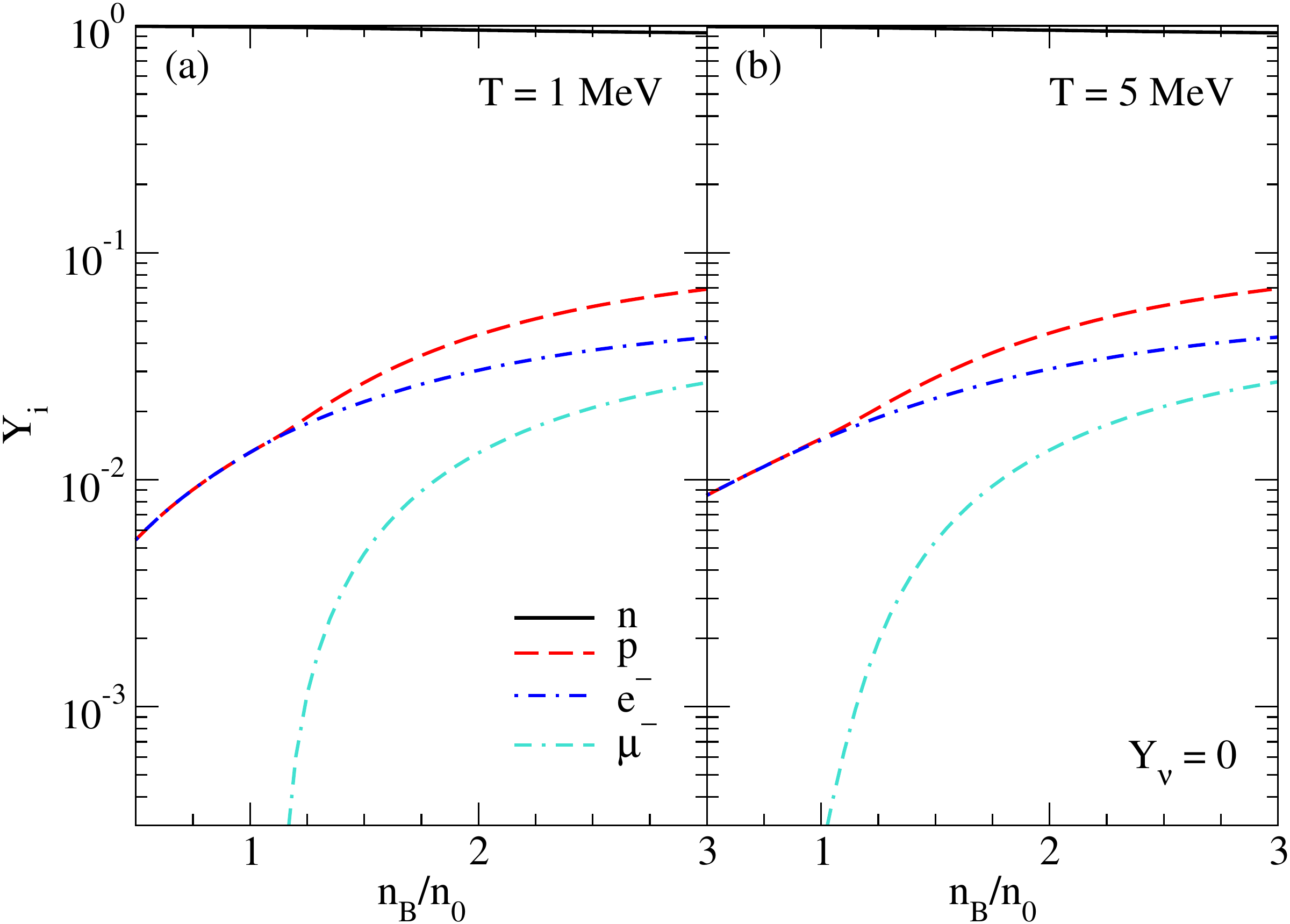}
\caption{ Particle fractions as functions of the baryon
density for neutrino-transparent matter $Y_\nu=0$
for two values of the temperature: (a) $T=1$ MeV; (b) $T=5$ MeV. }
\label{fig:fractions_dens3} 
\end{center}
\end{figure}

In this appendix we are interested in the domain of temperatures
$T \le T_{\rm tr}$ where the matter is neutrino transparent
($Y_\nu=0$). Below the temperature $T\simeq 1\,\MeV$, the modified
Urca process becomes important \cite{Alford:2018lhf} therefore our
results are strictly relevant for the temperature domain
$1 \le T\le 5$~MeV.  To account for uncertainty in the value of
$T_{\rm tr}$ the numerical results will be shown up to $T = 10$~MeV.
The particle fractions are shown in Fig.~\ref{fig:fractions_dens3}.
In this case muons appear only above a certain baryon density
$n_B\gtrsim n_0$, where the condition $\mu_e\geq m_\mu\simeq 106$ MeV
is satisfied. Below this threshold, the proton and electron fractions
are equal, as required by the charge neutrality condition, whereas
above the threshold the condition $Y_p=Y_e+Y_\mu$ is satisfied.

%-------------------------------------------------
%-------------------------------------------------
\begin{figure}[h!] 
\begin{center}
\includegraphics[width=0.8\columnwidth,keepaspectratio]{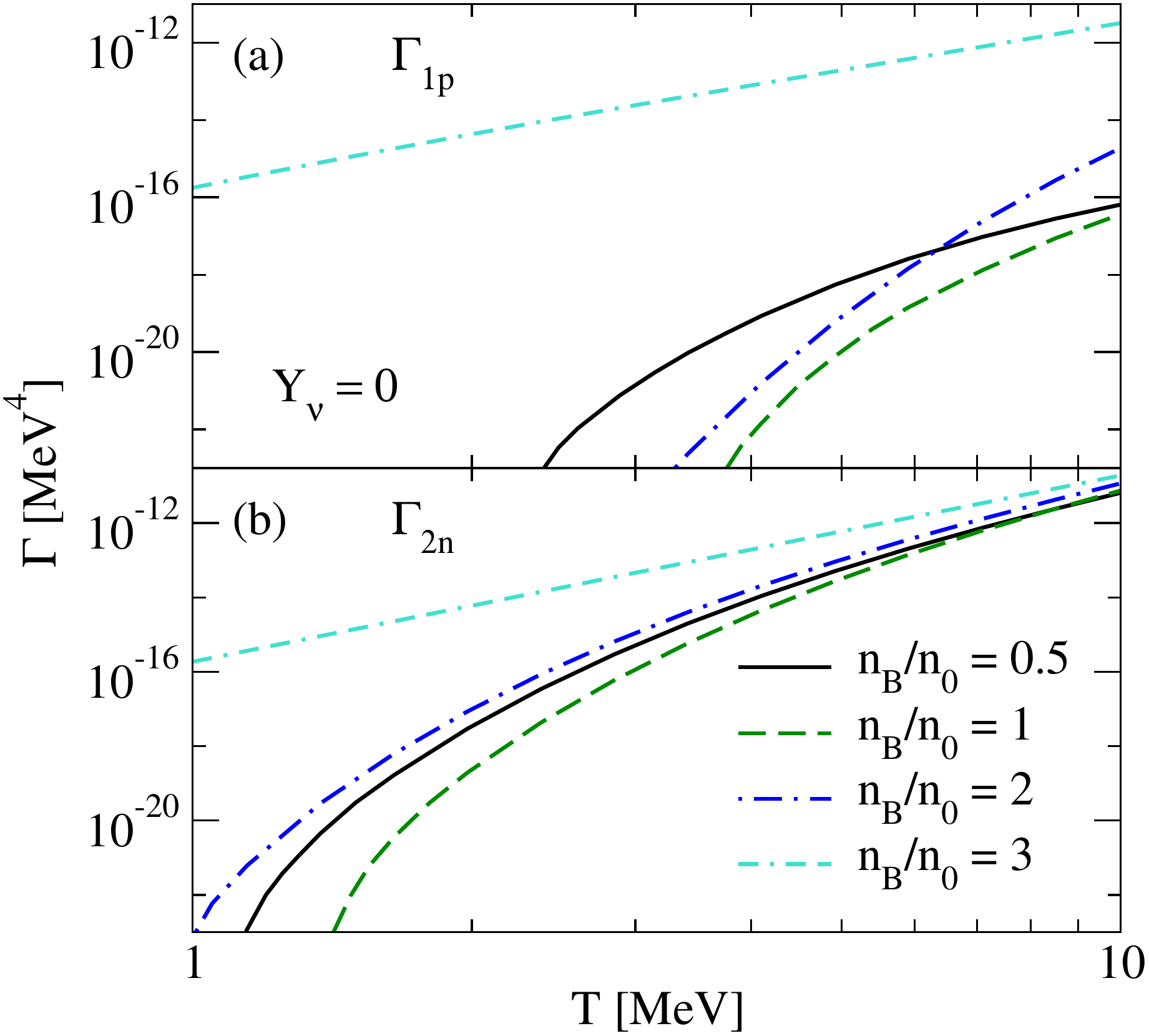}
\caption{ The neutron decay rate $\Gamma_{1p}$ (a) and
the electron capture rate $\Gamma_{2n}$ (b) as functions of the
 temperature for fixed values of the 
 baryon density in the neutrino-transparent case.}
\label{fig:gamma12_temp} 
\end{center}
\end{figure}
%-------------------------------------------------
\begin{figure}[t] 
\begin{center}
\includegraphics[width=\columnwidth, keepaspectratio]{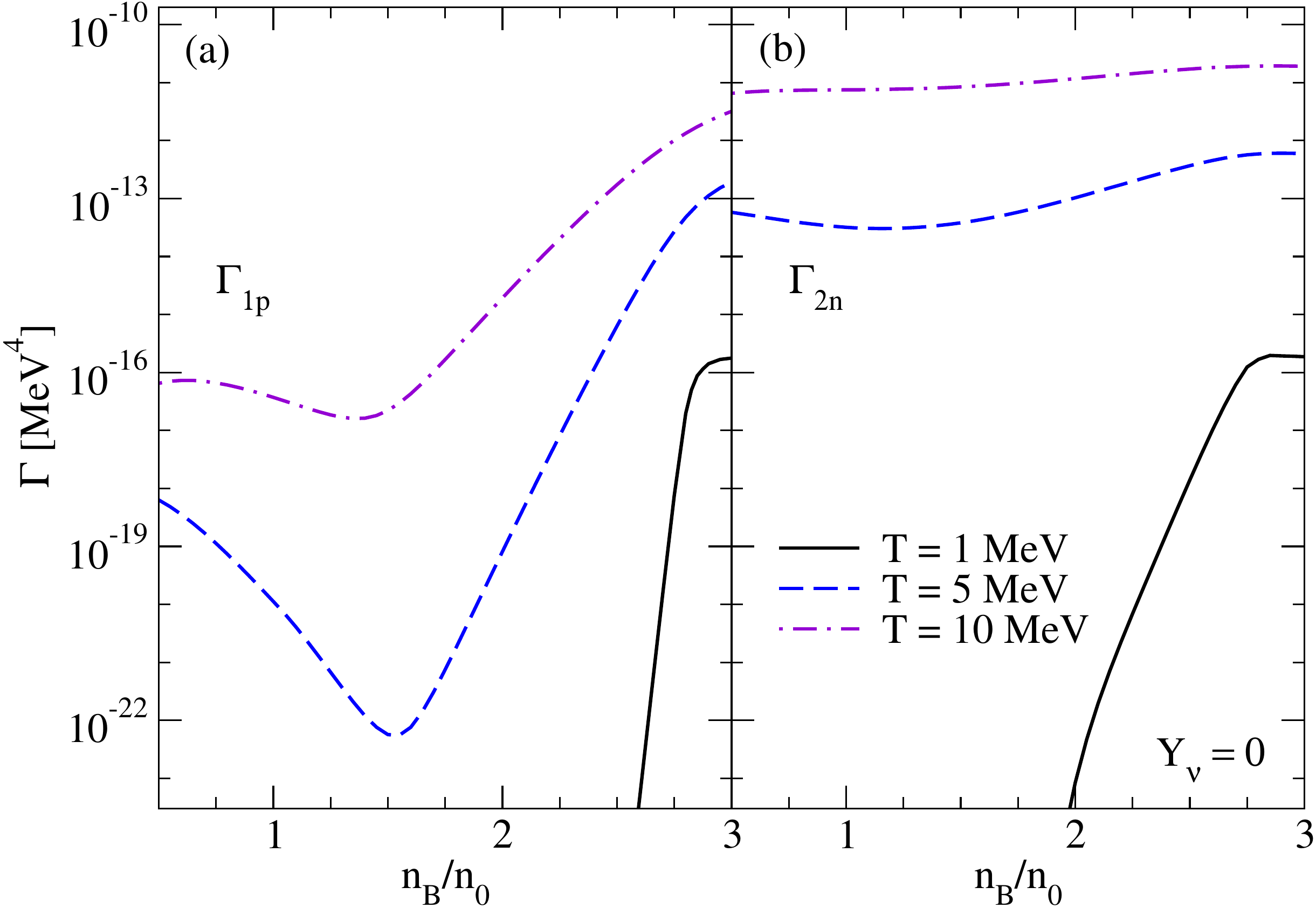}
\caption{ The neutron decay rate $\Gamma_{1p}$ (a) and
the electron capture rate $\Gamma_{2n}$ (b) as functions of the
  baryon density for fixed values of the temperature
  in the neutrino-transparent case $Y_\nu=0$.}
\label{fig:gamma12_dens0} 
\end{center}
\end{figure}
%-------------------------------------------------

The relevant $\beta$-equilibration rates $\Gamma_{1p}$ and
$\Gamma_{2n}$ are shown in Fig.~\ref{fig:gamma12_temp} and as
functions of the temperature.  Both quantities rapidly increase with
the temperature and are exponentially damped in the low-temperature
limit. The reason is that the condition $p_{Fp}+p_{Fe} \geq p_{Fn}$ is
never satisfied for the given model of hadronic matter because of very
small proton fraction $Y_p \leq 7$ \%, see
Fig.~\ref{fig:fractions_dens3}.  As a consequence, the direct Urca
processes for neutrino-transparent matter are always blocked at low
temperatures, therefore in that regime, the modified Urca processes
should be accurately taken into account \cite{Alford:2018lhf}.  

Quantitatively, the neutron decay rate $\Gamma_{1p}$ in the case of 
$Y_\nu=0$ is larger than in the case of $Y_L=0.1$, whereas the electron 
capture rate is smaller. This result can be anticipated
from Eqs.~\eqref{eq:Gamma_1_general} and \eqref{eq:Gamma_2_general},
where one should substitute $\alpha_\nu=0$ in the neutrino-transparent
case. However, the rate $\Gamma_{1p}$ again remains much smaller than
$\Gamma_{2n}$, and only at sufficiently high densities approaches
$\Gamma_{2n}$, as seen in Fig.~\ref{fig:gamma12_dens0}, since both quantities 
have the same low-temperature limit given by Eqs.~\eqref{eq:Gamma1_deg} and \eqref{eq:Gamma2_deg}.

We remark also, that the behavior of $\lambda_1$ and $\lambda_2$ is
very similar to that of $\Gamma_{1p}$ and $\Gamma_{2n}$,
respectively. At densities $n_B\geq n_0$ we have approximately
$\lambda_1\simeq \Gamma_{1p}/T$ and $\lambda_2\simeq \Gamma_{2n}/T$,
as it was the case of neutrino-trapped matter, see
Eqs.~\eqref{eq:lambda1} and \eqref{eq:lambda2}.

%-------------------------------------------------
\begin{figure}[t] 
\begin{center}
\includegraphics[width=0.8\columnwidth,keepaspectratio]{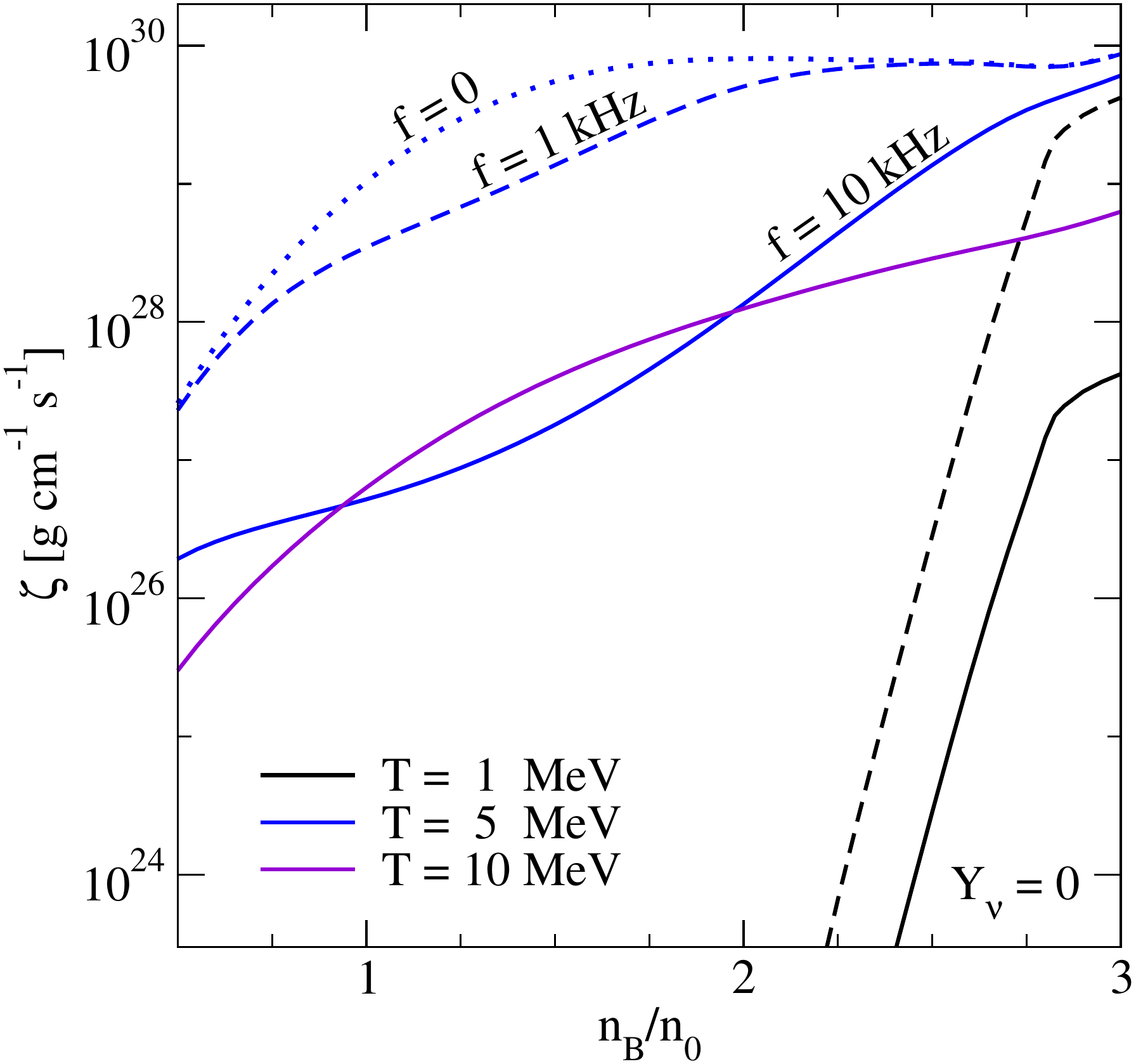}
\caption{ The density dependence of the bulk viscosity for various
  values of the temperature and the oscillation frequency in the case
  of $Y_\nu=0$.  In the case of high temperatures (the violet curve)
  $\zeta$ becomes independent of the frequency.}
\label{fig:zeta_dens0} 
\end{center}
\end{figure}
%-------------------------------------------------
\begin{figure}[h] 
\begin{center}
\includegraphics[width=0.8\columnwidth,keepaspectratio]{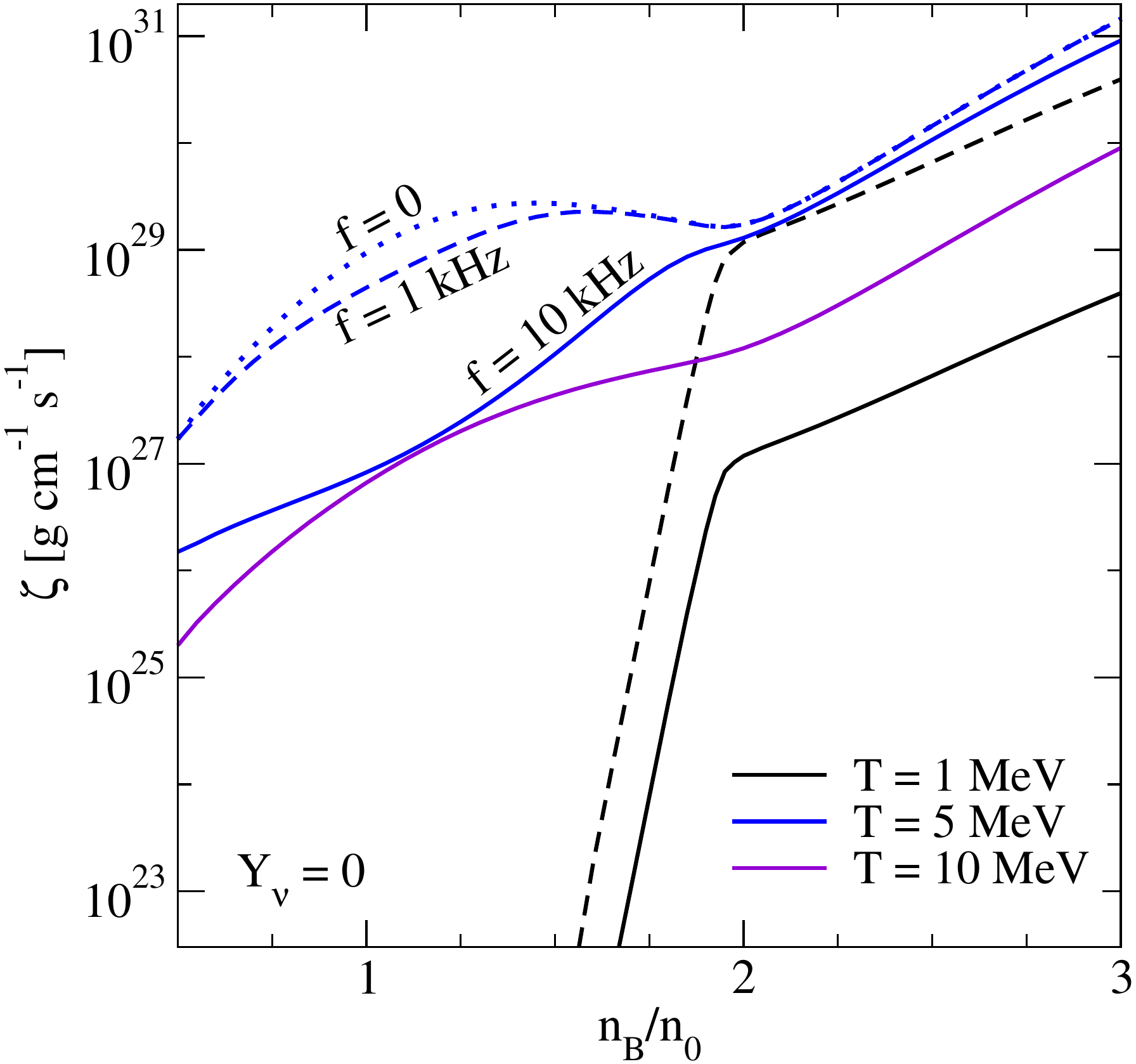}
\caption{ Same as Fig.~\ref{fig:zeta_dens0}, but for model NL3.}
\label{fig:zeta_dens0_NL3} 
\end{center}
\end{figure}
%-------------------------------------------------

Next, we show the density dependence of the bulk viscosity in
Fig.~\ref{fig:zeta_dens0}. As in the case of $Y_\nu\neq 0$, the
frequency dependence of $\zeta$ can be neglected at sufficiently high
temperatures $T\simeq 10$ MeV (see Fig.~\ref{fig:lambda_A_temp}), and,
because the product $\lambda A$ is almost density-independent, the
bulk viscosity as a function of density increases as $C^2$.  For
smaller temperatures $T\lesssim 5$ MeV the frequency dependence of
$\zeta$ becomes important, and the bulk viscosity drops rapidly with
$\omega$. We show the bulk viscosity for three values of the frequency
in Fig.~\ref{fig:zeta_dens0} for $T=5$ MeV. The general increase of
$\zeta$ with the density, in this case, is again caused by the factor
$C^2$, but the hump structures of $\zeta$ in the range arise from the 
weak density-dependence of the product $\lambda A$. In the case of 
$T=1$ MeV we have already the opposite limit $\omega\gg \lambda A$ (see
Fig.~\ref{fig:lambda_A_temp}), therefore the bulk viscosity scales as
$\zeta\propto \lambda C^2$, which rapidly increases with the density
and drops to zero at sufficiently low densities $n_B\lesssim 2n_0$.
We have checked that our low-temperature result for $T=0.1$ MeV agrees
with the result of the bulk viscosity shown in Fig.~2 of
Ref.~\cite{Haensel2000} obtained for direct Urca processes for the
same state of hadronic matter.
%-------------------------------------------------
\begin{figure}[t] 
\begin{center}
\includegraphics[width=\columnwidth, keepaspectratio]{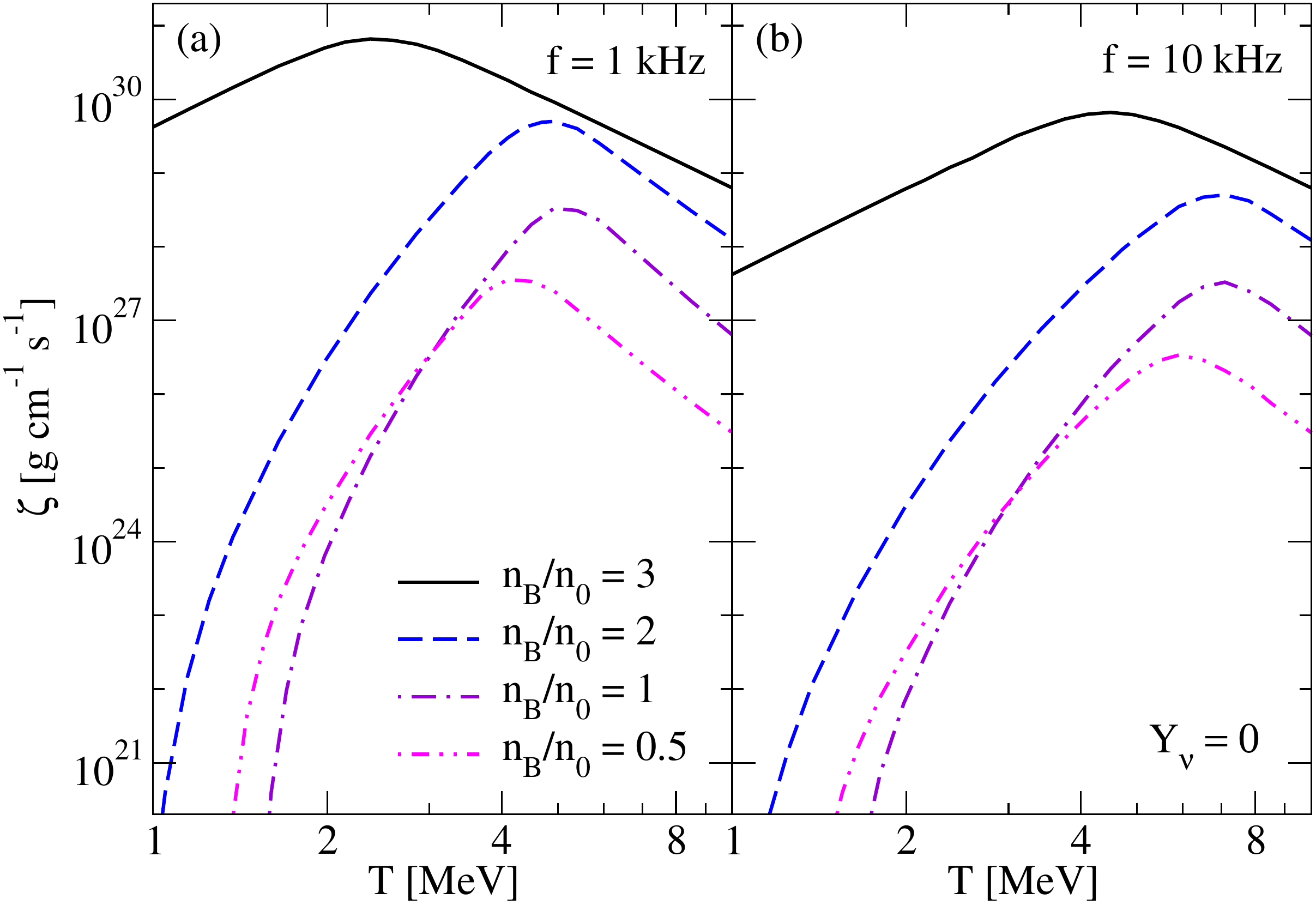}
\caption{ The temperature dependence of the bulk viscosity for model DD-ME2 
for various values of the density in the case of $Y_\nu=0$.
 The oscillation frequency is fixed at (a) $f=1$ kHz and (b) $f=10$ kHz.}
\label{fig:zeta_temp0} 
\end{center}
\end{figure}
For comparison, we show also the bulk viscosity of the
neutrino-transparent matter for the nuclear model NL3 in
Fig.~\ref{fig:zeta_dens0_NL3}. The results obtained within two models
DD-ME2 and NL3 differ mainly in the low-temperature region, where the
bulk viscosity is strongly suppressed at low densities because of
blocking of direct Urca processes. The suppression sets in at lower
densities in the case of the NL3 model because this model predicts
larger proton and electron fractions than the DD-ME2 model.

\begin{figure}[h] 
\begin{center}
\includegraphics[width=\columnwidth, keepaspectratio]{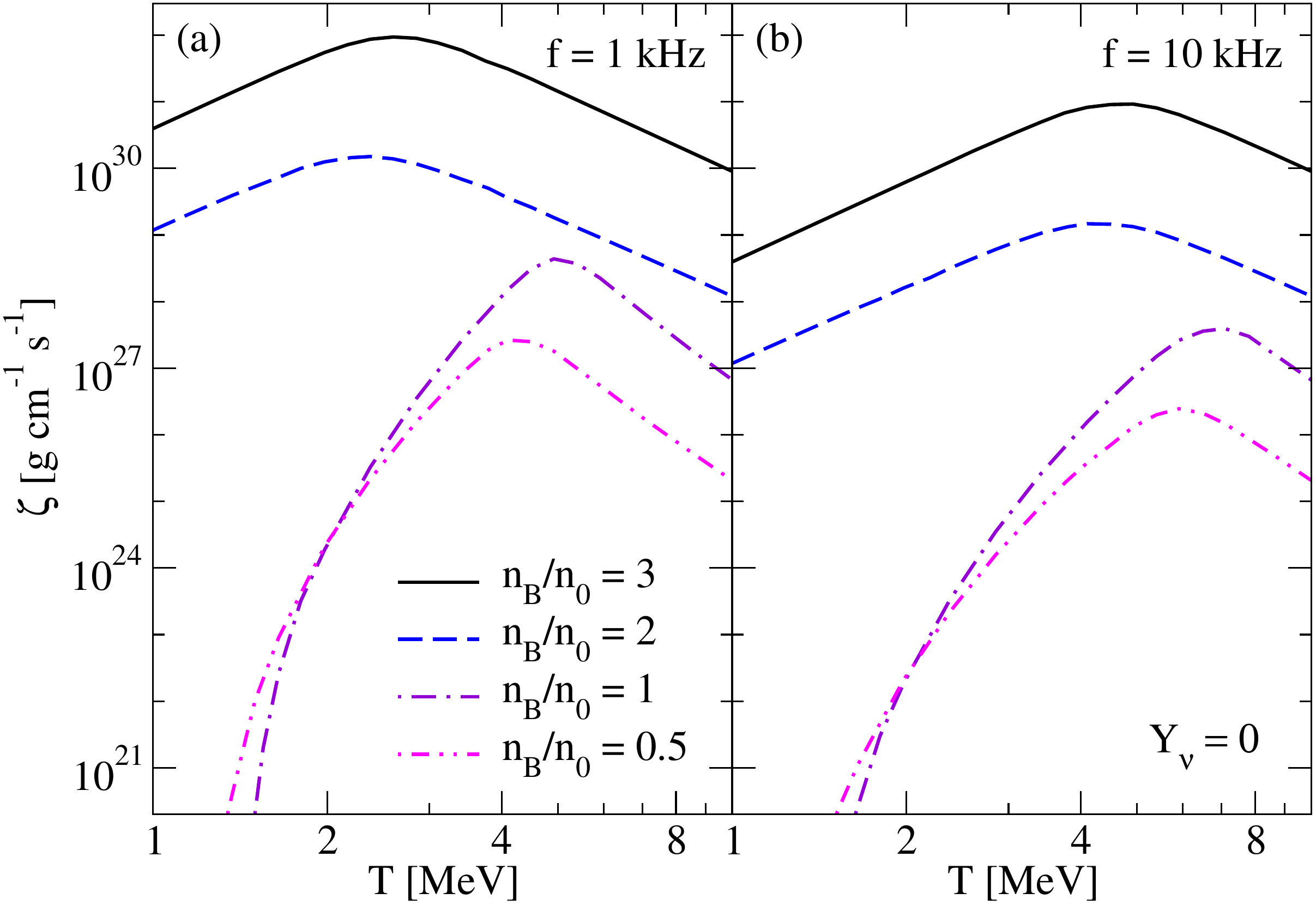}
\caption{ Same as Fig.~\ref{fig:zeta_temp0}, but for model NL3.} 
\label{fig:zeta_temp0_NL3} 
\end{center}
\end{figure}

Figures~\ref{fig:zeta_temp0} and \ref{fig:zeta_temp0_NL3} show the
temperature dependence of the bulk viscosity for models DD-ME2 and
NL3, respectively. It has a maximum at the temperature where
$\omega=\lambda A$, and the slope of the curve on the right side of
the maximum is larger than in the case of $Y_\nu\neq 0$ because of
stronger $\lambda=\lambda(T)$ dependence.  Using the approximate
scaling $\lambda\propto T^4$, we find that $T_{\rm
  max}=(\omega/\lambda_0 A)^{1/4}$, where $\lambda_0$ is the value of
$\lambda(T)$ at $T=1$ MeV, \ie, the maximum shifts to higher
temperatures for higher frequencies. As seen from a comparison of
Figs.~\ref{fig:zeta_dens} and \ref{fig:zeta_temp} with
Figs.~\ref{fig:zeta_dens0} and \ref{fig:zeta_temp0}, the bulk
viscosity is larger in the case of neutrino-transparent matter, as
expected.

%-------------------------------------------------
%------------------------------------------------- 

We also compare our results with the high-frequency (low-temperature)
result of Ref.~\cite{Haensel1992PhRvD} obtained from direct Urca processes
for the neutrino-transparent matter.  Choosing the parameters, \eg,
$T=1$ MeV, $f=1$ kHz, $n_B=2\, n_0$, we find from Figs.~1 and 2 and
Eq.~(13) of Ref.~\cite{Haensel1992PhRvD} the values
$\zeta\simeq 3.5\cdot 10^{28}$ g cm$^{-1}$ s$^{-1}$ for the model I,
and $\zeta\simeq 10^{29}$ g cm$^{-1}$ s$^{-1}$ for model II, whereas
our calculations give much lower result $\zeta\simeq 10^{21}$ g
cm$^{-1}$ s$^{-1}$.  The reason for this is the lower proton fraction
in our model, which leads to blocking of direct Urca processes at low
temperatures and, therefore, to lower bulk viscosity.

\providecommand{\href}[2]{#2}\begingroup\raggedright\endgroup

%\bibliographystyle{JHEP}                                                   
%\bibliography{urca_bulk}   

\begin{thebibliography}{10}

\bibitem{Abbott2017}
{\scshape LIGO Scientific Collaboration and Virgo Collaboration} collaboration,
  {The LIGO Scientific Collaboration} and {The Virgo Collaboration},
  \emph{Gw170817: Observation of gravitational waves from a binary neutron star
  inspiral},
  \href{http://dx.doi.org/10.1103/PhysRevLett.119.161101}{\emph{Phys. Rev.
  Lett.} {\bf 119} (2017) 161101}.

\bibitem{Perego:2019adq}
A.~Perego, S.~Bernuzzi and D.~Radice, \emph{{Thermodynamics conditions of
  matter in neutron star mergers}},
  \href{http://arxiv.org/abs/1903.07898}{{\tt 1903.07898}}.

\bibitem{Hanauske:2019qgs}
M.~Hanauske, J.~Steinheimer, A.~Motornenko, V.~Vovchenko, L.~Bovard, E.~R. Most
  et~al., \emph{{Neutron Star Mergers: Probing the EoS of Hot, Dense Matter by
  Gravitational Waves}},
  \href{http://dx.doi.org/10.3390/particles2010004}{\emph{Particles} {\bf 2}
  (2019) 44--56}.

\bibitem{Hanauske:2017oxo}
M.~Hanauske, J.~Steinheimer, L.~Bovard, A.~Mukherjee, S.~Schramm, K.~Takami
  et~al., \emph{{Concluding Remarks: Connecting Relativistic Heavy Ion
  Collisions and Neutron Star Mergers by the Equation of State of Dense Hadron-
  and Quark Matter as signalled by Gravitational Waves}},
  \href{http://dx.doi.org/10.1088/1742-6596/878/1/012031}{\emph{J. Phys. Conf.
  Ser.} {\bf 878} (2017) 012031}.

\bibitem{Kastaun:2016elu}
W.~Kastaun, R.~Ciolfi, A.~Endrizzi and B.~Giacomazzo, \emph{{Structure of
  Stable Binary Neutron Star Merger Remnants: Role of Initial Spin}},
  \href{http://dx.doi.org/10.1103/PhysRevD.96.043019}{\emph{Phys. Rev.} {\bf
  D96} (2017) 043019}, [\href{http://arxiv.org/abs/1612.03671}{{\tt
  1612.03671}}].

\bibitem{Bernuzzi:2015opx}
S.~Bernuzzi, D.~Radice, C.~D. Ott, L.~F. Roberts, P.~Moesta and F.~Galeazzi,
  \emph{{How loud are neutron star mergers?}},
  \href{http://dx.doi.org/10.1103/PhysRevD.94.024023}{\emph{Phys. Rev.} {\bf
  D94} (2016) 024023}, [\href{http://arxiv.org/abs/1512.06397}{{\tt
  1512.06397}}].

\bibitem{Foucart:2015gaa}
F.~Foucart, R.~Haas, M.~D. Duez, E.~O'Connor, C.~D. Ott, L.~Roberts et~al.,
  \emph{{Low mass binary neutron star mergers : gravitational waves and
  neutrino emission}},
  \href{http://dx.doi.org/10.1103/PhysRevD.93.044019}{\emph{Phys. Rev.} {\bf
  D93} (2016) 044019}, [\href{http://arxiv.org/abs/1510.06398}{{\tt
  1510.06398}}].

\bibitem{kiuchi:2012mk}
K.~Kiuchi, Y.~Sekiguchi, K.~Kyutoku and M.~Shibata, \emph{{Gravitational waves,
  neutrino emissions, and effects of hyperons in binary neutron star mergers}},
  \href{http://dx.doi.org/10.1088/0264-9381/29/12/124003}{\emph{Class. Quant.
  Grav.} {\bf 29} (2012) 124003}, [\href{http://arxiv.org/abs/1206.0509}{{\tt
  1206.0509}}].

\bibitem{sekiguchi:2011zd}
Y.~Sekiguchi, K.~Kiuchi, K.~Kyutoku and M.~Shibata, \emph{{Gravitational waves
  and neutrino emission from the merger of binary neutron stars}},
  \href{http://dx.doi.org/10.1103/PhysRevLett.107.051102}{\emph{Phys. Rev.
  Lett.} {\bf 107} (2011) 051102}, [\href{http://arxiv.org/abs/1105.2125}{{\tt
  1105.2125}}].

\bibitem{Ruiz:2016}
M.~{Ruiz}, R.~N. {Lang}, V.~{Paschalidis} and S.~L. {Shapiro}, \emph{{Binary
  Neutron Star Mergers: A Jet Engine for Short Gamma-Ray Bursts}},
  \href{http://dx.doi.org/10.3847/2041-8205/824/1/L6}{\emph{ApJ Lett.} {\bf
  824} (2016) L6}, [\href{http://arxiv.org/abs/1604.02455}{{\tt 1604.02455}}].

\bibitem{East:2016}
W.~E. {East}, V.~{Paschalidis}, F.~{Pretorius} and S.~L. {Shapiro},
  \emph{{Relativistic simulations of eccentric binary neutron star mergers:
  One-arm spiral instability and effects of neutron star spin}},
  \href{http://dx.doi.org/10.1103/PhysRevD.93.024011}{\emph{\prd} {\bf 93}
  (2016) 024011}, [\href{http://arxiv.org/abs/1511.01093}{{\tt 1511.01093}}].

\bibitem{Most2019}
E.~R. {Most}, L.~J. {Papenfort}, V.~{Dexheimer}, M.~{Hanauske}, S.~{Schramm},
  H.~{St{\"o}cker} et~al., \emph{{Signatures of Quark-Hadron Phase Transitions
  in General-Relativistic Neutron-Star Mergers}},
  \href{http://dx.doi.org/10.1103/PhysRevLett.122.061101}{\emph{\prl} {\bf 122}
  (2019) 061101}, [\href{http://arxiv.org/abs/1807.03684}{{\tt 1807.03684}}].

\bibitem{Bauswein2019}
A.~{Bauswein}, N.-U.~F. {Bastian}, D.~B. {Blaschke}, K.~{Chatziioannou}, J.~A.
  {Clark}, T.~{Fischer} et~al., \emph{{Identifying a First-Order Phase
  Transition in Neutron-Star Mergers through Gravitational Waves}},
  \href{http://dx.doi.org/10.1103/PhysRevLett.122.061102}{\emph{\prl} {\bf 122}
  (2019) 061102}, [\href{http://arxiv.org/abs/1809.01116}{{\tt 1809.01116}}].

\bibitem{Baiotti:2016qnr}
L.~Baiotti and L.~Rezzolla, \emph{{Binary neutron star mergers: a review of
  Einstein’s richest laboratory}},
  \href{http://dx.doi.org/10.1088/1361-6633/aa67bb}{\emph{Rept. Prog. Phys.}
  {\bf 80} (2017) 096901}, [\href{http://arxiv.org/abs/1607.03540}{{\tt
  1607.03540}}].

\bibitem{Faber2012:lrr}
J.~A. Faber and F.~A. Rasio, \emph{Binary neutron star mergers}, {\emph{Living
  Rev. Relativity} {\bf 15} (2012) }.

\bibitem{Alford2017}
M.~G. {Alford}, L.~{Bovard}, M.~{Hanauske}, L.~{Rezzolla} and K.~{Schwenzer},
  \emph{{Viscous Dissipation and Heat Conduction in Binary Neutron-Star
  Mergers}},
  \href{http://dx.doi.org/10.1103/PhysRevLett.120.041101}{\emph{Physical Review
  Letters} {\bf 120} (2018) 041101},
  [\href{http://arxiv.org/abs/1707.09475}{{\tt 1707.09475}}].

\bibitem{Harutyunyan:2016a}
A.~{Harutyunyan} and A.~{Sedrakian}, \emph{{Electrical conductivity of a warm
  neutron star crust in magnetic fields}},
  \href{http://dx.doi.org/10.1103/PhysRevC.94.025805}{\emph{Phys. Rev. C} {\bf
  94} (2016) 025805}, [\href{http://arxiv.org/abs/1605.07612}{{\tt
  1605.07612}}].

\bibitem{Harutyunyan:2018}
A.~{Harutyunyan}, A.~{Nathanail}, L.~{Rezzolla} and A.~{Sedrakian},
  \emph{{Electrical resistivity and Hall effect in binary neutron star
  mergers}}, \href{http://dx.doi.org/10.1140/epja/i2018-12624-1}{\emph{European
  Physical Journal A} {\bf 54} (2018) 191},
  [\href{http://arxiv.org/abs/1803.09215}{{\tt 1803.09215}}].

\bibitem{Roberts:2012um}
L.~F. Roberts, S.~Reddy and G.~Shen, \emph{{Medium modification of the charged
  current neutrino opacity and its implications}},
  \href{http://dx.doi.org/10.1103/PhysRevC.86.065803}{\emph{Phys. Rev.} {\bf
  C86} (2012) 065803}, [\href{http://arxiv.org/abs/1205.4066}{{\tt
  1205.4066}}].

\bibitem{Alford:2018lhf}
M.~G. Alford and S.~P. Harris, \emph{{Beta equilibrium in neutron star
  mergers}}, \href{http://dx.doi.org/10.1103/PhysRevC.98.065806}{\emph{Phys.
  Rev.} {\bf C98} (2018) 065806}, [\href{http://arxiv.org/abs/1803.00662}{{\tt
  1803.00662}}].

\bibitem{Sawyer1979ApJ}
R.~F. {Sawyer} and A.~{Soni}, \emph{{Transport of neutrinos in hot neutron-star
  matter}}, \href{http://dx.doi.org/10.1086/157146}{\emph{\apj} {\bf 230}
  (1979) 859--869}.

\bibitem{Sawyer1980ApJ}
R.~F. {Sawyer}, \emph{{Damping of neutron star pulsations by weak interaction
  processes}}, \href{http://dx.doi.org/10.1086/157858}{\emph{\apj} {\bf 237}
  (1980) 187--197}.

\bibitem{Sawyer1989}
R.~F. {Sawyer}, \emph{{Bulk viscosity of hot neutron-star matter and the
  maximum rotation rates of neutron stars}},
  \href{http://dx.doi.org/10.1103/PhysRevD.39.3804}{\emph{\prd} {\bf 39} (1989)
  3804--3806}.

\bibitem{Haensel1992PhRvD}
P.~{Haensel} and R.~{Schaeffer}, \emph{{Bulk viscosity of hot-neutron-star
  matter from direct URCA processes}},
  \href{http://dx.doi.org/10.1103/PhysRevD.45.4708}{\emph{\prd} {\bf 45} (1992)
  4708--4712}.

\bibitem{Dong2007}
H.~{Dong}, N.~{Su} and Q.~{Wang}, \emph{{Bulk viscosity in nuclear and quark
  matter}}, \href{http://dx.doi.org/10.1088/0954-3899/34/8/S63}{\emph{Journal
  of Physics G Nuclear Physics} {\bf 34} (2007) S643--S646},
  [\href{http://arxiv.org/abs/astro-ph/0702181}{{\tt astro-ph/0702181}}].

\bibitem{Alford2010JPhG}
M.~G. {Alford}, S.~{Mahmoodifar} and K.~{Schwenzer}, \emph{{Large amplitude
  behavior of the bulk viscosity of dense matter}},
  \href{http://dx.doi.org/10.1088/0954-3899/37/12/125202}{\emph{Journal of
  Physics G Nuclear Physics} {\bf 37} (2010) 125202},
  [\href{http://arxiv.org/abs/1005.3769}{{\tt 1005.3769}}].

\bibitem{Kolomeitsev2015}
E.~E. {Kolomeitsev} and D.~N. {Voskresensky}, \emph{{Viscosity of neutron star
  matter and r -modes in rotating pulsars}},
  \href{http://dx.doi.org/10.1103/PhysRevC.91.025805}{\emph{\prc} {\bf 91}
  (2015) 025805}, [\href{http://arxiv.org/abs/1412.0314}{{\tt 1412.0314}}].

\bibitem{Alford:2010jf}
M.~G. Alford and G.~Good, \emph{{Leptonic contribution to the bulk viscosity of
  nuclear matter}},
  \href{http://dx.doi.org/10.1103/PhysRevC.82.055805}{\emph{Phys. Rev.} {\bf
  C82} (2010) 055805}, [\href{http://arxiv.org/abs/1003.1093}{{\tt
  1003.1093}}].

\bibitem{Jones2001PhRvD}
P.~B. {Jones}, \emph{{Bulk viscosity of neutron-star matter}},
  \href{http://dx.doi.org/10.1103/PhysRevD.64.084003}{\emph{\prd} {\bf 64}
  (2001) 084003}.

\bibitem{Lindblom2002}
L.~{Lindblom} and B.~J. {Owen}, \emph{{Effect of hyperon bulk viscosity on
  neutron-star r-modes}},
  \href{http://dx.doi.org/10.1103/PhysRevD.65.063006}{\emph{\prd} {\bf 65}
  (2002) 063006}, [\href{http://arxiv.org/abs/astro-ph/0110558}{{\tt
  astro-ph/0110558}}].

\bibitem{Dalen2004PhRvC}
E.~N. {van Dalen} and A.~E. {Dieperink}, \emph{{Bulk viscosity in neutron stars
  from hyperons}},
  \href{http://dx.doi.org/10.1103/PhysRevC.69.025802}{\emph{\prc} {\bf 69}
  (2004) 025802}, [\href{http://arxiv.org/abs/nucl-th/0311103}{{\tt
  nucl-th/0311103}}].

\bibitem{Haensel2002}
P.~{Haensel}, K.~P. {Levenfish} and D.~G. {Yakovlev}, \emph{{Bulk viscosity in
  superfluid neutron star cores. III. Effects of $\Sigma^{-}$ hyperons}},
  \href{http://dx.doi.org/10.1051/0004-6361:20011532}{\emph{\aap} {\bf 381}
  (2002) 1080--1089}, [\href{http://arxiv.org/abs/astro-ph/0110575}{{\tt
  astro-ph/0110575}}].

\bibitem{Chatterjee2006}
D.~{Chatterjee} and D.~{Bandyopadhyay}, \emph{{Effect of hyperon-hyperon
  interaction on bulk viscosity and r-mode instability in neutron stars}},
  \href{http://dx.doi.org/10.1103/PhysRevD.74.023003}{\emph{\prd} {\bf 74}
  (2006) 023003}, [\href{http://arxiv.org/abs/astro-ph/0602538}{{\tt
  astro-ph/0602538}}].

\bibitem{Chatterjee2008ApJ}
D.~{Chatterjee} and D.~{Bandyopadhyay}, \emph{{Hyperon Bulk Viscosity in the
  Presence of Antikaon Condensate}},
  \href{http://dx.doi.org/10.1086/587956}{\emph{\apj} {\bf 680} (2008)
  686--694}, [\href{http://arxiv.org/abs/0712.3171}{{\tt 0712.3171}}].

\bibitem{Madsen1992}
J.~{Madsen}, \emph{{Bulk viscosity of strange quark matter, damping of quark
  star vibration, and the maximum rotation rate of pulsars}},
  \href{http://dx.doi.org/10.1103/PhysRevD.46.3290}{\emph{\prd} {\bf 46} (1992)
  3290--3295}.

\bibitem{Drago2005}
A.~{Drago}, A.~{Lavagno} and G.~{Pagliara}, \emph{{Bulk viscosity in hybrid
  stars}}, \href{http://dx.doi.org/10.1103/PhysRevD.71.103004}{\emph{\prd} {\bf
  71} (2005) 103004}, [\href{http://arxiv.org/abs/astro-ph/0312009}{{\tt
  astro-ph/0312009}}].

\bibitem{Alford:2006gy}
M.~G. Alford and A.~Schmitt, \emph{{Bulk viscosity in 2SC quark matter}},
  \href{http://dx.doi.org/10.1088/0954-3899/34/1/005}{\emph{J. Phys.} {\bf G34}
  (2007) 67--102}, [\href{http://arxiv.org/abs/nucl-th/0608019}{{\tt
  nucl-th/0608019}}].

\bibitem{Alford:2007rw}
M.~G. Alford, M.~Braby, S.~Reddy and T.~Schäfer, \emph{{Bulk viscosity due to
  kaons in color-flavor-locked quark matter}},
  \href{http://dx.doi.org/10.1103/PhysRevC.75.055209}{\emph{Phys. Rev.} {\bf
  C75} (2007) 055209}, [\href{http://arxiv.org/abs/nucl-th/0701067}{{\tt
  nucl-th/0701067}}].

\bibitem{Manuel:2007pz}
C.~Manuel and F.~J. Llanes-Estrada, \emph{{Bulk viscosity in a cold CFL
  superfluid}},
  \href{http://dx.doi.org/10.1088/1475-7516/2007/08/001}{\emph{JCAP} {\bf 0708}
  (2007) 001}, [\href{http://arxiv.org/abs/0705.3909}{{\tt 0705.3909}}].

\bibitem{SaD2007a}
B.~A. {Sa'd}, I.~A. {Shovkovy} and D.~H. {Rischke}, \emph{{Bulk viscosity of
  strange quark matter: Urca versus nonleptonic processes}},
  \href{http://dx.doi.org/10.1103/PhysRevD.75.125004}{\emph{\prd} {\bf 75}
  (2007) 125004}, [\href{http://arxiv.org/abs/astro-ph/0703016}{{\tt
  astro-ph/0703016}}].

\bibitem{SaD2007b}
B.~A. {Sa'd}, I.~A. {Shovkovy} and D.~H. {Rischke}, \emph{{Bulk viscosity of
  spin-one color superconductors with two quark flavors}},
  \href{http://dx.doi.org/10.1103/PhysRevD.75.065016}{\emph{\prd} {\bf 75}
  (2007) 065016}, [\href{http://arxiv.org/abs/astro-ph/0607643}{{\tt
  astro-ph/0607643}}].

\bibitem{Alford:2008pb}
M.~G. Alford, M.~Braby and A.~Schmitt, \emph{{Bulk viscosity in kaon-condensed
  color-flavor locked quark matter}},
  \href{http://dx.doi.org/10.1088/0954-3899/35/11/115007}{\emph{J. Phys.} {\bf
  G35} (2008) 115007}, [\href{http://arxiv.org/abs/0806.0285}{{\tt
  0806.0285}}].

\bibitem{Sinha2009}
M.~{Sinha} and D.~{Bandyopadhyay}, \emph{{Hyperon bulk viscosity in strong
  magnetic fields}},
  \href{http://dx.doi.org/10.1103/PhysRevD.79.123001}{\emph{\prd} {\bf 79}
  (2009) 123001}, [\href{http://arxiv.org/abs/0809.3337}{{\tt 0809.3337}}].

\bibitem{Wang2010PhRvD}
X.~{Wang} and I.~A. {Shovkovy}, \emph{{Bulk viscosity of spin-one color
  superconducting strange quark matter}},
  \href{http://dx.doi.org/10.1103/PhysRevD.82.085007}{\emph{\prd} {\bf 82}
  (2010) 085007}, [\href{http://arxiv.org/abs/1006.1293}{{\tt 1006.1293}}].

\bibitem{Huang2010}
X.-G. {Huang}, M.~{Huang}, D.~H. {Rischke} and A.~{Sedrakian},
  \emph{{Anisotropic hydrodynamics, bulk viscosities, and r-modes of strange
  quark stars with strong magnetic fields}},
  \href{http://dx.doi.org/10.1103/PhysRevD.81.045015}{\emph{\prd} {\bf 81}
  (2010) 045015}, [\href{http://arxiv.org/abs/0910.3633}{{\tt 0910.3633}}].

\bibitem{AlfordHan2015}
M.~G. {Alford}, S.~{Han} and K.~{Schwenzer}, \emph{{Phase conversion
  dissipation in multicomponent compact stars}},
  \href{http://dx.doi.org/10.1103/PhysRevC.91.055804}{\emph{\prc} {\bf 91}
  (2015) 055804}, [\href{http://arxiv.org/abs/1404.5279}{{\tt 1404.5279}}].

\bibitem{Schmitt2017}
A.~{Schmitt} and P.~{Shternin}, \emph{{Reaction rates and transport in neutron
  stars}}, {\emph{ArXiv e-prints} (2017) },
  [\href{http://arxiv.org/abs/1711.06520}{{\tt 1711.06520}}].

\bibitem{Haensel2000}
P.~{Haensel}, K.~P. {Levenfish} and D.~G. {Yakovlev}, \emph{{Bulk viscosity in
  superfluid neutron star cores. I. Direct Urca processes in npemu matter}},
  {\emph{\aap} {\bf 357} (2000) 1157--1169},
  [\href{http://arxiv.org/abs/astro-ph/0004183}{{\tt astro-ph/0004183}}].

\bibitem{Haensel2001}
P.~{Haensel}, K.~P. {Levenfish} and D.~G. {Yakovlev}, \emph{{Bulk viscosity in
  superfluid neutron star cores. II. Modified Urca processes in npe mu
  matter}}, \href{http://dx.doi.org/10.1051/0004-6361:20010383}{\emph{\aap}
  {\bf 372} (2001) 130--137},
  [\href{http://arxiv.org/abs/astro-ph/0103290}{{\tt astro-ph/0103290}}].

\bibitem{Gusakov2007}
M.~E. {Gusakov}, \emph{{Bulk viscosity of superfluid neutron stars}},
  \href{http://dx.doi.org/10.1103/PhysRevD.76.083001}{\emph{\prd} {\bf 76}
  (2007) 083001}, [\href{http://arxiv.org/abs/0704.1071}{{\tt 0704.1071}}].

\bibitem{Greiner2000gauge}
W.~Greiner and B.~M{\"u}ller, \emph{Gauge Theory of Weak Interactions}.
\newblock Physics and Astronomy Online Library. Springer, 2000.

\bibitem{glendenning2000compact}
N.~Glendenning, \emph{Compact Stars: Nuclear Physics, Particle Physics, and
  General Relativity}.
\newblock Astronomy and Astrophysics Library. Springer New York, 2000.

\bibitem{weber_book}
F.~Weber, \emph{Pulsars as astrophysical laboratories for nuclear and particle
  physics}.
\newblock Institute of Physics, Bristol, U.K., 1999.

\bibitem{Sedrakian2007}
A.~Sedrakian, \emph{The physics of dense hadronic matter and compact stars},
  {\emph{Prog. Part. Nucl. Phys.} {\bf 58} (2007) 168--246}.

\bibitem{Lalazissis2005}
G.~A. {Lalazissis}, T.~{Nik{\v s}i{\'c}}, D.~{Vretenar} and P.~{Ring},
  \emph{{New relativistic mean-field interaction with density-dependent
  meson-nucleon couplings}},
  \href{http://dx.doi.org/10.1103/PhysRevC.71.024312}{\emph{\prc} {\bf 71}
  (2005) 024312}.

\bibitem{Colucci2013}
G.~{Colucci} and A.~{Sedrakian}, \emph{{Equation of state of hypernuclear
  matter: Impact of hyperon-scalar-meson couplings}},
  \href{http://dx.doi.org/10.1103/PhysRevC.87.055806}{\emph{\prc} {\bf 87}
  (2013) 055806}, [\href{http://arxiv.org/abs/1302.6925}{{\tt 1302.6925}}].

\bibitem{Lalazissis1997}
G.~A. {Lalazissis}, J.~{K{\"o}nig} and P.~{Ring}, \emph{{New parametrization
  for the Lagrangian density of relativistic mean field theory}},
  \href{http://dx.doi.org/10.1103/PhysRevC.55.540}{\emph{\prc} {\bf 55} (Jan,
  1997) 540--543}, [\href{http://arxiv.org/abs/nucl-th/9607039}{{\tt
  nucl-th/9607039}}].

\bibitem{Prakash1997}
M.~{Prakash}, I.~{Bombaci}, M.~{Prakash}, P.~J. {Ellis}, J.~M. {Lattimer} and
  R.~{Knorren}, \emph{{Composition and structure of protoneutron stars}},
  \href{http://dx.doi.org/10.1016/S0370-1573(96)00023-3}{\emph{Physics Reports}
  {\bf 280} (Jan, 1997) 1--77},
  [\href{http://arxiv.org/abs/nucl-th/9603042}{{\tt nucl-th/9603042}}].

\bibitem{Malfatti2019}
G.~{Malfatti}, M.~G. {Orsaria}, G.~A. {Contrera}, F.~{Weber} and I.~F.
  {Ranea-Sand oval}, \emph{{Hot quark matter and (proto-) neutron stars}},
  \href{http://dx.doi.org/10.1103/PhysRevC.100.015803}{\emph{Physical Review C}
  {\bf 100} (Jul, 2019) 015803}, [\href{http://arxiv.org/abs/1907.06597}{{\tt
  1907.06597}}].

\bibitem{Weber2019}
F.~{Weber}, D.~{Farrell}, W.~M. {Spinella}, G.~{Malfatti}, M.~G. {Orsaria},
  G.~A. {Contrera} et~al., \emph{{Phases of Hadron-Quark Matter in (Proto)
  Neutron Stars}},
  \href{http://dx.doi.org/10.3390/universe5070169}{\emph{Universe} {\bf 5}
  (Jul, 2019) 169}, [\href{http://arxiv.org/abs/1907.06591}{{\tt 1907.06591}}].

\bibitem{Alford:2019zzz}
M.~Alford and S.~Harris, \emph{{Damping of density oscillations in
  neutrino-transparent nuclear matter}}, 
  \href{http://dx.doi.org/10.1103/PhysRevC.100.035803}{\emph{Phys. Rev. C}
  {\bf 100} (2019) 035803}, [\href{https://arxiv.org/abs/1907.03795}{{\tt
  1907.03795}}].

\bibitem{Yakovlev2018}
D.~G. {Yakovlev}, M.~E. {Gusakov} and P.~{Haensel}, \emph{{Bulk viscosity in a
  neutron star mantle}},
  \href{http://dx.doi.org/10.1093/mnras/sty2639}{\emph{\mnras} {\bf 481} (2018)
  4924--4930}, [\href{http://arxiv.org/abs/1809.08609}{{\tt 1809.08609}}].

\bibitem{Li2018_PLB}
J.~J.~{Li}, A.~{Sedrakian} and F.~{Weber}, \emph{{Competition between delta
  isobars and hyperons and properties of compact stars}}, {\emph{Phys. Lett. B}
  {\bf 783} (2018) 234--240}, [\href{http://arxiv.org/abs/1803.03661}{{\tt
  1803.03661}}].

\bibitem{Li2019ApJ}
J.~J. {Li} and A.~{Sedrakian}, \emph{{Implications from GW170817 for
  {\ensuremath{\Delta}}-isobar Admixed Hypernuclear Compact Stars}},
  \href{http://dx.doi.org/10.3847/2041-8213/ab1090}{\emph{\apjl} {\bf 874}
  (2019) L22}, [\href{http://arxiv.org/abs/1904.02006}{{\tt 1904.02006}}].

\bibitem{Chatterjee2007}
D.~{Chatterjee} and D.~{Bandyopadhyay}, \emph{{Bulk viscosity in kaon condensed
  matter}}, \href{http://dx.doi.org/10.1103/PhysRevD.75.123006}{\emph{\prd}
  {\bf 75} (Jun, 2007) 123006},
  [\href{http://arxiv.org/abs/astro-ph/0702259}{{\tt astro-ph/0702259}}].

\end{thebibliography}

\end{document}